\documentclass[twocolumn,twocolappendix,trackchanges]{aastex701}

\shorttitle{Optical GRB Redshift Classification}
\shortauthors{Sarkar et al.}

\usepackage{soul}
\usepackage{enumitem}
\usepackage{bm}
\usepackage{amsmath}
\usepackage{gensymb}
\usepackage{subfigure}
\usepackage{multirow}
\usepackage{array}
\usepackage{hhline}
\usepackage{hyperref}
\PassOptionsToPackage{hyphens}{url}\usepackage{hyperref}

\usepackage{xcolor} %
\definecolor{LightGray}{gray}{0.95}

\usepackage[normalem]{ulem}

\usepackage{CJK}

\begin{document}

\title{Redshift Classification of Optical Gamma-Ray Bursts using Supervised Learning}

\author[0009-0007-0050-9762]{Milind Sarkar}
\altaffiliation{These authors contributed equally to this work.}
\affiliation{Department of Physical Sciences, Indian Institute of Science Education and Research (IISER) Mohali, Punjab, India}
\email[]{milindsarkaryt@gmail.com}

\author[0000-0003-4442-8546]{Maria Giovanna Dainotti}
\altaffiliation{These authors contributed equally to this work.}
\affiliation{Division of Science, National Astronomical Observatory of Japan, 2-21-1 Osawa, Mitaka, Tokyo 181-8588, Japan}
\affiliation{Department of Astronomical Sciences, The Graduate University for Advanced Studies (SOKENDAI), Japan}
\affiliation{Space Science Institute, 4765 Walnut Street, Suite B, Boulder, CO 80301, USA}
\email[show]{maria.dainotti@nao.ac.jp}
\correspondingauthor{Maria Giovanna Dainotti}

\author[0009-0002-2068-3411]{Nikita S. Khatiya}
\affiliation{Department of Physics and Astronomy, Clemson University, Clemson, SC 29634, USA}
\email{}

\author[0009-0003-8022-8151]{Dhruv S. Bal}
\affiliation{Department of Physics and Astronomy, Clemson University, Clemson, SC 29634, USA}
\email{}

\author[0000-0002-0657-4342]{Malgorzata Bogdan}
\affiliation{University of Wroclaw, Wroclaw, Poland}
\email{}

\author[0000-0001-5931-2381]{Ye Li}
\affiliation{Purple Mountain Observatory, Chinese Academy of Sciences, Nanjing 210023, China}
\email{}

\author[0000-0003-3358-0665]{Agnieszka Pollo}
\affiliation{National Centre for Nuclear Research, 02-093 Warsaw, Poland}
\affiliation{Jagiellonian University, Krakow, Poland}
\email{}

\author[0000-0002-8028-0991]{Dieter H. Hartmann}
\affiliation{Department of Physics and Astronomy, Clemson University, Clemson, SC 29634, USA}
\email{}

\author[0000-0002-9725-2524]{Bing Zhang}
\affiliation{Nevada Center for Astrophysics, University of Nevada, Las Vegas, NV 89154, USA}
\email{}

\author[0000-0002-0173-6453]{Nissim Fraija}
\affiliation{Instituto de Astronomía, Universidad Nacional Autónoma de México, Circuito Exterior, C.U., A.P. 70-264, Mexico}
\email{}

\author[0000-0002-7738-6875]{J. Xavier Prochaska}
\affiliation{Division of Science, National Astronomical Observatory of Japan, 2-21-1 Osawa, Mitaka, Tokyo 181-8588, Japan}
\affiliation{University of California, Santa Cruz, 1156 High Street, Santa Cruz, CA 95064, USA}
\affiliation{Kavli Institute for the Physics and Mathematics of the Universe (Kavli IPMU), 5-1-5 Kashiwanoha, Kashiwa 277-8583, Japan}
\email{}

\begin{abstract}
Modern satellite missions localize many gamma-ray bursts (GRBs), but
spectroscopic redshifts remain scarce because rapid optical follow-up is
often unavailable. Optical plateau classification differs fundamentally
from its X-ray counterpart: heterogeneous ground-based follow-up, dust
extinction, and chromatic afterglow behavior introduce greater incompleteness than in the
X-ray data. We present a supervised machine-learning classification of
long-duration GRBs (duration $>$ 2 s)
with measured redshifts and possessing optical plateau emission. As
predictors, we combine plateau and prompt-emission observables.
Using 171 such GRBs, we evaluate binary classification
--- whether a GRB is
high- or low-$z$ --- at thresholds
$z_t = 2.0, 2.5$, and $3.0$. The best classifier, at $z_t = 2.0$,
achieves a training-set true positive rate (TPR) of 0.741 and an area under
the curve (AUC) of 0.841, with a train-to-test AUC
difference within 4 percent. The novelty of this work is a classifier built on an optical-plateau-selected sample. Prompt-emission and absorption variables are favored over plateau parameters at
$z_t = 2.0$ on the raw (complete-case) data with the Least Absolute Shrinkage and Selection Operator (LASSO), thus allowing a reliable classifier without waiting for the plateau signatures in the light curve. The X-ray absorption column density, $\log(\mathrm{NH})$, and
the prompt-emission peak flux, $\log(\mathrm{PeakFlux})$, are stably selected by both the optical and X-ray
classifiers. We publicly release a web-based application for the optical classifier, complementing the available X-ray classifier \citep{dainotti2025grbredshiftclassifierfollowup}, to help the community rapidly select high-$z$ GRBs for follow-up.

\end{abstract}


\section{Introduction}
\label{sec:intro}
Gamma-ray bursts (GRBs) are among the most spectacular events in high-energy astrophysics as they release in a few seconds the same amount of energy as our Sun releases in its entire lifetime.
From an observational perspective, a GRB consists of two distinct phases. 
The first is the prompt emission, which is typically detected in $\gamma$-rays, hard X-rays, soft X-rays, and occasionally in the optical band  \citep{Vestrand2005Natur,Oganesyan_2019, Shen_2009}.
The second phase is the afterglow, observed as the counterpart emission in hard and soft X-rays, optical wavelengths, and sometimes in the radio band \citep{Vestrand2005Natur,Beskin2010ApJ,2012MNRAS.421.1874G,2014Sci...343...38V,2014IJMPD..2330002Z, zhang_2018, costa1997,vanParadijs1997,Piro1998,2015PhR...561....1K}.
GRBs are classified into short-
and long-duration events on the basis of $T_{90}$, the interval over
which the burst emits from 5\% to 95\% of its total background-subtracted
$\gamma$-ray fluence \citep{Kouveliotou1993}. Short GRBs have
$T_{90} < 2$ s and are associated with the coalescence of compact
objects \citep{Eichler1989, 1992ApJ...395L..83N, Berger2014}, whereas long GRBs (LGRBs), with $T_{90} > 2$ s, are associated
with the core collapse of massive stars \citep{Woosley1993}. The 2 s division is an observational rather than a sharp
physical boundary, and depends on the energy band and sensitivity of the
detecting instrument; classifications based on progenitor type have
therefore also been proposed \citep{Zhang2004IJMPA..19.2385Z, Zhang2006}. All
bursts considered in this work are LGRBs.

Redshift is one of the most critical pieces of information in general for tackling population studies in extragalactic astronomy. For GRBs specifically, redshift enables studies of rest-frame energetics, luminosity functions, and rate evolution, and it is essential for any attempt to use GRBs as probes of the distant Universe. However, in practice, a large fraction of LGRBs still lack secure spectroscopic redshifts. This incompleteness is primarily driven by the difficulty of obtaining timely and sufficiently deep follow-up, especially when afterglows are faint, rapidly evolving, or strongly affected by extinction. As a result, an increasingly common goal is to infer whether a burst is likely to lie above a redshift threshold of interest using early-time observables, so that limited follow-up resources (spectroscopy and near-infrared imaging) can be prioritized toward the most promising high-$z$ candidates.

A prominent feature often observed in the light curves (LCs) of GRBs is the plateau phase, which is characterized by a relatively constant flux preceding the subsequent decay of the afterglow \citep{OBrien2006, Zhang2006, Nousek2006, Sakamoto2007, Liang2007, Dainotti2008, Zaninoni2013, Rowlinson2014}. 
Plateau phases are detected in approximately $42\%$ of X-ray afterglows \citep{Evans2009, Li2018b, Srinivasaragavan2020, 2021ApJS..255...13D,Narendra_2025}, $30\%$ of optical afterglows \citep{Vestrand2005Natur, Kann2006, Zeh2006, panaitescu2008taxonomy, panaitescu2011optical, Oates2012, dainotti2020b, Dainotti2022}, and in $6.6\%$ of a subset of GRBs with radio afterglows \citep{2022ApJ...925...15L}.

Importantly, optical afterglows show a wide range of temporal morphologies, including peaks (which are rare but, when observable, can shed light on the early emission processes in this band), plateaus, and breaks. The reported incidence of optical plateaus depends on the sample and on the criteria used to identify a plateau: the quoted $30\%$  above is drawn from the compilation of \citet{Dainotti2022}.
A central point of this work is that optical plateau classification constitutes a distinct classification problem from X-ray plateau classification due to both observational systematics and potential chromatic behavior of afterglow emission. Optical plateau measurements are shaped by heterogeneous follow-up cadence, depth, and instrument response, introducing strong selection effects and incomplete temporal coverage. These challenges are more pronounced than in X-ray afterglow samples, as demonstrated by studies of optical detectability and redshift bias~\citep{Coward_2009,Coward_2013}. 

In addition, dust extinction in GRB host environments can substantially
suppress optical fluxes, modify apparent temporal slopes, and produce
optically ``dark'' bursts despite detectable high-energy afterglows,
indicating that optical subsamples are subject to incompleteness and environment-dependent selection \citep{Covino_2013,Schady_2012,Greiner2011}.
Such bursts were originally identified simply as those
with an X-ray afterglow but no detected optical counterpart
\citep{Fynbo2001}, and are conventionally delimited by a brightness
limit at a fixed epoch, typically $R \gtrsim 23$ mag within 12--24 h of
the trigger \citep{Djorgovski2001}. This criterion is explicitly tied to
the depth and reaction time of the follow-up campaign, rather than to any
intrinsic property of the burst, and \citet{Fynbo2001} showed that
$\gtrsim 75\%$ of bursts with unsuccessful optical searches were
consistent with non-detection had they simply been as faint as the
dimmest detected afterglows. Darkness is, therefore, in large part a
statement about observational coverage, which is the same selection effect that governs whether an optical plateau can be characterized at
all, and which motivates the treatment of missingness adopted here.

Broadband analyses further show that optical/UV and X-ray afterglows often exhibit differing temporal behaviors, implying that optical plateau parameters do not necessarily trace the same emission component as X-ray plateaus in every event \citep{Oates_2011,Zaninoni2013}. Together, heterogeneous sampling, extinction-driven incompleteness, and chromatic evolution motivate treating optical plateau classification as a fundamentally optical-specific problem requiring dedicated modeling and validation.

\citet{morgan2012, ukwatta2016machine} have previously employed supervised machine learning (ML) techniques to identify high-$z$ GRBs using prompt $\gamma$-ray or early X-ray observables.
{
\citet{morgan2012} attempted to classify 135 \textit{Swift} \citep{Gehrels2004ApJ...611.1005G} LGRBs into high-$z$ ($z>4$) and low-$z$ events, using the Random Forest algorithm \citep{hastie2009elements, breiman2001randomforest}.}
Based on this, they claim that their top 40\% of predicted candidates capture $\approx$84\% of high-$z$ bursts.
\citet{ukwatta2016machine} performed a Random Forest regression analysis using \textit{Swift} LGRBs and achieved a Pearson correlation ($r$) of 0.57.

More recently, \citet{Dainotti_2024_ApJS} used a regression (estimation) ML-based analysis to determine the redshift of \textit{Swift} LGRBs, using a subset of those exhibiting an X-ray plateau, and achieved an $r=0.93$ including bias correction. 
Using the trained ML model, they predicted the redshift of 150 LGRBs.
\citet{Dainotti2024ApJ...967L..30D} used a subset of LGRBs exhibiting an optical plateau and also achieved an $r=0.93$, including bias correction. 
They also demonstrated the feasibility of using ML-based redshift estimates in determining the luminosity function and density rate evolution of LGRBs.
Most recently, \citet{dainotti2025grbredshiftclassifierfollowup} (hereafter MGD25) developed a SuperLearner-based binary redshift classifier for LGRBs using X-ray plateau and prompt-emission features, achieving area under the curve (AUC) values up to $\sim$0.85. Our work adopts the same statistical framework — outlier treatment, Multivariate Imputation by Chained Equations (MICE) imputation, Least Absolute Shrinkage and Selection
Operator (LASSO) feature selection, and SuperLearner ensembling — but applies it to the optical regime, which, as argued above, poses a qualitatively different missing-data problem. This shared methodology is a deliberate design choice: by holding the core pipeline largely fixed (see Section~\ref{subsec:comparison} for the remaining differences) and varying the wavelength regime, we can attribute differences in classification performance to the underlying observational physics rather than to modeling choices.
A central novelty of our study is the construction of a redshift classifier on a sample defined by the detection of an optical plateau, with optical plateau-phase variables included in the feature pool alongside prompt-emission and absorption observables. We develop and test a supervised ML framework to classify the redshift of these LGRBs with measured optical plateau and prompt-emission observables.
A similar technique has only been used in MGD25 for classifying high-$z$ and low-$z$ GRBs, and in \citet{Narendra_2025} for $z$-estimation via regression, both using X-ray plateau-phase features.
Our approach is designed to be robust to the challenges characteristic of optical plateau datasets: it addresses incomplete data via multivariate imputation, reduces dimensionality via feature selection, and combines multiple base learners through an ensemble strategy.  Our results provide an empirical assessment of how well optical plateau properties contribute to redshift classification when combined with prompt information, a characterization of which observables are most predictive in the optical plateau regime, and an evaluation of classifier behavior under practical constraints imposed by optical sample heterogeneity.

The paper is organized as follows. Section \ref{sec:sample} describes the LGRB sample with optical plateau measurements, defines the features used in this work, and summarizes the observational characteristics and missingness patterns of the dataset. Section \ref{sec:methods} formulates the classification problem and outlines the modeling approach and validation procedure. Section \ref{sec:results} presents the main classification results and robustness checks. Section \ref{sec:discussion} discusses implications for follow-up prioritization, limitations, and the relationship of optical plateau classification to broader multi-wavelength GRB studies. Section \ref{sec:generalisation} applies the companion X-ray classifier of MGD25 to an independent generalization set (263 LGRBs from \citet{Narendra_2025} that lacked spectroscopic redshifts in that compilation and were not used to train either classifier), as a check on the shared analysis pipeline rather than on the optical classifier itself. Section \ref{sec:conclusion} summarizes our conclusions.

\section{Optical Data Sample}
\label{sec:sample}

In this study, we use a sample of 171 LGRBs compiled by \citet{Dainotti2024ApJ...967L..30D}, comprising bursts with measured optical plateau properties and spectroscopic redshifts. 
These LGRBs were observed by the \textit{Swift} Ultraviolet/Optical Telescope (UVOT) as well as by a large network of ground-based facilities (455 telescopes and detectors), including the Subaru Telescope, the Gamma-Ray Burst Optical/Near-infrared Detector (GROND), the Reionization And Transients InfraRed camera (RATIR), and the Multicolor Imaging Telescopes for Survey and Monstrous Explosions (MITSuME). 
Optical plateau datasets are affected by heterogeneous follow-up strategies, varying photometric depth, extinction-driven attenuation, and gaps in temporal coverage, which complicate the direct application of standard supervised learning techniques \citep[e.g.,][]{Coward_2009,Coward_2013,Covino_2013,Greiner2011}. 
The multi-instrument nature of this dataset reflects the typical heterogeneity of optical GRB follow-up campaigns, where cadence, depth, and filter coverage vary substantially from burst to burst.

Optical afterglow detectability and measurement quality also depend on follow-up response time, observing conditions, and host dust extinction, which introduce selection effects and incomplete coverage \citep[e.g.,][]{Coward_2009,Coward_2013,Covino_2013,Greiner2011}. 
These characteristics motivate explicit treatment of missing values and robustness to extreme measurements in the modeling framework.

The dataset contains the following observed quantities:

\begin{enumerate}
    \item $z$ --- spectroscopic redshift of the GRB. This is the target variable from which the binary class labels are derived, and is not used as a predictor.
    \item $T_{90}$ --- duration over which 90\% of the prompt $\gamma$-ray fluence is emitted, characterizing the prompt emission timescale.
    \item $F_{a}$ --- observed optical flux density at the end of the plateau phase, derived from optical LC fitting and subject to bandpass and extinction effects.
    \item $T_{a}$ --- observer-frame time marking the termination of the optical plateau phase.
    \item $\alpha$ --- optical temporal decay index describing the post-plateau afterglow decline.
    \item $\beta$ --- optical spectral index during the plateau phase, assuming a power-law spectral energy distribution.
    \item NH --- the line-of-sight neutral hydrogen column density, obtained from time-averaged spectral fitting of the Swift X-ray Telescope (XRT) \citep{2005SSRv..120..165B} data rather than from the optical LC. We adopt the same quantity as MGD25: the X-ray Galactic column in the direction of the burst is computed with NHFtool and frozen, and the spectrum is fitted with phabs*phabs*pow, the second absorption component being free to vary. Both components are evaluated in the observer frame, so $N_H$ is determined without reference to the burst redshift and is available for bursts whose redshift is unknown — a prerequisite for applying the classifier. It serves as a proxy for absorption along the line of sight and has been reported to increase with redshift in \emph{Swift} GRB samples \citep{Grupe_2006,Campana_2012}.
    \item Fluence --- prompt-emission energy fluence measured over $T_{90}$ in erg cm$^{-2}$.
    \item PeakFlux --- prompt-emission peak photon flux measured by the \textit{Swift} Burst Alert Telescope (BAT) in photons cm$^{-2}$ s$^{-1}$ \citep{barthelmy2005burst}.
    \item PhotonIndex --- photon index of the prompt $\gamma$-ray spectrum obtained from power-law modeling of BAT data.
\end{enumerate}

Nine of these quantities serve as predictive features and fall into three groups based on how they are measured. Four ($T_{90}$, Fluence, PeakFlux, PhotonIndex) describe the prompt $\gamma$-ray emission and are derived from BAT data. Four ($T_a$, $F_a$, $\alpha$, $\beta$) characterize the optical plateau and afterglow evolution and are derived from ground-based and UVOT optical LCs. The remaining quantity, NH, is an absorption measurement obtained from X-ray spectroscopy, and therefore belongs to neither of the first two groups: although it bears directly on optical detectability, it is not itself an optical measurement and is not subject to the bandpass, extinction, and cadence effects that shape the optical plateau parameters. This distinction becomes relevant in Section \ref{subsec:comparison2}, where we find that the features retained most consistently by feature selection are precisely those measured from high-energy rather than optical data.
The optical plateau parameters are taken from \citet{Dainotti2022}, where the plateau phase is identified through broken power-law modeling of the optical LCs.

Given the wide dynamic range of GRB observables, we apply base-10 logarithmic transformations to $T_{90}$, Fluence, PeakFlux, NH, $F_a$, and $T_a$. 
Logarithmic scaling improves numerical stability and reflects the approximately power-law nature of many GRB distributions.

Following the procedure of \citet{Dainotti_2024_ApJS}, we assign N/A to the values where $\log(\text{NH})<20$, $\text{PhotonIndex} < 0$, $\text{PeakFlux} = 0$, $\alpha>3$, and $\beta>3$.
These thresholds exclude extreme or poorly constrained values that lie in the tails of the observed distributions and may arise from heterogeneous follow-up or low signal-to-noise measurements \citep[e.g.,][]{Coward_2009,Greiner2011}. 
Rather than discarding entire bursts, we treat these values as missing and later address them with multivariate imputation (Appendix~\ref{sec:mice}), thereby retaining the maximum possible sample size while mitigating the influence of outliers.
Imputing these values with
MICE increases the sample by 63\% (from 90 to 147 GRBs when the M-estimator is applied).
For the response variable, we adopt $\log(z+1)$, which provides a natural parametrization of redshift and improves learning performance relative to raw $z$. 
Binary class labels are subsequently defined using redshift thresholds $z_t$ (see Sections~\ref{sec:Metrics} and \ref{sec:results}). 

The preprocessing choices described above generate the four datasets analyzed in this work, which differ in size because they retain different subsets of the original 171 bursts. Requiring every feature to be measured and discarding measurements with relative error exceeding unity leaves 97 GRBs; applying the M-estimator to this complete-case sample removes a further 7, leaving 90. Applying MICE to the filtered sample, rather than discarding the bursts with missing values, yields 157 GRBs, and applying MICE after the M-estimator yields 147 (Figure~\ref{Fig:flowchart}). These four sample sizes are constant across the three redshift thresholds; only the division into high- and low-$z$ classes changes with $z_t$, as set out in Table \ref{tab:GRBdata}.

\section{Methodology}
\label{sec:methods}

The observational effects described in Section~\ref{sec:sample} introduce additional scatter and incompleteness in optical plateau measurements and motivate the use of a robust statistical framework for redshift classification.

To address the challenges of incompleteness and heterogeneity, we employ a supervised-learning pipeline consisting of four main stages: robust outlier treatment, multivariate imputation of missing values, feature selection, and ensemble classification. The overall framework follows the statistical learning methodology adopted for the X-ray plateau classifier presented in MGD25. By intentionally maintaining a consistent ML framework across the optical and X-ray studies, we can attribute differences in classification performance primarily to the underlying observational properties of the two wavelength regimes rather than to methodological variations. This approach enables a direct assessment of the extent to which optical plateau observables provide independent information for identifying high-redshift GRBs.

\begin{figure*}[htbp]
\centering
   \includegraphics[width=0.9\textwidth,height=0.9\textheight]{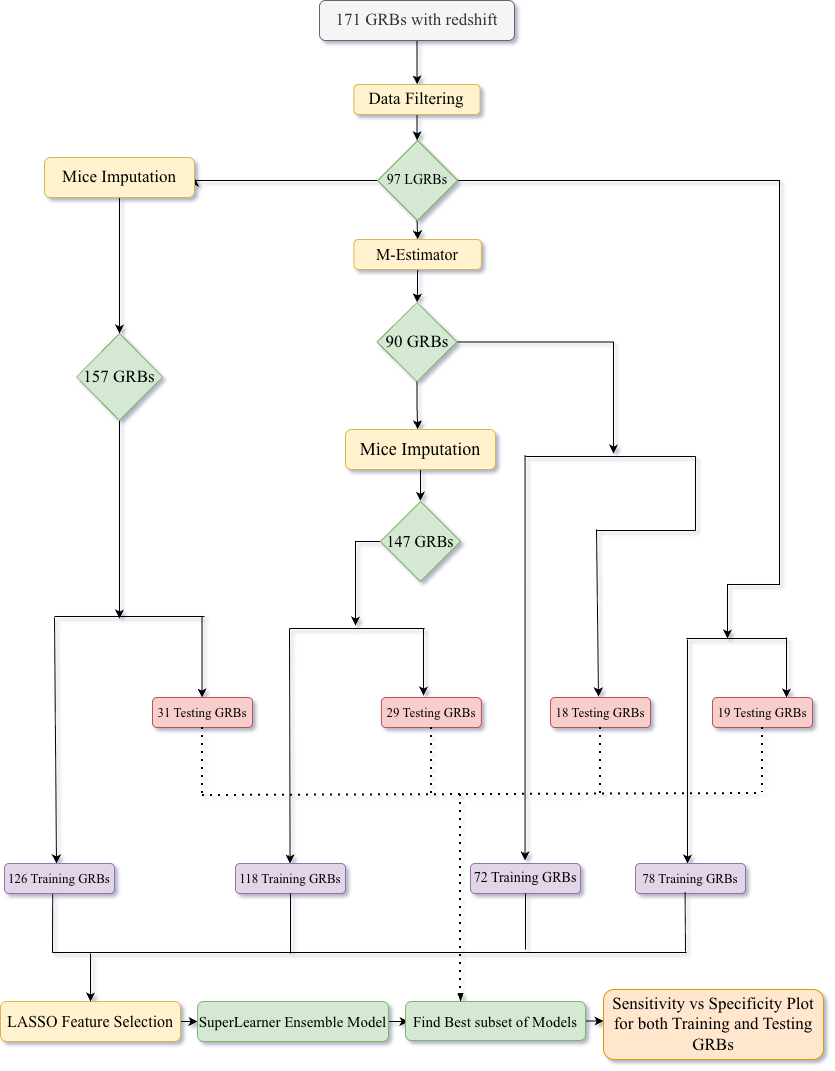}
\caption{Flowchart detailing each pipeline step, from the initial data to the SuperLearner ensemble model. 
Yellow boxes show the data engineering technique applied in the pipeline.
Green diamonds display the total number of GRBs after each step. 
Violet boxes and red boxes show the number of GRBs in the training and test sets, respectively. 
Green boxes highlight the steps involved in the model construction. 
The orange box represents the outcome of the constructed model applied to both the training and test sets. }
\label{Fig:flowchart}
\end{figure*}

Figure~\ref{Fig:flowchart} summarizes the complete analysis pipeline. First, extreme or poorly constrained measurements are identified using a robust M-estimator procedure. Missing values are subsequently treated through MICE, allowing us to retain a larger fraction of the available sample. Relevant predictors are then selected using the LASSO, which reduces dimensionality and mitigates overfitting. Finally, the selected features are used to train a SuperLearner ensemble that combines multiple parametric, semi-parametric, and non-parametric algorithms through cross-validation.

This framework is designed to maximize sample retention. Since the primary goal of this work is to investigate the redshift-classification capabilities of optical plateau observables, only a brief overview of each methodological component is presented in the main text. Complete technical descriptions, mathematical formulations, algorithmic implementations, and validation procedures are provided in Appendix~\ref{sec:methodology}.

\subsection{Outlier Treatment}

Optical plateau measurements are affected by heterogeneous observational conditions and occasionally contain extreme values that can disproportionately influence supervised-learning models. To mitigate their impact, we employ a robust M-estimator approach \citep{huber1964,huber1996robust,DEMENEZES2021107254}, following the methodology adopted in \citet{Dainotti2024ApJ...967L..30D}. The M-estimator iteratively downweights observations with large residuals relative to a robust regression model, thereby reducing sensitivity to outliers while preserving the majority of the sample.

The regression model is constructed using a subset of the prompt-emission, absorption, and optical plateau observables (Equation~\ref{eq:mestimator}) and is used to assign a weight to each GRB. Objects with weights below a predefined threshold are classified as outliers and removed from the sample before the train--test split. To assess the impact of outlier removal on classification performance, all analyses are performed both with and without the M-estimator filtering.

The robust regression model used for outlier identification is

\begin{equation}
\log(1+z)=f(\mathrm{PeakFlux},\mathrm{NH},F_a,T_a,\alpha,\beta,\mathrm{PhotonIndex}),
\label{eq:mestimator}
\end{equation}

where the explicit functional form is given in Appendix~\ref{sec:Mestimator}. Additional implementation details, weight distributions, threshold selection, and diagnostic tests are also provided in Appendix~\ref{sec:Mestimator}.

\subsection{Data Imputation}

Unlike X-ray observations obtained with relatively uniform space-based instrumentation, optical plateau measurements often come from a large network of facilities with varying sensitivities and temporal coverage, leading to missing data. Discarding GRBs with incomplete measurements would substantially reduce the available sample and potentially bias the inferred relationships between optical observables and redshift.

To preserve the largest possible sample while minimizing biases associated with complete-case analyses, we employ MICE \citep{schafer2002missing,van2011mice}, following previous applications to GRB and active galactic nuclei (AGN) datasets \citep{gibson2022,10.1093/mnras/stad2593,Dainotti_2024_ApJS}. MICE iteratively reconstructs missing values using information contained in the observed variables under the assumption that the data are missing at random (MAR) \citep{rubin1976inference}, an assumption which is based on the fact that the orbital gaps appear randomly and consequently the time and flux of the plateau can be randomly affected by them. The imputation procedure is performed using predictive mean matching (PMM) with the \texttt{midastouch} implementation and iterated until convergence. 
The consistency between the observed and imputed distributions is evaluated using Kolmogorov--Smirnov (KS) tests. The tests indicate no statistically significant differences between the original and reconstructed data distributions; as discussed in Appendix \ref{sec:mice}, this serves as a diagnostic that the imputation does not distort the distribution of each individual feature. 
The dataset is randomly split into 80\% training and 20\% test subsets to evaluate out-of-sample predictive performance, while cross-validation within the training set is used for model fitting and ensemble weighting. 

We also investigated Synthetic Minority Over-sampling Technique (SMOTE) methods \citep{chawla2002smote} to mitigate class imbalance. However, the introduction of synthetic GRBs did not improve classifier performance and altered the intrinsic properties of the optical sample. Consequently, SMOTE-based datasets were not adopted in the final analysis. Complete details regarding the MICE implementation, distribution tests, and SMOTE experiments are provided in Appendix~\ref{sec:mice}.

\subsection{Feature Selection}

Optical plateau samples contain a diverse set of prompt-emission, afterglow, and plateau observables, many of which are correlated and may not contribute equally to redshift discrimination. Identifying the most informative variables is therefore essential for constructing a physically interpretable classifier and reducing the risk of overfitting in a limited-sample regime. Similar considerations have motivated feature-selection approaches in previous GRB ML studies (e.g., MGD25).

To identify the parameters most strongly associated with redshift, we employ the LASSO \citep{TibshiraniLasso}, which performs simultaneous coefficient estimation and variable selection through regularization. Features consistently assigned non-negligible importance are retained and used as inputs for the classification models. This procedure enables the classifier to focus on the subset of optical plateau and prompt-emission properties that carry the strongest redshift-discriminatory information while reducing redundancy among predictors.

The resulting feature set forms the basis of the SuperLearner ensemble described below. Complete details regarding the LASSO implementation, threshold selection, stability tests, and feature rankings are provided in Appendix~\ref{sec:LASSO}.

\subsection{Algorithms}

The relationship between optical plateau observables and redshift is unlikely to be adequately described by a single statistical model owing to the heterogeneous nature of optical follow-up observations and the potentially nonlinear dependence of redshift on prompt-emission and afterglow properties. We therefore adopt an ensemble-learning approach based on the SuperLearner framework \citep{vanderLaanPolleyHubbard+2007}, which combines multiple algorithms spanning parametric, semi-parametric, and non-parametric model classes.

The ensemble includes linear models, generalized linear models, discriminant analysis methods, regularized regression, tree-based learners, kernel methods, and support vector machines. By combining models with different inductive biases, the SuperLearner framework can capture both linear and nonlinear relationships while reducing sensitivity to the assumptions of any individual algorithm. This approach has been successfully applied in previous GRB ML studies, including the X-ray plateau classifier of MGD25.

Model weights are determined through cross-validation, and the final prediction is obtained as a weighted combination of the individual learners. The algorithms included in the ensemble are listed in Appendix~\ref{sec:algorithms}, and their full descriptions are given in MGD25.

\subsection{SuperLearner}
\label{sec:SL}

\begin{figure*}[htbp]
\centering
\includegraphics[width = 1\textwidth]{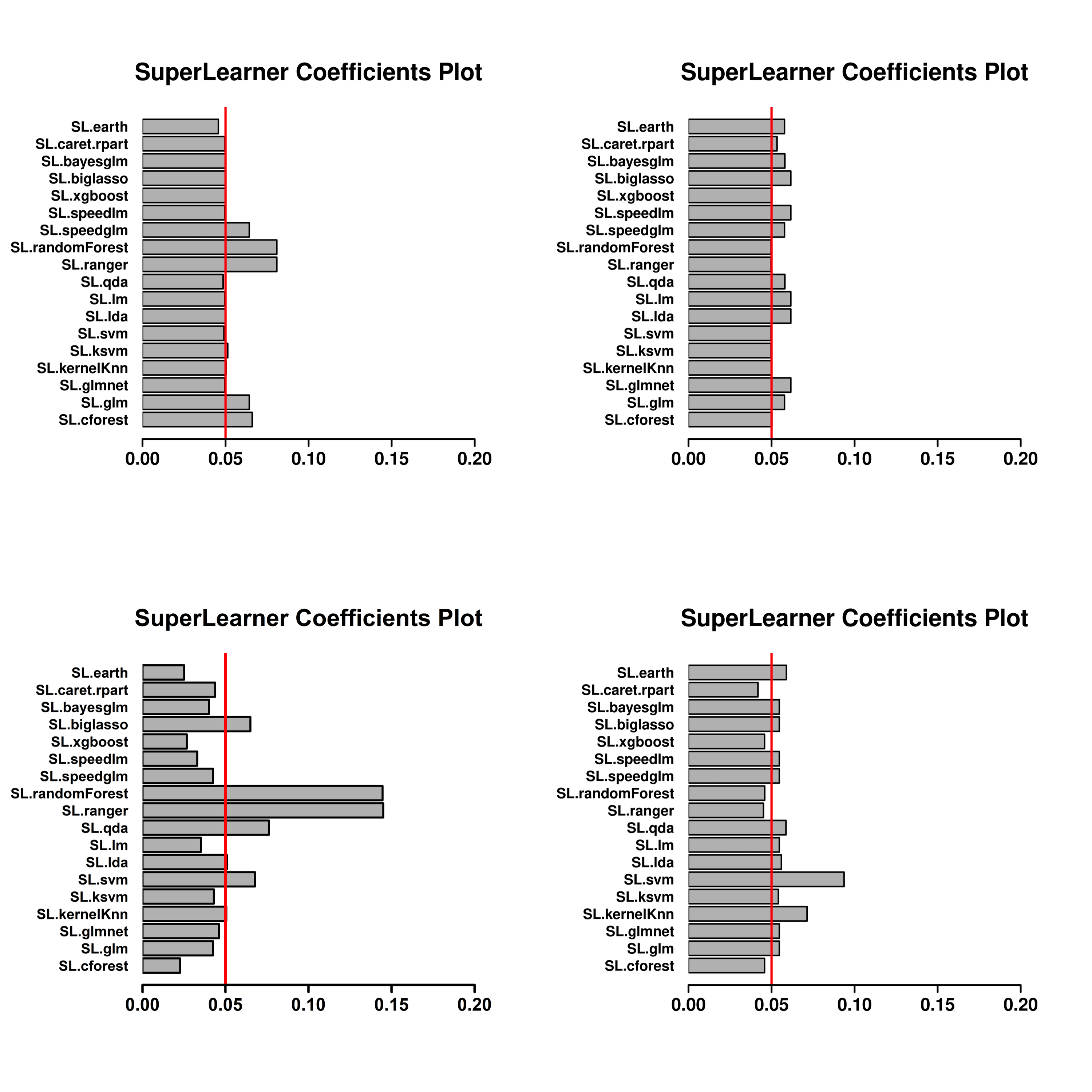}
\caption{Weights assigned to each algorithm by SuperLearner for $z_{t}$ = 3.0 with the raw dataset without the M-estimator (top left), the raw dataset with the M-estimator (top right), the MICE-imputed dataset without the M-estimator (bottom left), and the MICE-imputed dataset with the M-estimator (bottom right). Algorithms with weights $>$ 0.05 are selected as the best algorithms.}
\label{Fig:SL weights}
\end{figure*}

SuperLearner \citep{vanderLaanPolleyHubbard+2007} is an ensemble learning algorithm that combines multiple models to improve predictive performance. 
It uses 10-fold Cross-Validation (10fCV) to assess the model’s performance by partitioning the dataset into ten subsets, training the model on nine and testing on the remaining one, then averaging the results. Each base learner algorithm is assigned a weight based on its Root Mean Squared Error, with the total sum of weights equaling 1 \citep{polley2010super}.
We implemented SuperLearner with 10fCV repeated $n$ times ($n$ = 100, 200, 300, 500, and 1000). Our results showed a stable training-set AUC across runs, so we chose $n = 100$ to be consistent with prior studies \citep{dainotti2021predicting, narendra2022predicting, Dainotti_2024_ApJS}. 
We discarded models with weights below 0.05 after 100 iterations to ensure stability and enhance performance, as models with weights above this threshold exhibited the strongest predictive power. 
Figure \ref{Fig:SL weights} shows an example of the weights for $z_t = 3.0$, where $z_t$ is the threshold redshift.

\subsection{Metrics}
\label{sec:Metrics}

\begin{figure*}[htbp]
\centering
    \includegraphics[width = 0.7\textwidth]{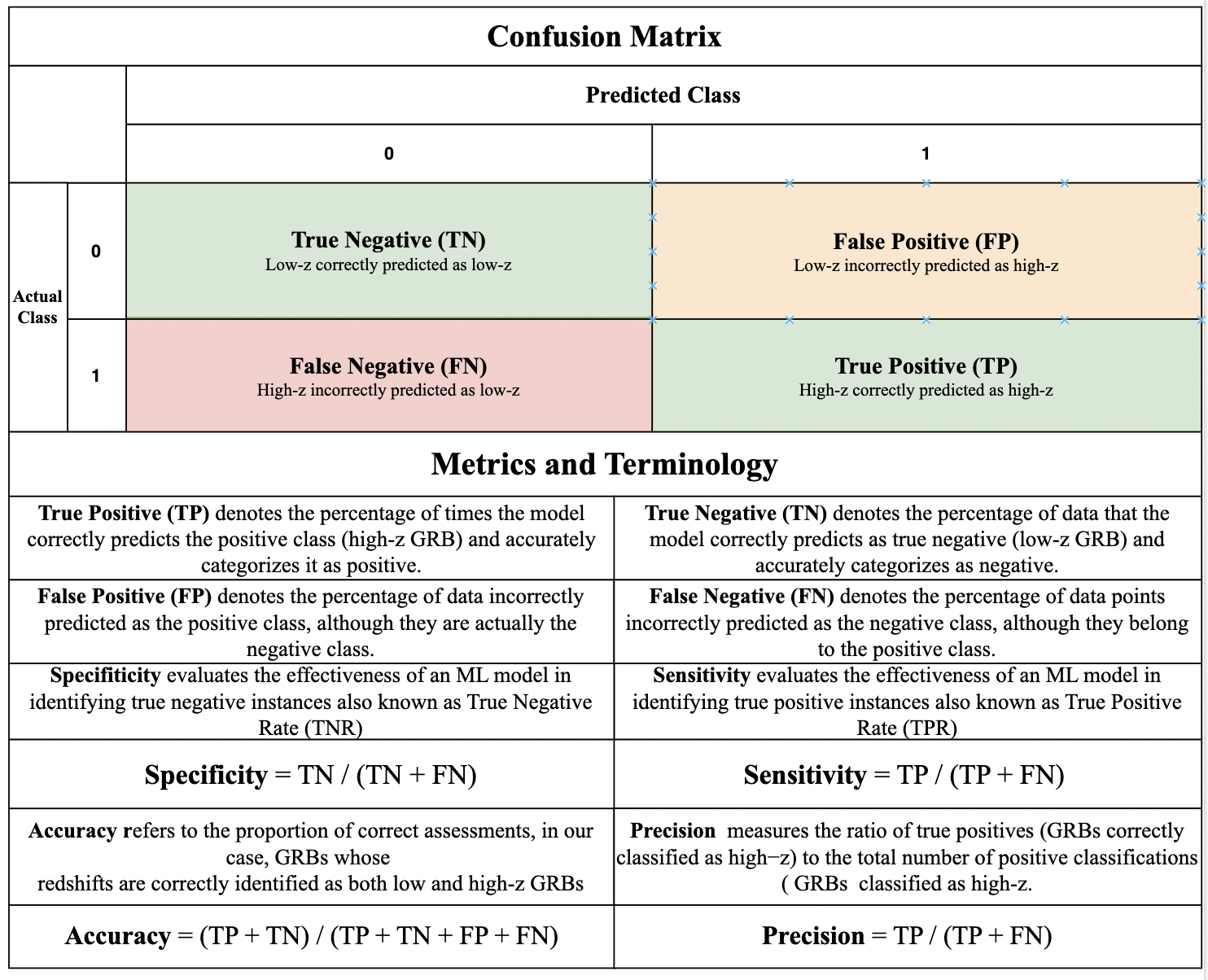}
\caption{Confusion matrix in relation to our data, where class 0 refers to low-$z$ and class 1 to high-$z$ GRBs.}
\label{Fig:LegConfusionMatrix}
\end{figure*}

This section outlines the terminology and classification metrics used to evaluate results and classifier performance. In this work, GRBs with high-$z$ are assigned to class ``1" and GRBs with low-$z$ to class ``0". 
The confusion matrix, metrics, and terminology plot in Figure \ref{Fig:LegConfusionMatrix} summarizes four key definitions for these classes.
The receiver operating characteristic (ROC) curve provides a visual representation of a classifier's performance across varying classification thresholds. In this ROC space, the x-axis represents 1$-$specificity (the false positive rate), while the y-axis represents sensitivity. 

Sensitivity, also known as the true positive rate (TPR), is defined as
\begin{equation}
    \mathrm{TPR} = \frac{\mathrm{TP}}{\mathrm{TP} + \mathrm{FN}},
\end{equation}
where TP and FN are the number of true positives and false negatives, respectively. Specificity, also known as the true negative rate (TNR), is defined as
\begin{equation}
    \mathrm{TNR} = \frac{\mathrm{TN}}{\mathrm{TN} + \mathrm{FP}},
\end{equation}
where TN and FP are the number of true negatives and false positives, respectively. Here, a true positive (TP) is a high-$z$ GRB (class ``1") that the classifier correctly labels as high-$z$; a false negative (FN) is a high-$z$ GRB that the classifier incorrectly labels as low-$z$; a true negative (TN) is a low-$z$ GRB (class ``0") that the classifier correctly labels as low-$z$; and a false positive (FP) is a low-$z$ GRB that the classifier incorrectly labels as high-$z$. See Figure~\ref{Fig:LegConfusionMatrix}.

Each point along the ROC curve corresponds to a different decision threshold. The AUC quantifies the model's ability to distinguish between high-$z$ and low-$z$ GRBs. A higher AUC value signifies better classification performance, with a maximum of 1 indicating perfect separation between the two classes. The AUC values shown in Figures~\ref{fig:combined_raw_z2.0}--\ref{fig:combined_raw_z3.0} reflect both individual model performance and the combined effectiveness of SuperLearner. The percentage difference $\Delta$ is defined as:

\begin{equation}
\Delta=\frac{|AUC_{training}-AUC_{test}|}{\frac{(AUC_{training}+AUC_{test})}{2}}
\label{Eq:percentage-difference}
\end{equation}

SuperLearner Algorithm (SLA) denotes the SuperLearner algorithm with all models included and picked by the SuperLearner. AUC-SLA is defined as the AUC value obtained with the models selected by the SuperLearner. The Least Performing Algorithms Removed (LPR) denotes the subset of SuperLearner algorithms with the least performing algorithms removed from SLA. AUC-LPR is defined as the AUC obtained with these specific models after removing the lowest-performing ones.

\section{Results}
\label{sec:results}

Our results are presented with the aim of providing a binary classifier for optically selected GRBs, which the community can freely use in the form of a web-app for follow-up observations of potentially high-$z$ GRBs. Optical observations play a central role in GRB redshift determination because most spectroscopic redshifts are ultimately obtained through optical and near-infrared follow-up campaigns. Consequently, a classifier calibrated on the population of bursts with detected optical plateaus is well matched to the sample within which follow-up decisions are actually taken. In our analysis, correctly classified high-$z$ GRBs are the true positives (TP, see Figure \ref{Fig:LegConfusionMatrix}) in the classifier. Thus, we highlight those models that achieve a high TPR, since missing a genuinely high-$z$ burst may result in the loss of a valuable target for spectroscopic follow-up.

Throughout this section, the TPR and TNR values we quote are computed on the training set, and are labeled as such wherever they appear. We adopt this convention for comparability: the X-ray classifier of MGD25 reports its confusion matrices and quoted sensitivities on the training set, as do the earlier prompt-emission classifiers of \citet{morgan2012} and \citet{ukwatta2016machine} with which we compare in Section \ref{subsec:comparison3}, so training-set sensitivities are the matched quantity for every comparison drawn in this work. Generalization is reported separately and for every configuration through the quantity $\Delta$, the fractional difference between the training and test AUC values given in Table \ref{tab:AUC_for_SuperLearner}; a small $\Delta$ indicates that the training-set performance carries over to the test set, i.e., to GRBs not used for training, and we treat $\Delta$ rather than the training metrics themselves as the evidence bearing on generalization.

Three different redshift cutoffs ($z_t$) of 2.0, 2.5, and 3.0 are used to assess the performance of our classifier. Changing the cutoff does not alter the size of a given dataset, but it does change how that dataset divides between the two classes, and in particular how many high-$z$ events are available for the classifier to learn from: raising $z_t$ progressively depletes the high-$z$ class while enlarging the low-$z$ one. Table \ref{tab:GRBdata} presents the resulting class counts for each dataset and each $z_t$. These counts refer to each full dataset before the train--test split; the dataset totals (97, 90, 157, and 147 GRBs) are given in Section~\ref{sec:sample} and do not depend on $z_t$. Each dataset is then divided into an $80\%$ training set and a $20\%$ test set, so the training-set class counts are approximately $0.8$ times the tabulated values; the exact training-set counts can be read from the confusion matrices in Table \ref{tab:confusion_matrices_2.0}. Using three thresholds helps us assess the most suitable $z_t$ such that our classifier performs best in terms of sensitivity and AUC, while also making explicit the increasing class imbalance that accompanies the higher thresholds. For each $z_t$ case, we have used four datasets in SuperLearner: the raw dataset with and without the M-estimator, and the MICE-imputed dataset with and without the M-estimator. Thus, in total, we have twelve cases that allow us to investigate how different preprocessing choices affect the recovery of redshift information from optical plateau observables.

Table \ref{tab:LASSO} summarizes the features selected by LASSO with weights $>$ 2\% for the twelve cases. Figure \ref{Fig:LASSO} shows the weights assigned by LASSO to the features, for the cases of $z_t = 3.0$. Table \ref{tab:SL-algo} lists the top-performing algorithms chosen by SuperLearner for all twelve cases, where only models with weights above 0.05 are considered. Between 5 and 13 algorithms are retained in each case. For our best-performing configuration (raw data with the M-estimator at $z_t = 2.0$), the ensemble consists of glm, glmnet, ksvm, lda, speedglm, and bayesglm, i.e., five parametric learners and one support vector machine. Across all twelve cases, the most frequently retained algorithms are kernelKnn (nine cases), glm, ksvm, and lda (eight cases each), and glmnet (seven cases), so the ensembles combine linear and nonlinear learners rather than relying on a single type of model. Figure \ref{Fig:SL weights} shows the weights assigned by SuperLearner for the cases of $z_t = 3.0$.

Figures \ref{fig:combined_raw_z2.0}, \ref{fig:combined_raw_z2.5}, and \ref{fig:combined_raw_z3.0} show the ROC curves for the twelve cases. Each figure consists of four rows: the first and second rows correspond to the raw dataset without and with the M-estimator, respectively. The third and fourth rows represent the MICE-imputed dataset without and with the M-estimator, respectively.

In these figures, the left panels highlight the performance of the best models selected by SuperLearner, whereas the right panels showcase the outcomes after removing the least effective algorithms. Given the stochastic nature of the ensemble-fitting procedure, we assessed its variability by performing multiple iterations of our models on the MICE-imputed data for each ($z_t$) value, specifically by running 200, 300, 500, and 1000 iterations in addition to the original run. As in all four datasets, GRBs with $\Delta x / x > 1$ in any variable are excluded from these runs (Section~\ref{sec:sample}). The results, as presented in Table \ref{tab:combined_auc_scores_mice}, remain highly consistent with those from the initial 100-loop analysis, as shown in Table \ref{tab:AUC_for_SuperLearner}: across all loop counts the training AUC varies by less than $0.05$. This establishes the stability of the fitting procedure for a fixed dataset; the precision of the reported metrics is set by the sizes of the training and test samples, discussed below. The corresponding test AUC values in Table \ref{tab:combined_auc_scores_mice} vary more across loop counts than the training values.

Depending on the chosen model and dataset, the best metric among TPR, TNR, and AUC varies. This means that a model that attains the highest TPR may not have the highest AUC. Table \ref{tab:confusion_matrices_2.0} presents the full confusion matrices for all twelve cases. As stated above, TPR and TNR are training-set quantities, and train-to-test transfer is reported through $\Delta$ (Table \ref{tab:AUC_for_SuperLearner}). The SuperLearner model trained with the raw dataset with M-estimator at $z_{t} = 2.0$ is the best-performing model, as discussed in Section~\ref{sec:results_auc_overall}.

\subsection{The general trend of the results across the datasets, with and without MICE}
The sample sizes underlying these metrics should be kept in mind when comparing configurations. The four datasets contain 97, 90, 157, and 147 GRBs, of which $20\%$ are held out, giving test sets of 19, 18, 31, and 29 objects respectively.
One property of these datasets should be borne in mind when interpreting the difference in training-set TPR between the raw and MICE-imputed datasets (Sections~\ref{sec:results_zt2.0}--\ref{sec:results_zt3.0}). The raw and MICE-imputed datasets are not the same sample processed in two ways: the raw datasets are complete-case subsets, retaining only those bursts for which all features are measured with relative error below unity, whereas the MICE-imputed datasets are correspondingly larger because they include the bursts in which at least one feature, most often $\log(\mathrm{NH})$, is missing (Section~\ref{sec:sample}). Those bursts are, by construction, the more sparsely sampled ones. A lower TPR on the imputed datasets is therefore consistent with the greater difficulty of reconstructing the noisier optical features, or with the enlarged sample being intrinsically harder to classify because of the objects added to it. 
Because $z_t = 2.0$ and $z_t = 2.5$ are evaluated on the same four underlying datasets, the repeated raw-versus-imputed TPR ordering at these two thresholds (Sections~\ref{sec:results_zt2.0} and \ref{sec:results_zt2.5}) reflects the same sample-composition difference discussed above rather than an independent imputation effect, and shows that the raw-versus-imputed ordering does not depend on the particular choice of $z_t$.
For the practical purpose of this work, the distinction between these explanations does not affect the recommendation that follows from the results: for this classifier, the raw datasets yield higher training-set TPR than their imputed counterparts at every threshold examined, and are accordingly the configurations we recommend. Establishing which mechanism is responsible would require a dedicated study of the imputation of optical plateau parameters, which we regard as a separate problem from the classification task addressed here.
\subsection{Results for \texorpdfstring{$z_t = 2.0$}{zt = 2.0}}

\label{sec:results_zt2.0}

We can see that the SuperLearner model performs best in terms of training-set TPR for $z_{t} = 2.0$ in the raw dataset with M-estimator (TPR = 0.741), followed by the raw dataset without M-estimator (TPR = 0.656; Table \ref{tab:confusion_matrices_2.0}). We can also see that for $z_{t} = 2.0$, training-set TPR falls to 0.571 (a decrease of 23\%) in the MICE-imputed dataset with M-estimator and further falls to 0.451 (a decrease of 39\%) in the MICE-imputed dataset without M-estimator.

\subsection{Results for \texorpdfstring{$z_t = 2.5$}{zt = 2.5}}
\label{sec:results_zt2.5}

For $z_{t} = 2.5$, the SuperLearner model also attains the highest training-set TPR in the raw dataset with M-estimator (TPR = 0.590), followed by the raw dataset without M-estimator (TPR = 0.565). TPR drops to 0.444 (a decrease of 25\%) for the MICE-imputed dataset with the M-estimator and further falls to 0.289 (a decrease of 51\%) for the MICE-imputed dataset without the M-estimator. The same ordering, therefore, holds at this threshold.

\subsection{Results for \texorpdfstring{$z_t = 3.0$}{zt = 3.0}}
\label{sec:results_zt3.0}

Finally, for $z_{t} = 3.0$, the SuperLearner model attains the highest training-set TPR on the raw dataset with M-estimator (TPR = 0.333), followed by the MICE-imputed dataset with M-estimator (TPR = 0.260), the raw dataset without the M-estimator (TPR = 0.153), and the MICE-imputed dataset without the M-estimator (TPR = 0.120), corresponding to decreases of 22\%, 54\%, and 64\% relative to the best configuration. We note that the ordering at this threshold shows that the effect of outlier rejection is larger than that of imputation; the pairwise comparison between each raw dataset and its imputed counterpart nonetheless preserves the ordering seen at the lower thresholds. The decline in sensitivity at the highest redshift threshold is expected because the number of available high-$z$ optical GRBs decreases progressively, making the classification problem more challenging. The stronger decline at $z_t=3.0$ is also consistent with expectations from optical observations. High-redshift bursts are intrinsically more difficult to observe because of cosmological dimming and increasing absorption blueward of Ly$\alpha$ \citep{Prochaska2007, Prochaska2008, Tanvir2009Nature, Greiner2011}. As a result, the number of well-sampled optical plateaus decreases with redshift, reducing the amount of information available for classifier training. The lower TPR at this threshold therefore likely reflects both the smaller high-$z$ sample and the observational challenges of characterizing distant optical afterglows. Thus, in terms of training-set TPR, SuperLearner performs best on the raw dataset with the M-estimator at the $z_t=2.0$ cutoff.

\subsection{AUC, \texorpdfstring{$\Delta$}{Delta}, and Overall Assessment}
\label{sec:results_auc_overall}

In terms of AUC, the SuperLearner model performs the best on the raw dataset with the M-estimator across all redshift cutoffs. For $z_t = 2.0$, $2.5$ and $3.0$, the AUC values are 0.841, 0.877 and 0.875, respectively. All the other cases shown in Table \ref{tab:AUC_for_SuperLearner} attain lower training-set AUC values, with the exception of the test-set AUC at $z_t = 3.0$, where the MICE-imputed data without the M-estimator reaches 0.860 and 0.866 against 0.833 for the raw data with the M-estimator; as discussed below, differences of this size at $z_t = 3.0$ are within the resolution of the test sample.

The persistence of AUC values above 0.84 across all three redshift thresholds shows that the classifier retains substantial discriminatory power even as the classification task becomes increasingly challenging. The features carrying this discriminatory power depend on the configuration. On the raw data with the M-estimator, LASSO retains only prompt-emission and absorption features at $z_t = 2.0$ and $z_t = 3.0$, and these together with the optical plateau parameter $\beta$ at $z_t = 2.5$, while under MICE imputation with M-estimator one or more plateau parameters are retained at all three thresholds (Table \ref{tab:LASSO}). We return to this behavior in Sections \ref{subsec:comparison} and \ref{subsec:comparison2}, where we argue that the systematic offset of $F_{a,\mathrm{opt}}$ below its X-ray counterpart, which we attribute to suppression of the measured optical flux by host-galaxy dust extinction and heterogeneous observing cadence, gives a physical reason to expect $F_{a,\mathrm{opt}}$ to be selected less often than the prompt-emission and absorption features.

That AUC values of this magnitude are attained in this regime is notable, since optical plateau measurements are more strongly affected by extinction, chromatic evolution, and heterogeneous observing cadence than their X-ray counterparts \citep{Covino_2013, Greiner2011, Oates_2011, Zaninoni2013, Coward_2009}, and the resulting sample is correspondingly less complete. These metrics alone do not separate intrinsic redshift-dependent physics from the observational selection effects that shape the optical sample; disentangling the two would require a model of the optical selection function, which is beyond the scope of this analysis.

In terms of $\Delta$, at $z_t=2.0$ the raw dataset with M-estimator has the lowest $\Delta$ of the four datasets, with $1\% < \Delta < 4\%$, together with the highest training-set TPR of 0.741; at $z_t = 2.5$ and $3.0$ other datasets attain a lower $\Delta$ (Table \ref{tab:AUC_for_SuperLearner}).

 The high-$z$ class is the smaller one throughout and becomes progressively sparser as $z_t$ rises: at $z_t = 3.0$ each of the four test sets contains six high-$z$ bursts, and the corresponding training sets contain between 12 and 25. Two consequences follow. First, several of the quoted metrics are ratios of small integers --- the training-set TPR of 0.333 at $z_t = 3.0$ on the raw data with the M-estimator, for instance, corresponds to four correctly classified bursts out of twelve --- and their precision is limited by these small counts. Second, with six high-$z$ and twelve or thirteen low-$z$ bursts in the raw-data test sets, the test AUC at $z_t = 3.0$ can take only values separated by roughly $0.014$, which is why the same value of $0.833$ appears four times in Table \ref{tab:AUC_for_SuperLearner}.
 When we repeat the fit with 100 to 1000 loops on the MICE-imputed dataset with the M-estimator, the test-set AUC values at $z_t = 3.0$, shown in Table \ref{tab:combined_auc_scores_mice}, range from $0.768$ to $0.833$. This shows that a change of $0.065$ or less is inherent to the uncertainties of the machine learning analysis rather than to the algorithms or sample considered.
For this reason, we base our conclusions on results that hold at every threshold rather than on individual differences of this size: each raw dataset attains a higher training-set TPR than its MICE-imputed counterpart, and the raw dataset with the M-estimator attains the highest training-set TPR.

Thus, based on the above results, we propose that the SuperLearner model trained on the raw dataset using an M-estimator at $z_{t} = 2.0$ is the best-performing model. It achieves the highest training-set TPR and the lowest $\Delta$ at its threshold, and its training-set AUC after removing the least-performing algorithms (0.841) is exceeded only by the same dataset at $z_t = 2.5$ and $3.0$. From an observational perspective, this configuration is of particular practical interest because optical observations remain the primary route by which most GRB redshifts are measured, so a classifier calibrated on an optically selected sample is matched to the population in which follow-up decisions are actually made. At this threshold, the retained features are $\log(\mathrm{PeakFlux})$, PhotonIndex, and $\log(\mathrm{NH})$ --- prompt $\gamma$-ray and absorption properties rather than plateau parameters --- and the result accordingly attributes the redshift discrimination at this threshold to these observables rather than to the optical plateau phase alone. It establishes that, within the population of bursts possessing a detected optical plateau, high-$z$ candidates can be recovered with high sensitivity from prompt-emission and absorption observables, and that the resulting classifier generalizes well. Such classifiers may therefore provide a practical tool for prioritizing spectroscopic resources and optimizing follow-up strategies for potentially distant GRBs while also complementing existing X-ray-based approaches.

The appendices provide a detailed breakdown of AUC performance across different $z_t$ values (see Appendices~\ref{sec:z2.0}, \ref{sec:z2.5}, and \ref{sec:z3.0}), highlighting the best- and least-performing models in the training set for each threshold.

\begin{table*}
    \centering
    \begin{tabular}{lcc}
    \hline
    \multicolumn{3}{c}{\textbf{Number of GRBs in the different redshift ranges}} \\
       \hline
    \hline
        \bf Dataset & \bf Low-$z$ GRBs & \bf High-$z$ GRBs \\
    \hline \hline 
        \multicolumn{3}{c}{$\bm{z_{t}=2.0}$} \\
        \hline
        Raw dataset without M-estimator: & 55 & 42 \\
        Raw dataset with M-estimator: & 49 & 41 \\
        MICE-imputed dataset without M-estimator: & 93 & 64 \\
        MICE-imputed dataset with M-estimator: & 85 & 62 \\
        \hline
        \hline
        \multicolumn{3}{c}{$\bm{z_{t}=2.5}$} \\
        \hline
        Raw dataset without M-estimator: & 64 & 33 \\
        Raw dataset with M-estimator: & 58 & 32 \\
        MICE-imputed dataset without M-estimator: & 107 & 50 \\
        MICE-imputed dataset with M-estimator: & 99 & 48 \\  
        \hline
        \hline
        \multicolumn{3}{c}{$\bm{z_{t}=3.0}$} \\
        \hline
        Raw dataset without M-estimator: & 78 & 19 \\
        Raw dataset with M-estimator: & 72 & 18 \\
        MICE-imputed dataset without M-estimator: & 126 & 31 \\
        MICE-imputed dataset with M-estimator: & 118 & 29 \\
        \hline
        \hline
    \end{tabular}
    \caption{Numbers of low- and high-$z$ GRBs (based on spectroscopic redshifts) in each of the four datasets, for redshift thresholds $z_{t} = 2.0$, $2.5$, and $3.0$ (from top to bottom).}
    \label{tab:GRBdata}
\end{table*}

\begin{table*}
  \centering
  \resizebox{!}{.165\paperheight}{%
    \begin{tabular}{lcc}
      \hline
      \multicolumn{2}{c}{\textbf{Features selected by LASSO in different redshift ranges with weights $>2\%$}} \\
      \hline\hline
      \textbf{Dataset} & \textbf{Features Selected} \\ 
      \hline\hline
      \multicolumn{2}{c}{$\bm{z_{t}=2.0}$} \\
      \hline
      Raw dataset without M-estimator: & $\log(\mathrm{PeakFlux})$, PhotonIndex, and $\log(\mathrm{NH})$ \\
      Raw dataset with M-estimator: & $\log(\mathrm{PeakFlux})$, PhotonIndex, and $\log(\mathrm{NH})$ \\
      MICE-imputed dataset without M-estimator: & $\log(\mathrm{PeakFlux})$ and $\log(\mathrm{NH})$ \\
      MICE-imputed dataset with M-estimator: & $\log(T_{90})$, $\log(\mathrm{PeakFlux})$, PhotonIndex, $\log(\mathrm{NH})$, \\ &
      $\log(F_{a})$, and $\log(T_{a})$ \\
      \hline\hline
      \multicolumn{2}{c}{$\bm{z_{t}=2.5}$} \\
      \hline
      Raw dataset without M-estimator: & $\log(T_{90})$, $\log(\mathrm{PeakFlux})$, $\log(\mathrm{NH})$, PhotonIndex, and $\beta$ \\
      Raw dataset with M-estimator: &  $\log(\mathrm{PeakFlux})$, $\log(\mathrm{NH})$, PhotonIndex, and $\beta$ \\
      MICE-imputed dataset without M-estimator:& $\log(\mathrm{NH})$ and $\beta$ \\
      MICE-imputed dataset with M-estimator: & $\log(T_{90})$, $\log(\mathrm{PeakFlux})$, $\log(\mathrm{Fluence})$, PhotonIndex, \\ &
      $\log(\mathrm{NH})$, $\log(F_{a})$, and $\log(T_{a})$ \\
      \hline\hline
      \multicolumn{2}{c}{$\bm{z_{t}=3.0}$} \\
      \hline
      Raw dataset without M-estimator: & $\log(T_{90})$, $\log(\mathrm{PeakFlux})$, and $\log(\mathrm{NH})$ \\
      Raw dataset with M-estimator: & $\log(T_{90})$, $\log(\mathrm{PeakFlux})$, and $\log(\mathrm{NH})$ \\
      MICE-imputed dataset without M-estimator: & $\log(\mathrm{PeakFlux})$, $\log(\mathrm{NH})$ and $\beta$ \\
      MICE-imputed dataset with M-estimator: & $\log(T_{90})$, $\log(\mathrm{PeakFlux})$, $\log(\mathrm{Fluence})$, PhotonIndex, \\
      & $\log(\mathrm{NH})$, $\log(F_{a})$, $\log(T_{a})$, $\alpha$, and $\beta$ \\
      \hline\hline
    \end{tabular}%
  } %
  \caption{Features selected by LASSO with weights $>2\%$ for $z_{t}$ of 2.0, 2.5, and 3.0 (top to bottom panels) across four different datasets (left column).}
  \label{tab:LASSO}
\end{table*}

\begin{table*}
    \centering
    {\resizebox{!}{.165\paperheight}{%
    \begin{tabular}{lc}
    \hline
    \multicolumn{2}{c}{\bf Datasets and algorithms selected by the SuperLearner in the \bm{$z_{t}$} ranges with weights \bm{$>$} 0.05} \\
    \hline
    \hline
    \bf Dataset & \bf Algorithms Selected \\ 
    \hline
    \hline
        \multicolumn{2}{c}{$\bm{z_{t}=2.0}$} \\
        \hline
        Raw dataset without M-estimator: & cforest, glm, glmnet, ksvm, lda, lm, qda, speedglm, \\
         & speedlm, xgboost, biglasso, bayesglm, earth \\
        Raw dataset with M-estimator: & glm, glmnet, ksvm, lda, speedglm, bayesglm \\
        MICE-imputed dataset without M-estimator: & glm, glmnet, lda, lm, ranger, randomForest, \\
         & speedglm, speedlm, biglasso \\
        MICE-imputed dataset with M-estimator: & kernelKnn, ksvm, svm, lm, qda, speedlm \\
        \hline
        \hline
        \multicolumn{2}{c}{$\bm{z_{t}=2.5}$} \\
        \hline
        Raw dataset without M-estimator: & glm, glmnet, kernelKnn, lda, lm, speedglm, speedlm, \\
         & biglasso, bayesglm, caret.rpart, earth \\
        Raw dataset with M-estimator: & cforest, glm, glmnet, kernelKnn, ksvm, svm \\
        MICE-imputed dataset without M-estimator: & cforest, kernelKnn, ksvm, qda, ranger, \\
         & randomForest, xgboost, earth \\
        MICE-imputed dataset with M-estimator: & cforest, kernelKnn, ksvm, svm, qda, ranger, \\
         & randomForest, earth \\
        \hline
        \hline
        \multicolumn{2}{c}{$\bm{z_{t}=3.0}$} \\
        \hline
        Raw dataset without M-estimator: & cforest, glm, kernelKnn, ksvm, lda, ranger, \\
         & randomForest, speedglm, xgboost, biglasso, bayesglm \\
        Raw dataset with M-estimator: & glm, glmnet, kernelKnn, lda, lm \\
        MICE-imputed dataset without M-estimator: & kernelKnn, svm, lda, qda, ranger, randomForest, biglasso \\
        MICE-imputed dataset with M-estimator: & glm, glmnet, kernelKnn, ksvm, svm, lda, lm, \\
         & qda, speedglm, speedlm \\
        \hline
    \end{tabular}}}
    \caption{Base-learner algorithms assigned weights $>$ 0.05 by the SuperLearner ensemble, for $z_{t}$ of 2.0, 2.5, and 3.0 (from top to bottom panel) across the four datasets (left column). The corresponding features retained by LASSO are given in Table \ref{tab:LASSO}. The AUC values of the best- and least-performing individual algorithms, and of the ensemble after removing the least-performing ones, are given in Appendices \ref{sec:z2.0}, \ref{sec:z2.5}, and \ref{sec:z3.0}.}
    \label{tab:SL-algo}
\end{table*}

\begin{table*}[!ht]
    \centering
    {\resizebox{!}{.15\paperheight}{%
    \begin{tabular}{lcccc}
    \hline
    \multicolumn{5}{c}{\bf AUC for SuperLearner for training and test sets} \\
       \hline
    \hline
        \bf Dataset & \bf AUC - SLA & \bf $\Delta_{SLA}$ (\%) & \bf AUC - LPR & \bf $\Delta_{LPR}$ (\%)  \\
        \hline \hline
        \multicolumn{5}{c}{$\bm{z_{t}=2.0}$} \\
        \hline
        Raw dataset without M-estimator: & 0.762, 0.711 & 7 \% & 0.775, 0.722 & 7 \% \\
        Raw dataset with M-estimator: & 0.826, 0.837 & 1 \% & 0.841, 0.812 & 4 \%  \\
        MICE-imputed without M-estimator: & 0.696, 0.606 & 14 \% & 0.706, 0.645 & 9 \% \\
        MICE-imputed with M-estimator: & 0.802, 0.692 & 15 \% &  0.781, 0.649 & 18 \% \\
        \hline \hline
        \multicolumn{5}{c}{\bm{$z_{t}=2.5$}} \\
        \hline
        Raw dataset without M-estimator: & 0.804, 0.722 & 11 \% & 0.807, 0.744 & 8 \%\\
        Raw dataset with M-estimator: & 0.872, 0.753 & 15 \% & 0.877, 0.790 & 10 \%  \\
        MICE-imputed without M-estimator: & 0.700, 0.581 & 18 \% & 0.698, 0.652  & 7 \% \\
        MICE-imputed with M-estimator: & 0.834, 0.731 & 13 \% & 0.828, 0.700  & 17 \% \\
        \hline   \hline
        \multicolumn{5}{c}{\bm{$z_{t}=3.0$}} \\
        \hline
        Raw dataset without M-estimator: & 0.800, 0.833 & 4 \% & 0.809, 0.833 & 3 \%  \\
        Raw dataset with M-estimator: & 0.876, 0.833 & 5 \% & 0.875, 0.833  & 5 \% \\
        MICE-imputed without M-estimator: & 0.735, 0.860 & 16 \% & 0.752, 0.866 & 14 \%  \\
        MICE-imputed with M-estimator: & 0.819, 0.826 & 1 \% & 0.813, 0.789 & 3 \% \\
        \hline
        
        \hline
    \end{tabular}}}
   \caption{Training- and test-set AUC values for the SuperLearner ensemble at each redshift threshold. AUC -- SLA is the AUC from the SuperLearner selected algorithms, and AUC -- LPR is the AUC obtained after removing the least-performing models. The second and fourth columns each contain a pair of AUC values, quoted as training followed by test. The third and fifth columns show the percentage difference $\Delta$ between the training and test AUC, calculated as the absolute deviation between the training and test values divided by their mean. All values are quoted to three decimal digits. Note that the TPR and TNR values quoted in the text and in Table \ref{tab:confusion_matrices_2.0} are training-set quantities; $\Delta$ is the quantity through which train-to-test transfer is reported in this work.}
    \label{tab:AUC_for_SuperLearner}
\end{table*}

\begin{figure*}[htbp]
    \centering
    \includegraphics[width = 0.8\textwidth]{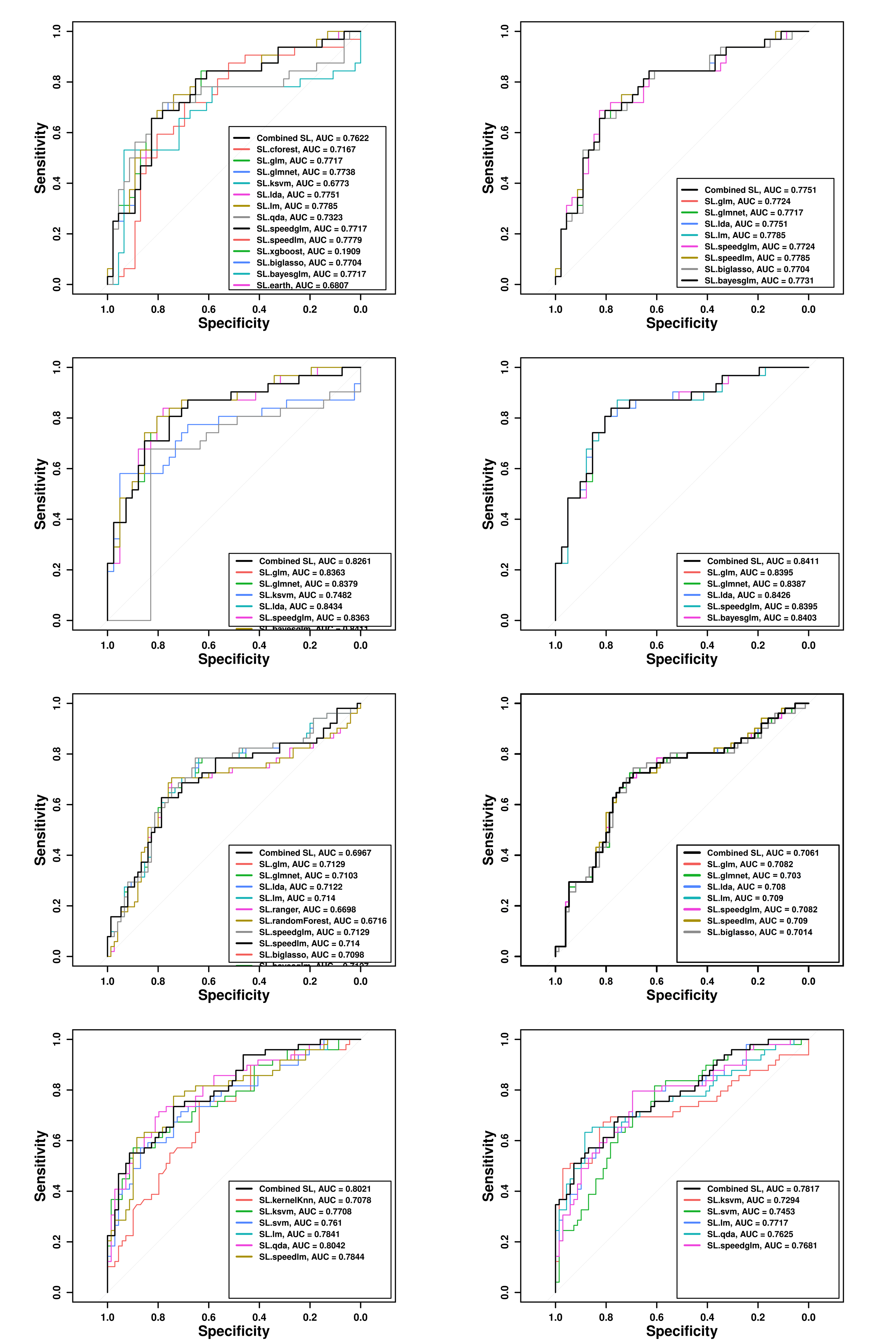}
    \caption{For $z_{t}=2.0$. Top row: raw without M-estimator. Second row: raw with M-estimator. Third row: MICE-imputed without M-estimator. Bottom row: MICE-imputed with M-estimator. First column: ROC curve with best algorithms chosen by SuperLearner with weights $>$ 0.05 in each dataset. Top right: ROC curve without `cforest', `ksvm', `qda', `xgboost', and `earth'. Second right: ROC curve without `ksvm'. Third right: ROC curve without `ranger' and `randomForest'. Bottom right: ROC curve without `kernelKnn'.}
    \label{fig:combined_raw_z2.0}
\end{figure*}

\begin{table*}[!ht]
\centering
\begin{tabular}{c | c c | c c | c }
\hline
\multicolumn{6}{c}{\bf Confusion Matrix for the several datasets including TPR and TNR} \\
\hline
\multicolumn{6}{c}{\bm{$z_{t}=2.0$}} \\ 
\hline
\multicolumn{1}{c}{TNR and TPR} & \multicolumn{2}{|c|}{\textbf{Raw without M-estimator}} & \multicolumn{2}{c|}{\textbf{Raw without M-estimator - algos removed}} & \multicolumn{1}{c}{TNR and TPR} \\

\hline
0.826 & 38 & 12 & 38 & 11 & 0.826\\
0.625 & 8 & 20 & 8 & 21 & 0.656 \\
\hline
\multicolumn{1}{c}{TNR and TPR} & \multicolumn{2}{|c|}{\textbf{Raw with M-estimator}}  &  \multicolumn{2}{c|}{\textbf{Raw with M-estimator - algos removed}} & \multicolumn{1}{c}{TNR and TPR} \\
\hline
0.829 & 34 & 9 & 33 & 8 & 0.804\\
0.709 & 7 & 22 & 8 & 23 & 0.741\\
\hline
\multicolumn{1}{c}{TNR and TPR} & \multicolumn{2}{|c|}{\textbf{MICE without M-estimator}} & \multicolumn{2}{c|}{\textbf{MICE without M-estimator - algos removed}} & \multicolumn{1}{c}{TNR and TPR} \\
\hline
0.786 & 59 & 24 & 60 & 28 & 0.800\\
0.529 & 16 & 27 & 15 & 23 & 0.451\\
\hline
\multicolumn{1}{c}{TNR and TPR} & \multicolumn{2}{|c|}{\textbf{MICE with M-estimator}} & \multicolumn{2}{c|}{\textbf{MICE with M-estimator- algos removed}} & \multicolumn{1}{c}{TNR and TPR}\\
\hline
0.840 & 58 & 21 & 60 & 21 & 0.869 \\
0.571 & 11 & 28 & 9 & 28 & 0.571\\
\hline
\multicolumn{6}{c}{\bm{$z_{t}=2.5$}} \\
\hline
\multicolumn{1}{c}{TNR and TPR} & \multicolumn{2}{|c|}{\textbf{Raw without M-estimator}} & \multicolumn{2}{c|}{\textbf{Raw without M-estimator - algos removed}} & \multicolumn{1}{c}{TNR and TPR} \\
\hline
0.890 & 49 & 10 & 48 & 10 & 0.872\\
0.565 & 6 & 13 & 7 & 13 & 0.565 \\
\hline
\multicolumn{1}{c}{TNR and TPR} & \multicolumn{2}{|c|}{\textbf{Raw with M-estimator}}  &  \multicolumn{2}{c|}{\textbf{Raw with M-estimator - algos removed}} & \multicolumn{1}{c}{TNR and TPR} \\
\hline
0.900 & 45 & 9 & 47 & 9 & 0.940\\
0.590 & 5 & 13 & 3 & 13 & 0.590\\
\hline
\multicolumn{1}{c}{TNR and TPR} & \multicolumn{2}{|c|}{\textbf{MICE without M-estimator}} & \multicolumn{2}{c|}{\textbf{MICE without M-estimator - algos removed}} & \multicolumn{1}{c}{TNR and TPR} \\
\hline
0.897 & 79 & 30 & 82 & 27 & 0.931\\
0.210 & 9 & 8 & 6 & 11 & 0.289\\
\hline
\multicolumn{1}{c}{TNR and TPR} & \multicolumn{2}{|c|}{\textbf{MICE with M-estimator}} & \multicolumn{2}{c|}{\textbf{MICE with M-estimator - algos removed}} & \multicolumn{1}{c}{TNR and TPR}\\
\hline
0.926 & 76 & 20 & 75 & 20 & 0.914\\
0.444 & 6 & 16 & 7 & 16 & 0.444 \\
\hline
\multicolumn{6}{c}{\bm{$z_{t}=3.0$}} \\
\hline
\multicolumn{1}{c}{TNR and TPR} & \multicolumn{2}{|c|}{\textbf{Raw without M-estimator}} & \multicolumn{2}{c|}{\textbf{Raw without M-estimator - algos removed}} & \multicolumn{1}{c}{TNR and TPR} \\
\hline
0.984 & 64 & 12 & 63 & 11 & 0.969\\
0.076 & 1 & 1 & 2 & 2 & 0.153\\
\hline
\multicolumn{1}{c}{TNR and TPR} & \multicolumn{2}{|c|}{\textbf{Raw with M-estimator}}  &  \multicolumn{2}{c|}{\textbf{Raw with M-estimator - algos removed}} & \multicolumn{1}{c}{TNR and TPR} \\
\hline
0.983 & 59 & 8 & 59 & 8 & 0.983\\
0.333 & 1 & 4 & 1 & 4 & 0.333\\
\hline
\multicolumn{1}{c}{TNR and TPR} & \multicolumn{2}{|c|}{\textbf{MICE without M-estimator}} & \multicolumn{2}{c|}{\textbf{MICE without M-estimator - algos removed}} & \multicolumn{1}{c}{TNR and TPR} \\
\hline
0.980 & 99 & 25 & 98 & 22 & 0.970 \\
0.000 & 2 & 0 & 3 & 3 & 0.120 \\
\hline
\multicolumn{1}{c}{TNR and TPR} & \multicolumn{2}{|c|}{\textbf{MICE with M-estimator}} & \multicolumn{2}{c|}{\textbf{MICE with M-estimator - algos removed}} & \multicolumn{1}{c}{TNR and TPR}\\
\hline
0.957 & 91 & 17 & 90 & 17 & 0.947\\
0.260 & 4 & 6 & 5 & 6 & 0.260\\
\hline

\end{tabular}
\caption{Confusion matrices for the \textbf{training set} of each dataset, with the corresponding TNR (every first row) and TPR (every second row). In each matrix, rows correspond to the predicted class and columns to the true class, with the first row and column referring to the low-$z$ class and the second to the high-$z$ class (see Figure \ref{Fig:LegConfusionMatrix}); the entries are therefore TN and FN in the first row and FP and TP in the second. TNR is computed as $\mathrm{TN}/(\mathrm{TN}+\mathrm{FP})$ and TPR as $\mathrm{TP}/(\mathrm{TP}+\mathrm{FN})$. All TPR and TNR values quoted for the optical classifier are training-set quantities, for consistency with MGD25 and with the earlier prompt-emission classifiers against which we compare; generalization is reported separately through $\Delta$ in Table \ref{tab:AUC_for_SuperLearner}.}
\label{tab:confusion_matrices_2.0}
\end{table*}

\begin{figure*}[htbp]
\centering
    \includegraphics[width = 0.8\textwidth]{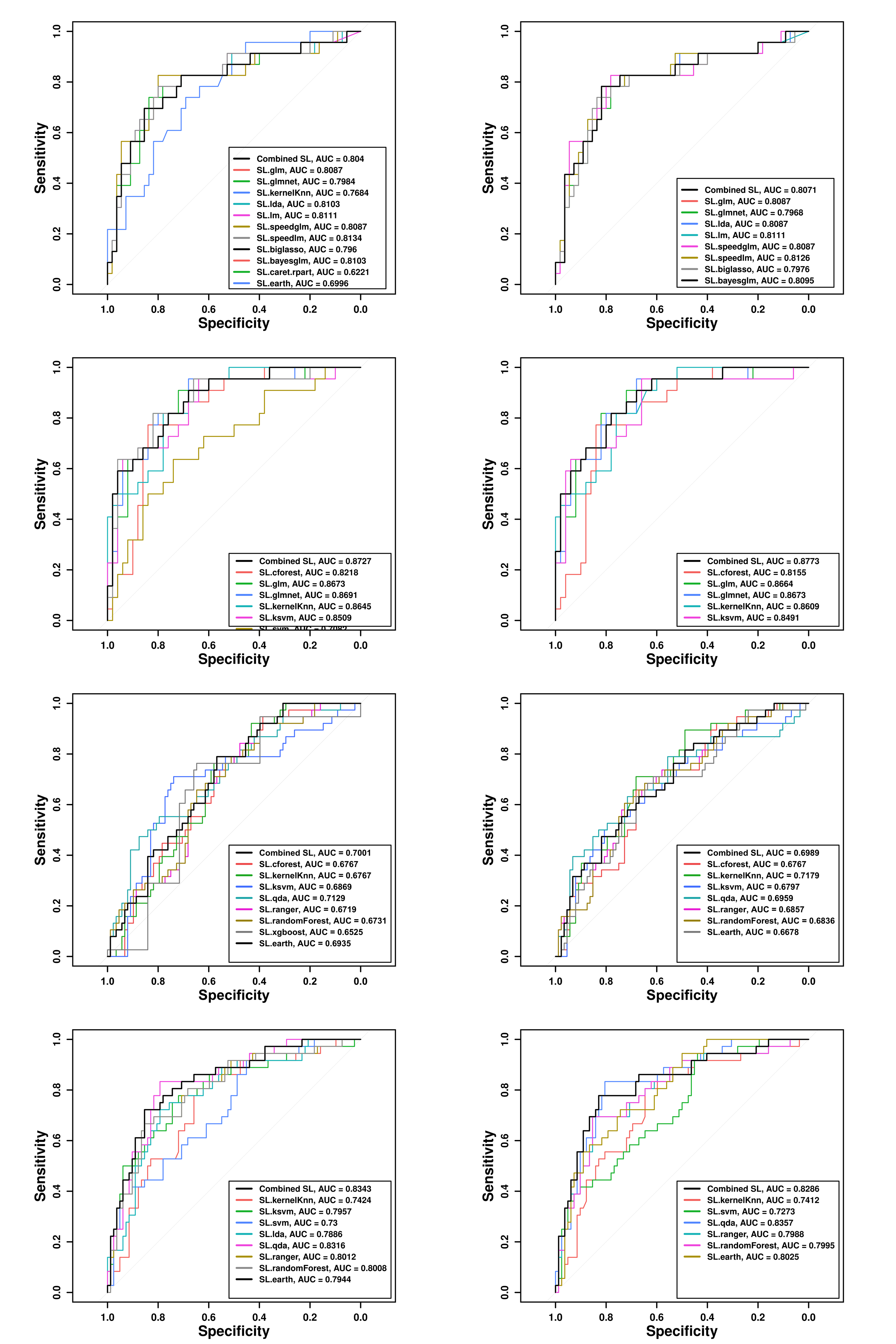}
\caption{For $z_{t}=2.5$. Top row: raw without M-estimator. Second row: raw with M-estimator. Third row: MICE-imputed without M-estimator. Bottom row: MICE-imputed with M-estimator. First column: ROC curve with best algorithms chosen by SuperLearner with weights $>$ 0.05 in each dataset. Top right: ROC curve without `kernelKnn', `caret.rpart', and `earth'. Second right: ROC curve without `svm'. Third right: ROC curve without `xgboost'. Bottom right: ROC curve without `cforest' and `ksvm'.}
\label{fig:combined_raw_z2.5}
\end{figure*}

\begin{figure*}[htbp]
\centering
    \includegraphics[width = 0.8\textwidth]{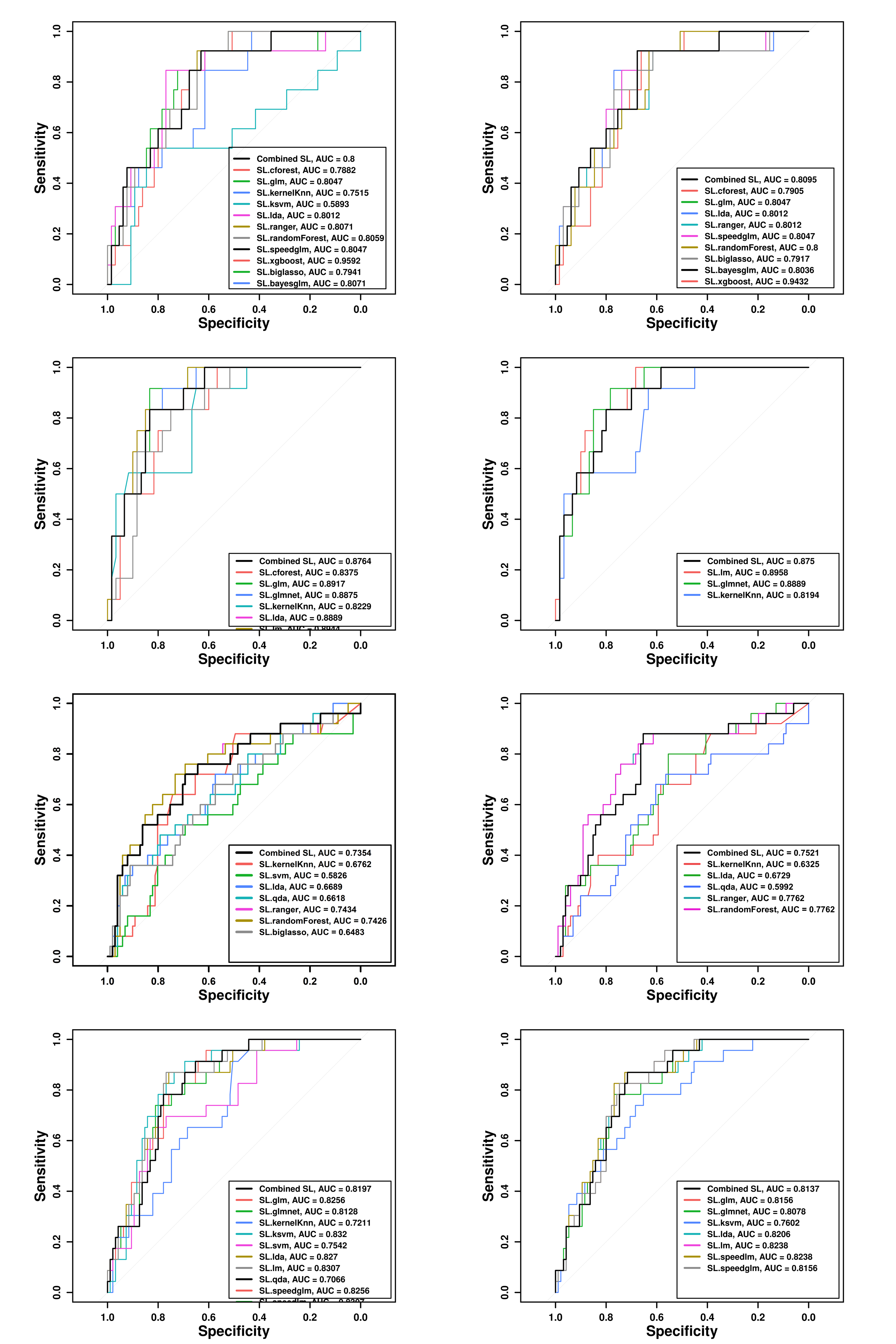}
\caption{For $z_{t}=3.0$. Top row: raw without M-estimator. Second row: raw with M-estimator. Third row: MICE-imputed without M-estimator. Bottom row: MICE-imputed with M-estimator. First column: ROC curve with best algorithms chosen by SuperLearner with weights $>$ 0.05 in each dataset. Top right: ROC curve without `kernelKnn' and `ksvm'. Second right: ROC curve without `glm' and `lda'. Third right: ROC curve without `svm' and `biglasso'. Bottom right: ROC curve without `kernelKnn', `svm', and `qda'.}
\label{fig:combined_raw_z3.0}
\end{figure*}

\begin{table*}
    \centering
    \begin{tabular}{lcccccc}
    \hline
    \multicolumn{6}{c}{\bf Loop for each \bm{$z_{t}$} repeated twice for the MICE-imputed data with the M-estimator} \\
    \hline
    \hline
    \textbf{Run} & \textbf{100 loops} & \textbf{200 loops} & \textbf{300 loops} & \textbf{500 loops} & \textbf{1000 loops} \\
    \hline
    \hline
    & \textbf{Training Test} & \textbf{Training Test} & \textbf{Training Test} & \textbf{Training Test} & \textbf{Training Test} \\
    \hline
    \hline
    \multicolumn{6}{c}{\bm{$z_{t}=2.0$}} \\
    \hline
    \hline
        1 & 0.7972\quad 0.6923 & 0.8007\quad 0.6923 & 0.8048\quad 0.7067 & 0.8039\quad 0.7115 & 0.8098\quad 0.7067 \\
        2 & 0.8045\quad 0.7115 & 0.8137\quad 0.7067 & 0.8007\quad 0.6971 & 0.8001\quad 0.6971 & 0.8027\quad 0.6982 \\
    \hline
    \hline
    \multicolumn{6}{c}{\bm{$z_{t}=2.5$}} \\
    \hline
    \hline
        1 & 0.8283\quad 0.7421 & 0.8367\quad 0.7368 & 0.8279\quad 0.7001 & 0.8296\quad 0.6895 & 0.8404\quad 0.7474 \\
        2 & 0.8313\quad 0.7316 & 0.7913\quad 0.5737 & 0.8313\quad 0.7211 & 0.8316\quad 0.7211 & 0.7984\quad 0.6737\\
    \hline
    \hline
    \multicolumn{6}{c}{\bm{$z_{t}=3.0$}} \\
    \hline
    \hline
        1 & 0.8252\quad 0.8116  & 0.8096\quad 0.7899 & 0.8101\quad 0.7826 & 0.8243\quad 0.8043 & 0.8119\quad 0.7826 \\
        2 & 0.8261\quad 0.8333  & 0.8275\quad 0.8333 & 0.8146\quad 0.7754 & 0.8092\quad 0.7681 & 0.8316\quad 0.8333 \\
    \hline
    \hline
    \end{tabular}
    \caption{Combined training and test AUC scores for the MICE-imputed dataset with the M-estimator for $z_{t}$ = 2.0, 2.5, and 3.0 (from top to bottom). The left value in each loop column shows the AUC value for the training dataset, while the right value in each loop column shows the AUC value for the testing dataset. 
   }

    \label{tab:combined_auc_scores_mice}
\end{table*}

\section{Discussion}\label{sec:discussion}

In this section, we interpret the performance and implications of our optical redshift classifier, compare it with the companion X-ray classifier of MGD25 and earlier ML-based efforts, and describe practical tools for deployment and validation.  We highlight that the optical model—trained on raw data with the M-estimator—shows the greatest sensitivity at $z_t=2.0$, whereas the X-ray classifier shows the smallest difference $\Delta$ between its training- and test-set AUC, and therefore the most reliable performance on GRBs not used for training, at a higher threshold, $z_t = 3.5$.  We then examine how differing input distributions and preprocessing choices, such as the application of the M-estimator and MICE, affect feature selection and classification outcomes. Finally, we validate the pipeline on an independent generalization set, and compare classifier and regression predictions against known redshifts.  The following subsections present these comparisons, the differences in methods, and the web-app implementation in detail; the validation is presented in Section~\ref{sec:generalisation}.

\subsection{Lyman-\texorpdfstring{$\alpha$}{alpha} Break, Dropout Techniques, and the Scope of the Present Analysis}
The standard dropout (Lyman-break) technique is the canonical method for identifying high-redshift galaxies and GRB afterglows using broadband photometry. It relies on strong absorption by neutral hydrogen in the intergalactic medium blueward of the Lyman-$\alpha$ transition at 1216\,\AA, which produces a sharp, wavelength-dependent suppression of flux in observer-frame bands shortward of the redshifted Ly$\alpha$ transition. This discontinuity leads to characteristic non-detections (``dropouts'') in specific filters and underpins the majority of high-redshift galaxy surveys and early GRB afterglow redshift identifications.

Works by \citet{Prochaska2007,Prochaska2008} have demonstrated both the effectiveness and the limitations of applying dropout-based techniques to GRB afterglows. In particular, these studies showed that reliable redshift identification through afterglow photometry or spectroscopy requires early optical observations in several separate filters, which preserve wavelength-dependent information. They also showed that a substantial fraction of GRBs are optically faint or ``dark'' due to dust extinction in their host galaxies or absorption at high redshift. They further established that, when such multiband information is unavailable, the redshift must instead be obtained from host-galaxy identification and late-time spectroscopy rather than from dropout-based afterglow methods. Subsequent GRB host-galaxy surveys reinforced this conclusion \citep{Perley2009,Perley2013}.

The optical plateau sample employed in this work often lacks these early, multi-filter observations, so the Lyman-$\alpha$ break technique cannot be applied to it. The plateau LCs are constructed from highly heterogeneous observations obtained with Swift/UVOT and 455 ground-based telescopes and detectors, often using different filters, unfiltered imaging, and non-uniform observing cadences. To analyze all these GRBs together, the optical plateau fluxes are homogenized by rescaling to a common R-band reference using the observed or inferred optical spectral index $\beta$ \citep{Dainotti2022}.

This R-band rescaling converts the measurements in different filters into a single equivalent R-band flux, so by construction the original wavelength-dependent information is lost. As a direct consequence, any filter-specific signature -- most notably the flux suppression caused by intergalactic absorption blueward of the redshifted Lyman-$\alpha$ transition -- cannot be recovered afterwards without returning to a full multi-wavelength analysis. Such an analysis would have to be restricted to the substantially smaller subset of GRBs with plateau emission observed in a single filter.

\subsection{The Model Performance}

Rather than identifying specific spectral features such as the Lyman-$\alpha$ break, the classifier uses statistical correlations between the measured observables and redshift across the GRB sample as a whole. In particular, the model exploits prompt $\gamma$-ray and X-ray absorption properties and, in some configurations, the observer-frame properties of the optical plateau emission.
Taken together, these variables correlate with redshift across the GRB sample, and the classifier can learn these correlations. This behavior means that the model relies on trends shared by many bursts in the sample, not on a signature measured in any single burst.
We emphasize that the inferred redshift estimates are inherently probabilistic and should be interpreted as statistical indicators rather than precise measurements. As such, they are not intended to replace spectroscopic or host-galaxy redshift determinations, but to provide complementary information on the redshift distributions of GRBs that lack secure measurements. The reported feature importances and performance metrics describe the subsample of well-observed GRBs with a detected optical plateau rather than the optical GRB population as a whole; because a complete sample is not attainable for this study, the predictions are specific to this sample.

\subsection{Comparing Parameters of Optical and X-ray Samples}
\label{subsec:comparison}

Our optical classifier and the X-ray classifier presented in MGD25 are constructed from different datasets that contain the same prompt-emission features and largely the same afterglow features, but differ in observational wavelength, sample size, and data-processing strategies.

The X-ray classifier uses a larger sample of 251 GRBs with well-characterized plateau phases in the X-ray band, primarily from the BAT and XRT instruments onboard \textit{Swift}. It draws on ten predictive features spanning the prompt, plateau, and afterglow phases. In contrast, the optical classifier relies on 171 GRBs with identified optical plateaus, compiled from both UVOT and 455 ground-based telescopes and detectors. It uses nine of these ten features (all except $\Gamma$), and is restricted to GRBs with optical afterglow and plateau measurements. While both datasets share core variables (e.g., $T_{90}$, $F_a$, $T_a$, NH), the X-ray sample includes $\Gamma$, the spectral index for the XRT detection of the burst source obtained from an automated time-averaged spectral fit of the Photon Counting (PC) mode data. If the PC-mode data are not available, the Window Timing mode data are used instead. This variable is not available in the optical dataset.

In terms of preprocessing, both classifiers analyze four main data variants, though not the same four. The X-ray classifier uses raw data with and without the M-estimator, a single MICE-imputed dataset, and a SMOTE-balanced dataset; the present work uses raw data with and without the M-estimator and MICE-imputed data with and without the M-estimator, and does not employ SMOTE. The optical model avoids SMOTE to preserve physically realistic distributions, as over-sampling can distort early-time optical LC properties. In addition, because we analyze the imputed data both with and without the M-estimator, we can separate the effects of outlier rejection and imputation, which are combined in the single MICE-imputed dataset of the X-ray analysis. Because the imputed datasets are not prepared identically, comparisons between the two classifiers on imputed data are approximate. The X-ray analysis contains a single MICE-imputed dataset, to which outlier removal is also applied, whereas we analyze two (with and without the M-estimator); only our MICE-imputed dataset with the M-estimator therefore has a direct counterpart in the X-ray analysis.
MICE imputation also affects the optical and X-ray classifiers differently. With imputation, the X-ray classifier performs better than the optical one. The optical classifier reaches its highest training-set TPR on the raw data; on the imputed data, its training-set TPR is 36--78\% of this best value, depending on threshold and outlier treatment, whereas the X-ray classifier is essentially unchanged at $z_t = 2.0$ and improves modestly at higher thresholds. 
We consider two possible explanations for why the optical classifier performs best on the raw data. First, the optical features are more variable and the optical LCs are more sparsely sampled than their X-ray counterparts, which makes their missing values more challenging to reconstruct.
Second, the imputed datasets contain 60 (without the M-estimator) and 57 (with the M-estimator) more GRBs than the corresponding raw datasets (157 versus 97, and 147 versus 90; Section~\ref{sec:sample}). These additional bursts are excluded from the raw datasets because at least one of their features is not measured, and for many of them the missing feature is $\log(\mathrm{NH})$ (Figure~\ref{Fig:mices}), the only feature that LASSO selects in every configuration (Table~\ref{tab:LASSO}). For these bursts, the classifier therefore relies on an imputed rather than a measured value of one of its most informative features, which makes them more challenging to classify (Section \ref{sec:results}).

\begin{figure*}[htbp]
    \centering
    \includegraphics[width=0.45\textwidth]{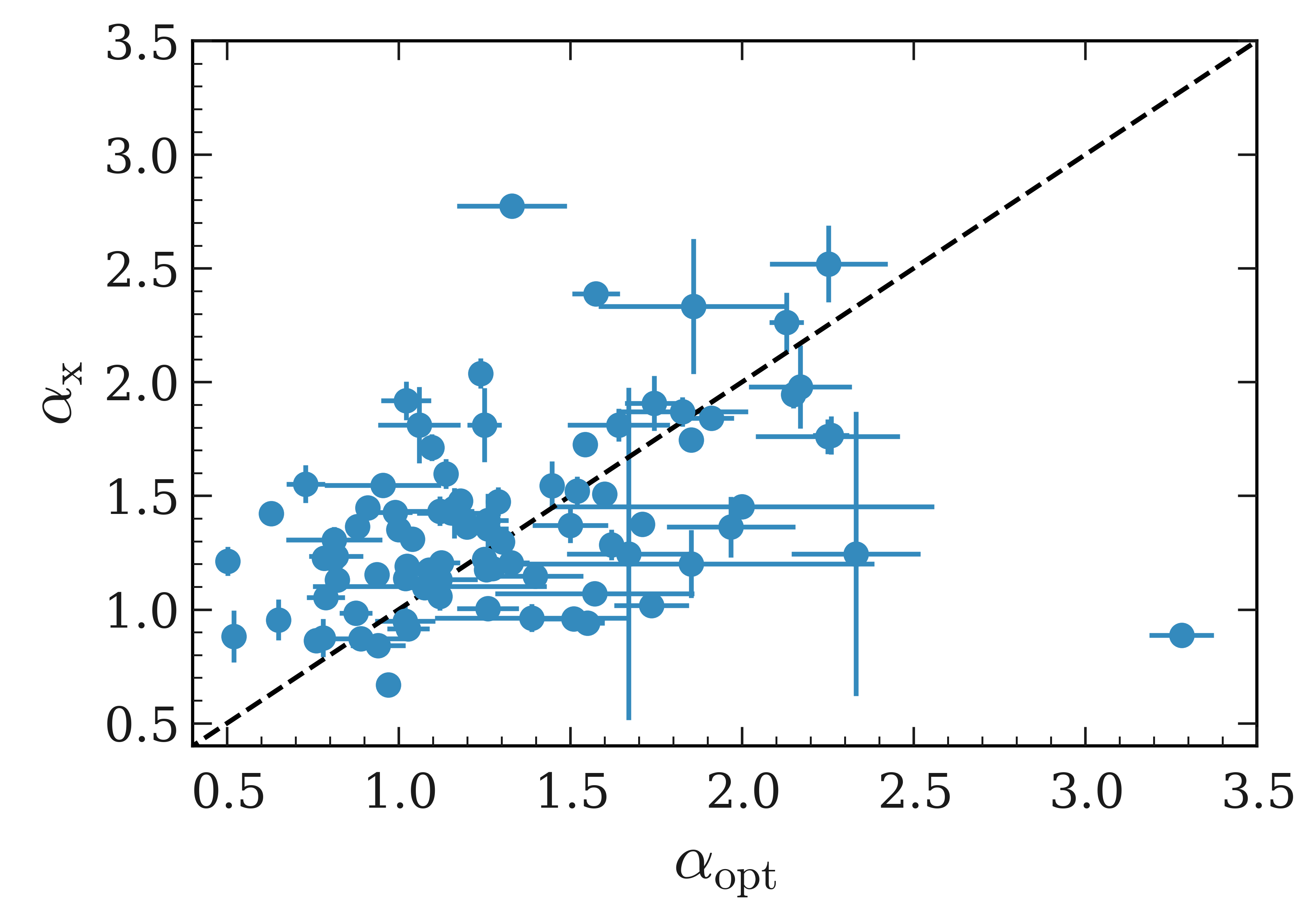}
    \includegraphics[width=0.45\textwidth]{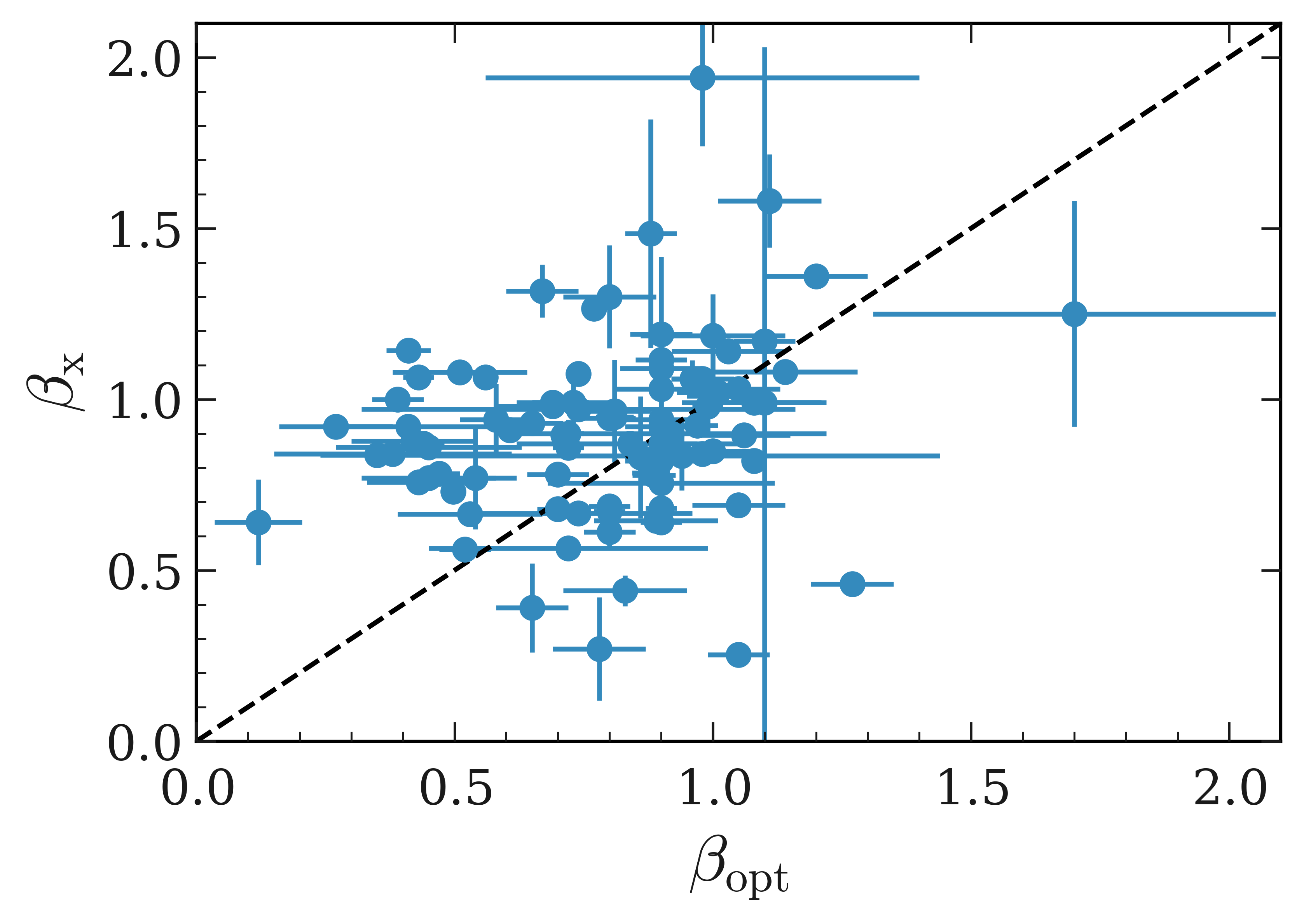}
    \includegraphics[width=0.45\textwidth]{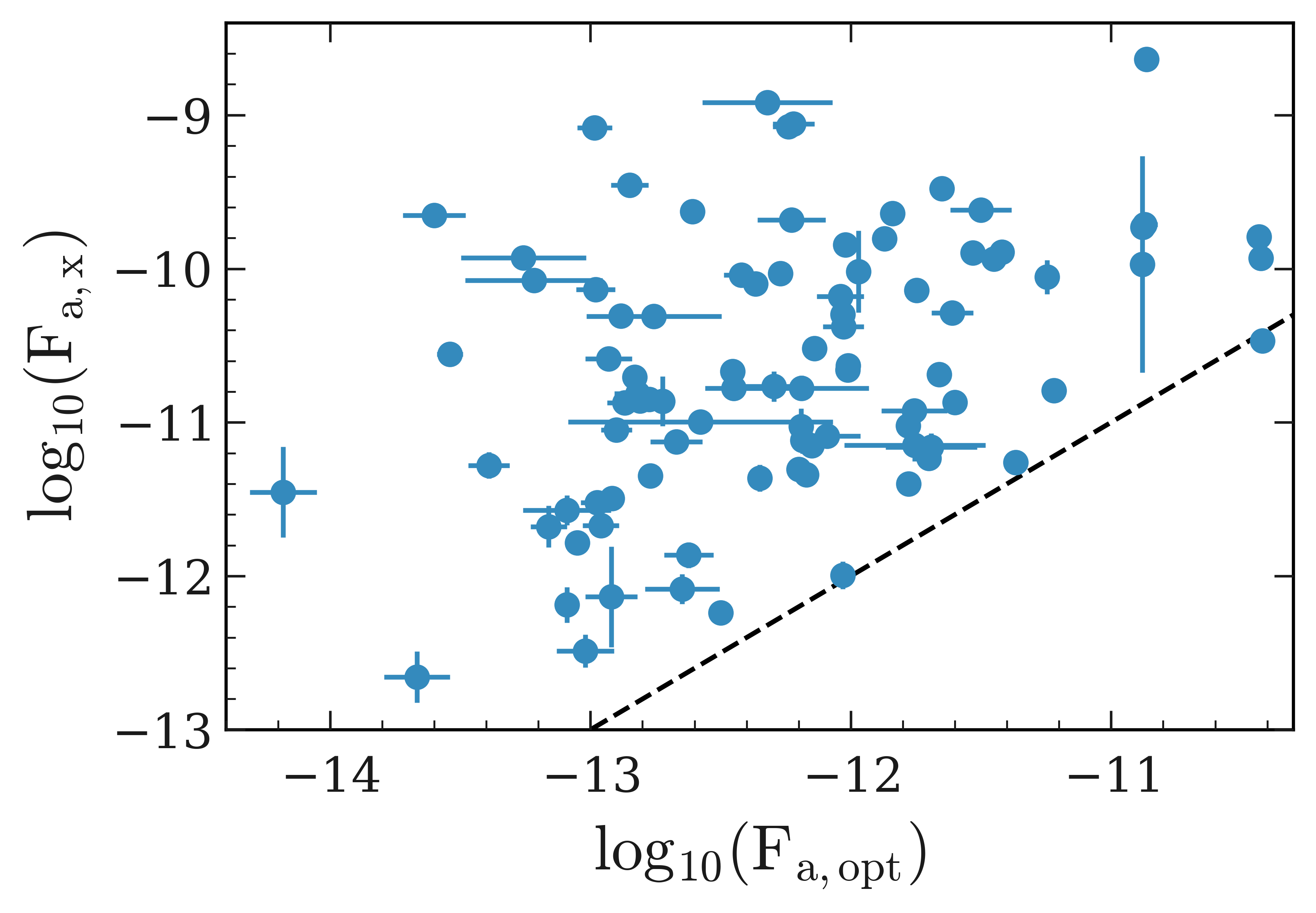}
    \includegraphics[width=0.45\textwidth]{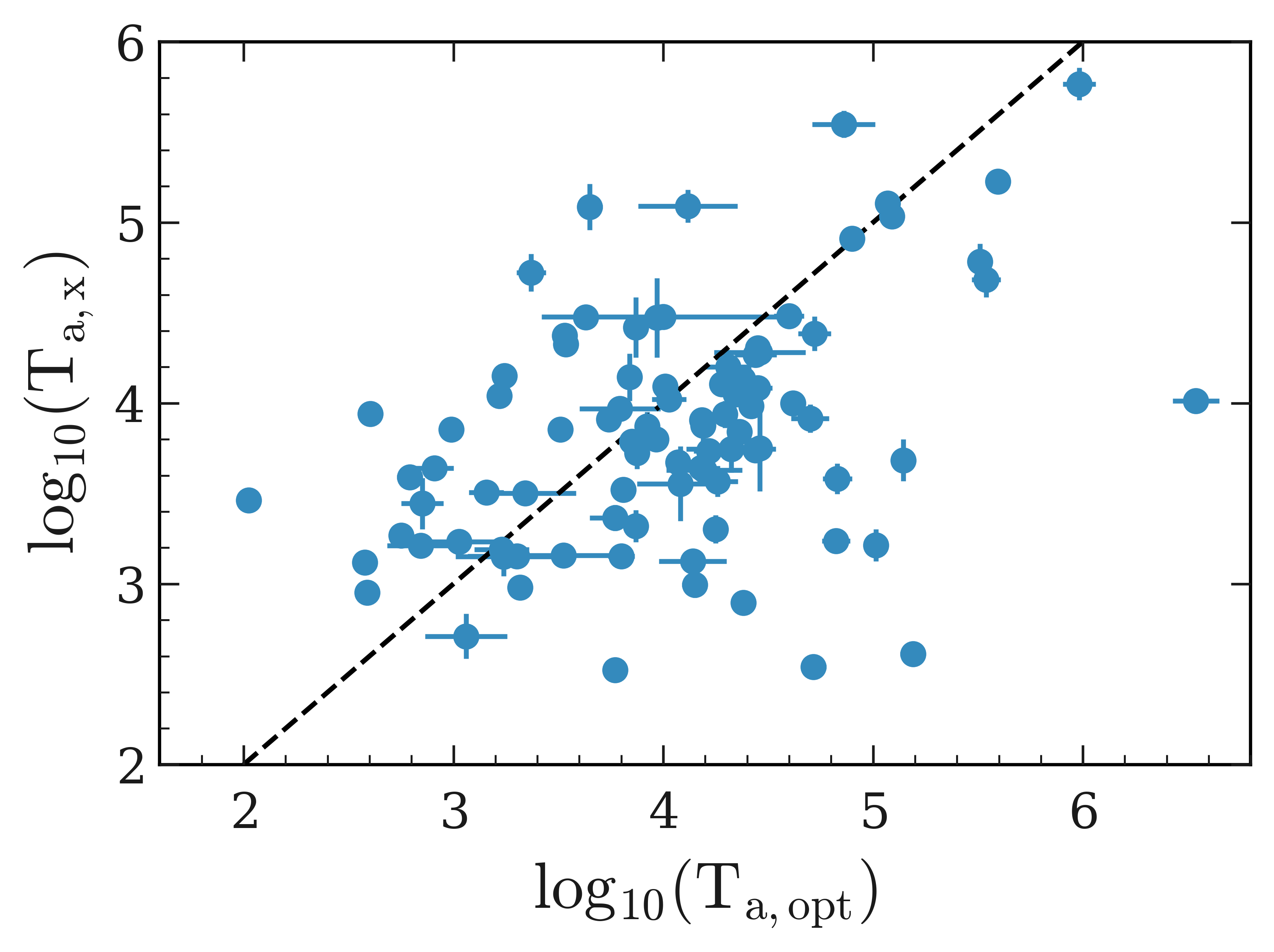}
    \caption{Optical vs X-ray classifier input parameter correlation plots. The black dashed line is the 1:1 line of correlation.}
    \label{fig:opt_x_cor}
\end{figure*}

We also compare the input parameters of the two classifiers using a scatter matrix plot in Figure \ref{fig:opt_x_cor}, showing the four parameters ($F_a$, $T_a$, $\alpha$, $\beta$). The other parameters, such as NH and $T_{90}$, have the same values in both samples, so we do not show them. For $T_a$, $\alpha$, and $\beta$, the values scatter symmetrically around the equality line. However, in the $F_a$ plot (bottom left), we can see that $F_{a,\mathrm{x}}$ is higher than $F_{a,\mathrm{opt}}$ by more than an order of magnitude.

\begin{figure*}[htbp]
    \centering
    \includegraphics[width=0.41\textwidth]{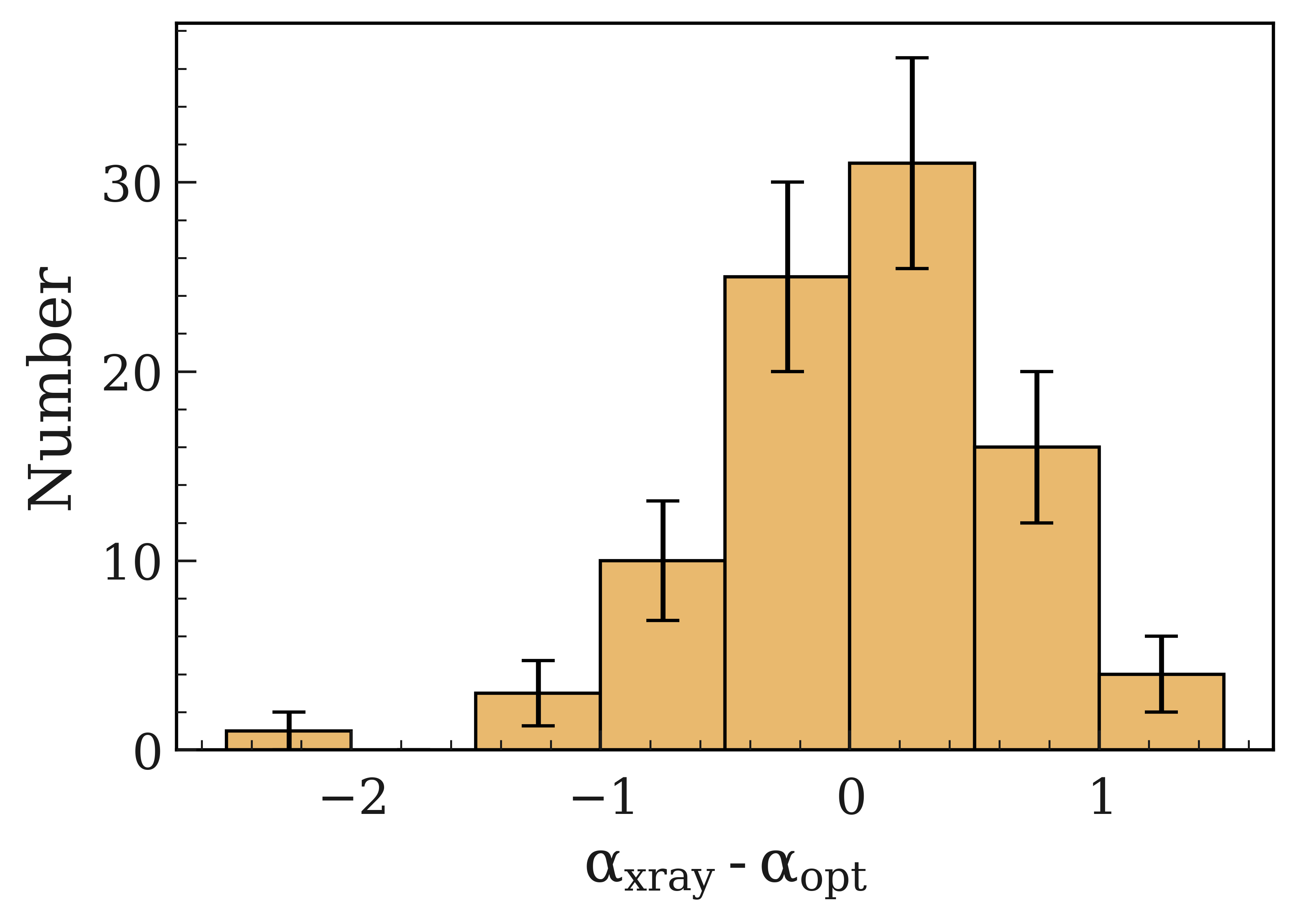}
    \includegraphics[width=0.41\textwidth]{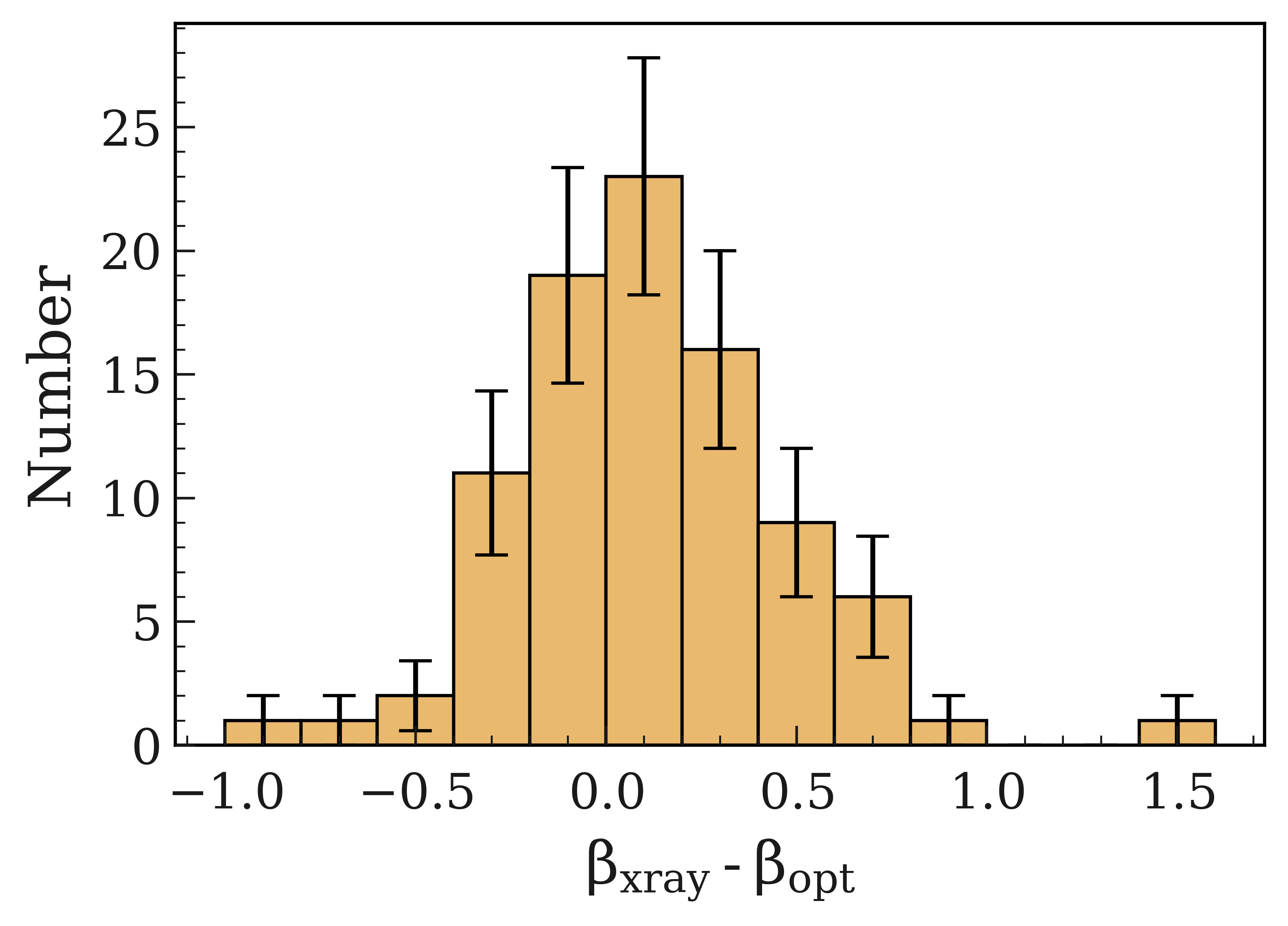}
    \includegraphics[width=0.41\textwidth]{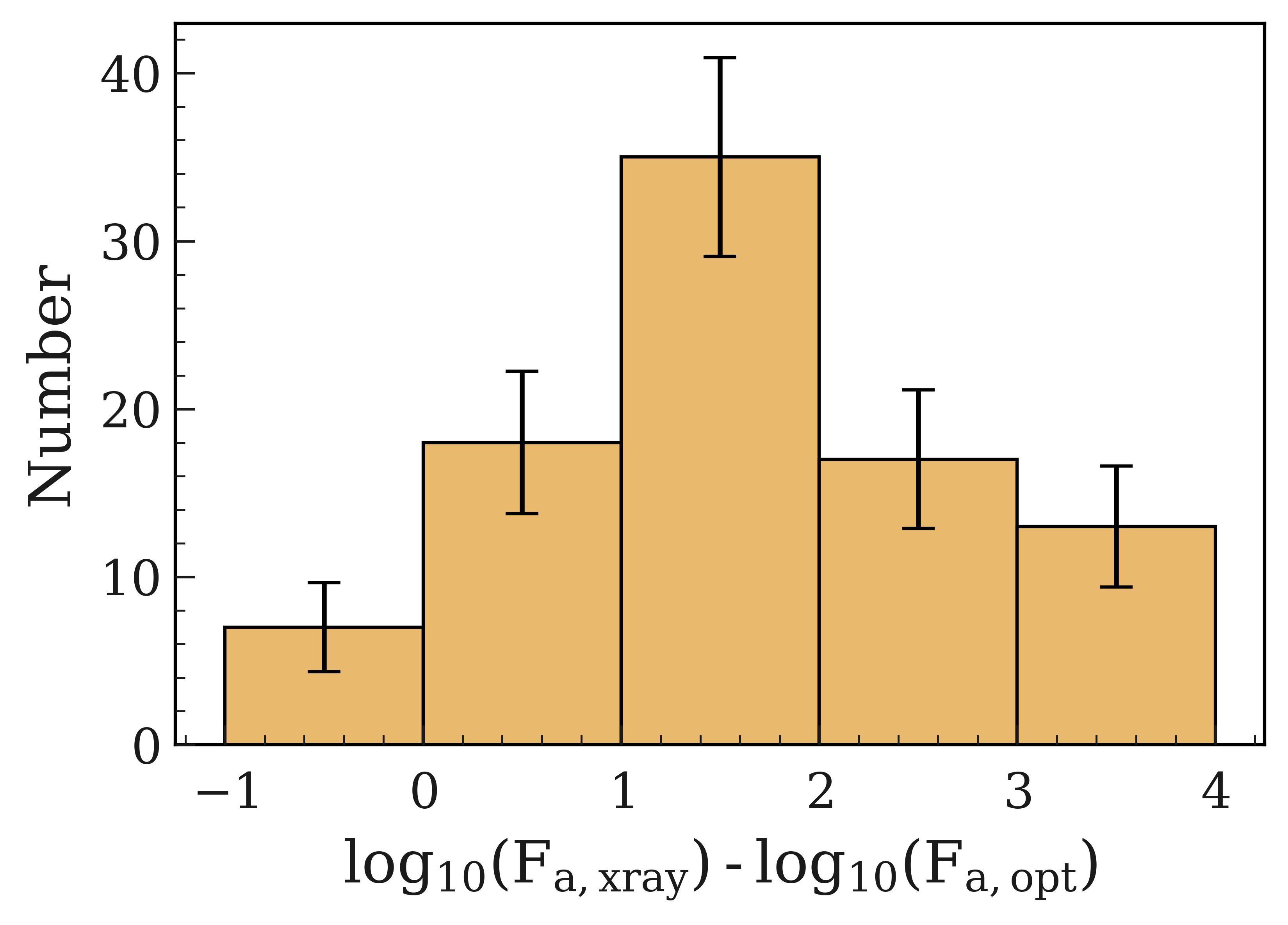}
    \includegraphics[width=0.41\textwidth]{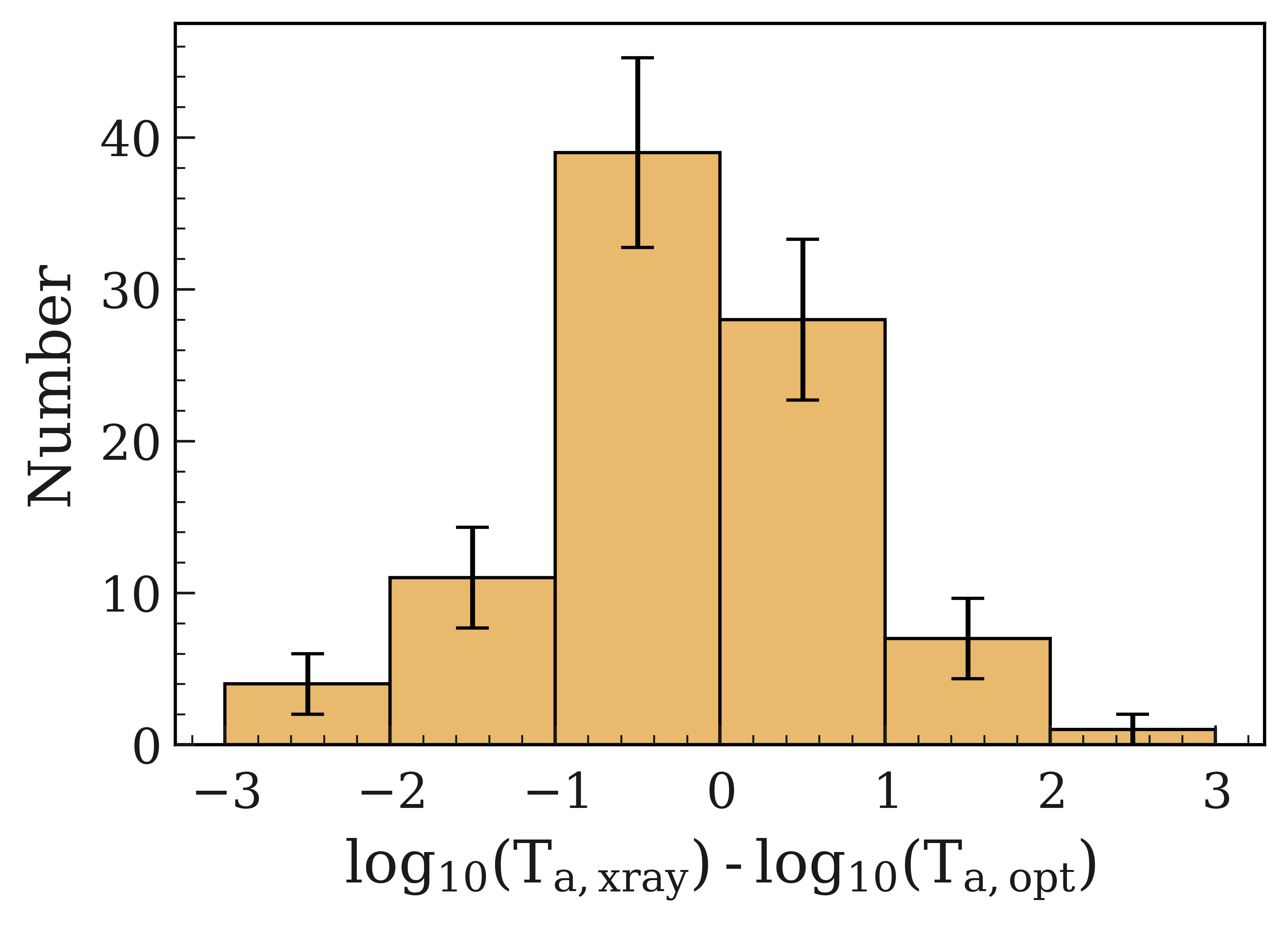}
    \caption{Optical vs X-ray classifier input parameter histograms. The histograms are constructed to compare the two parameter sets. We subtract the optical parameters from the X-ray parameters. The error bars of the histogram are Poisson errors ($\sqrt{N}$).}
    \label{fig:opt_x_hist}
\end{figure*}

Furthermore, Figure \ref{fig:opt_x_hist} shows the distribution of the difference in the input parameters ($F_a$, $T_a$, $\alpha$, $\beta$) between the X-ray and optical classifiers. These distributions show how much each parameter differs between the X-ray and optical samples. The largest spread is seen in the $\log_{10}(F_{a,\mathrm{x}}) - \log_{10}(F_{a,\mathrm{opt}})$ distribution, which is skewed toward positive values, with a mean of 1.62 and a standard deviation of 0.89 -- more than $1\sigma$ away from zero. This is consistent with the correlation plot of $F_a$ in Figure \ref{fig:opt_x_cor}, where $F_{a,\mathrm{x}}$ is systematically higher than $F_{a,\mathrm{opt}}$. The remaining parameter differences are broadly centered around zero, with means within $1\sigma$ of zero, in contrast to the clear offset seen in $F_a$.

Among the shared plateau parameters, only $F_a$ shows such an offset. This offset is unlikely to be a modeling artifact, because both classifiers were built using the same M-estimator, imputation, and feature-selection framework (Appendix~\ref{sec:methodology}). Instead, the offset points to a genuine physical or observational distinction between the optical and X-ray plateau fluxes. Two explanations are plausible and not mutually exclusive. First, the optical and X-ray plateaus may trace physically distinct emission components with different intrinsic flux normalizations, such as the two bands lying in different segments of the synchrotron spectrum, or a long-lived central-engine contribution that is more prominent in X-rays than in the optical band. Second, the optical flux may be systematically reduced relative to the X-ray flux by host-galaxy dust extinction and by selection effects that depend on the observing band, consistent with the observational characteristics of the optical sample. Optical plateau measurements are drawn from heterogeneous ground-based follow-up with variable depth and filter coverage, and are therefore more affected by extinction, and by biases in which plateaus are bright enough to be detected, than the more uniformly sampled X-ray measurements \citep[e.g.,][]{Covino_2013, Greiner2011, Oates_2011, Zaninoni2013}. Because the optical sample is already known to be incomplete owing to dust extinction and irregular observing cadence, we consider the second explanation (observational suppression) the more likely dominant cause, while a genuine difference in emission mechanism remains possible.

Under the second interpretation, which we favor, $F_{a,\mathrm{opt}}$ is a less reliable tracer of the intrinsic afterglow flux than $F_{a,\mathrm{x}}$. This leads to a testable prediction for the LASSO feature-selection step: a feature that is systematically offset and suppressed by observational effects should be a less reliable predictor of redshift than a feature without such an offset. LASSO fits the data with a penalty on the sum of the absolute values of the feature coefficients (Appendix~\ref{sec:methodology}); this penalty sets to zero the coefficients of features that add little information about redshift, so that these features are not selected. A feature whose values are shifted by observational effects should therefore be more likely to receive a zero coefficient than features that are not affected by the same optical-band extinction and selection effects, such as the prompt-emission and absorption observables.
The LASSO results are consistent with this prediction: at $z_t = 2.0$ on the raw data, both with and without the M-estimator (the raw data with the M-estimator yields our best-performing classifier, with training-set TPR = 0.741 and training-set AUC = 0.841), LASSO selects only the prompt-emission properties $\log(\mathrm{PeakFlux})$ and PhotonIndex together with $\log(\mathrm{NH})$, and retains no optical plateau parameter (Table \ref{tab:LASSO}). In this configuration, the redshift information identified by LASSO is therefore carried by these prompt-emission and absorption observables. This does not imply that the optical plateau carries no redshift information, because, by construction, every GRB in this sample possesses an optical plateau. The contrast with the X-ray classifier is marked: $\log(F_a)$ is retained in thirteen of the sixteen configurations examined by MGD25, but in three of the twelve examined here, all three of which are MICE-imputed rather than raw. 

Optical and X-ray plateau data therefore differ in their missing-data patterns and observational biases, and the $F_a$ offset is a quantitative example of this difference.

\subsection{Comparison with Similar X-ray Analysis}
\label{subsec:comparison2}

The objective of both our optical classifier and the X-ray classifier presented in MGD25 is to support high-$z$ GRB identification for timely follow-up observations using ensemble learning across various redshift thresholds and data treatments. 
Although they use a similar methodology, the two classifiers perform best at different redshift thresholds, as detailed below.

The optical model achieves its best performance at $z_t = 2.0$ using the raw dataset with the M-estimator, reaching a training-set TPR of 0.741, a training-set AUC of 0.841, and a train-to-test deviation of $1\% < \Delta < 4\%$. The X-ray classifier, by contrast, reaches its best values at higher thresholds than the optical one. Excluding its SMOTE-balanced configurations, which have no counterpart here, its largest training AUC is 0.849, obtained at $z_t = 3.0$ on MICE-imputed data, while its most favorable generalization (smallest $\Delta$) occurs at $z_t = 3.5$. There, for the raw data with the M-estimator, $\Delta$ is between $0.5\%$ and $10\%$ and the training-set TPR is 0.667.

A meaningful comparison between the two classifiers requires matching the redshift threshold, since sensitivity depends on $z_t$ for both classifiers. At $z_t = 2.0$ on raw data with the M-estimator, the two classifiers are almost indistinguishable in training-set sensitivity: the optical model reaches a TPR of 0.709 before removal of the least-performing algorithms and 0.741 after, against 0.742 and 0.712 respectively for the X-ray model. They differ substantially, however, in how well that performance transfers to the test set. The optical classifier attains test AUC values of 0.837 and 0.812 against training values of 0.826 and 0.841, giving $1\% < \Delta < 4\%$, whereas the X-ray classifier attains test AUC values of 0.491 and 0.470 against training values of 0.798 and 0.809, giving $\Delta$ of $48\%$ and $53\%$. At this threshold, the optical classifier therefore generalizes considerably better, whereas MGD25 identifies $z_t = 3.5$, where the X-ray $\Delta$ is smallest, as the preferred configuration of the X-ray classifier.

This comparison applies to this single threshold. The X-ray classifier generalizes well at $z_t = 3.5$, a threshold suited to its larger sample; for the optical classifier, modeling a high-$z$ class above $z = 3.5$ would require more bursts with detected optical plateaus than are currently available. Conversely, the optical classifier generalizes well at $z_t = 2.0$. Each classifier therefore generalizes well at a different threshold. This matters for follow-up planning: a campaign targeting $z > 2$ candidates is best served by the optical classifier, and one targeting $z > 3.5$ candidates by the X-ray classifier.

The optical classifier attains its highest training-set TPR on the raw datasets; at $z_t = 2.0$, its TPR on the MICE-imputed dataset without the M-estimator is 61\% of that obtained on the raw dataset with the M-estimator. Possible reasons for this difference are given in Section~\ref{subsec:comparison}. For the X-ray classifier, MICE imputation leaves the training-set TPR essentially unchanged at $z_t = 2.0$ (0.742 to 0.730) and raises it modestly at $z_t = 2.5$ and $z_t = 3.0$ (0.583 to 0.667 and 0.625 to 0.667, respectively), with larger gains under SMOTE balancing. This difference may be related to the smoother temporal and spectral behavior of X-ray afterglows and their more uniform sampling cadence.

The features selected by LASSO partly overlap between the two classifiers. Two are retained in nearly every configuration: $\log(\mathrm{NH})$ in all twelve optical and all sixteen X-ray configurations (MGD25), and $\log(\mathrm{PeakFlux})$ in eleven of twelve and all sixteen, respectively. Both therefore help separate high-$z$ from low-$z$ bursts regardless of the plateau emission and the missing-data treatment. $\mathrm{PhotonIndex}$ is also shared but selected less often, in seven of the twelve optical and eleven of the sixteen X-ray configurations; it is absent from the X-ray raw datasets at $z_t = 2.5$ and $z_t = 3.5$ and from three of the four optical configurations at $z_t = 3.0$. We therefore count only $\log(\mathrm{NH})$ and $\log(\mathrm{PeakFlux})$ as consistently selected by both classifiers.

It is physically plausible that both classifiers select these features: $\log(\mathrm{NH})$ traces line-of-sight neutral hydrogen absorption, which has long been reported to increase with redshift in Swift GRB samples \citep{Grupe_2006}; larger samples confirmed an intrinsic $N_H(z)$ trend, attributed to denser host and circumburst environments at earlier cosmic epochs \citep{Campana_2012}. $\mathrm{PhotonIndex}$, the hardness of the prompt $\gamma$-ray spectrum, was also a useful high-$z$ discriminator in earlier prompt-emission classifiers \citep{morgan2012, ukwatta2016machine}, though its less consistent selection here suggests a weaker or more sample-dependent link with redshift. All three features ($\log(\mathrm{NH})$, $\log(\mathrm{PeakFlux})$, and PhotonIndex) are measured from high-energy prompt or X-ray spectral data rather than from optical or X-ray plateau observations, so they are largely unaffected by the band-specific selection effects discussed in Sections~\ref{sec:sample} and \ref{subsec:comparison}, consistent with their selection by both classifiers.

For the optical classifier, LASSO selects $\log(T_{90})$, $\log(F_a)$, $\log(T_a)$, $\alpha$, and $\beta$ only in some datasets and at some thresholds, suggesting that their selection depends more on which bursts are in each dataset and that their correlation with redshift varies across datasets and thresholds. 

LASSO selects features somewhat more consistently for the X-ray classifier, but only for specific features. The clearest difference is in the plateau flux: $\log(F_a)$ is retained in thirteen of the sixteen X-ray configurations but in only three of the twelve optical ones, and all three use the MICE-imputed dataset with the M-estimator, one at each redshift threshold; none of the raw datasets retains it (Table~\ref{tab:LASSO} and Section~\ref{subsec:comparison}). $\log(T_{90})$ shows a milder version of the same pattern (thirteen of sixteen against six of twelve). $\log(T_a)$, in contrast, appears in only three of the sixteen X-ray and three of the twelve optical configurations, so it does not distinguish the two classifiers and is not part of a stable X-ray core. Overall, the X-ray classifier retains more features in most of its dataset variants, implying that X-ray temporal and spectral properties correlate with redshift somewhat more consistently across its training sample.

The optical classifier is an important, independent complement to the X-ray model: it is calibrated on the optically selected population, in which most secure GRB redshifts are measured through optical and near-infrared spectroscopy (Section~\ref{sec:results}), and it can be applied when X-ray plateau coverage is unavailable or incomplete.

In summary, among our optical configurations, the raw data with the M-estimator at $z_t = 2.0$ give the best combination of sensitivity (training-set TPR) and generalization ($\Delta$). The X-ray classifier reaches comparable generalization at a higher threshold, on a substantially larger sample that makes this redshift range accessible. These findings motivate combining the optical and X-ray classifiers, so that bursts can be classified reliably at every threshold relevant to follow-up prioritization.

As shown in Section~\ref{subsec:comparison}, at $z_t = 2.0$ on the raw data, LASSO selects only $\log(\mathrm{PeakFlux})$, PhotonIndex, and $\log(\mathrm{NH})$, and no optical plateau parameter. Because two of these three features are prompt $\gamma$-ray properties, this configuration could also serve for a rapid first ranking of new bursts for follow-up, before an optical plateau can be measured. In practice, $\log(\mathrm{PeakFlux})$ and PhotonIndex come from BAT prompt-emission data and are available within minutes of the trigger, whereas $\log(\mathrm{NH})$ comes from time-averaged XRT spectral fitting of accumulated X-ray follow-up, so how soon the classifier can be applied to a given burst depends on how quickly its $N_{\mathrm{H}}$ can be constrained.

This absence of optical plateau parameters is specific to this configuration: at $z_t = 2.5$ on the raw data and at all three thresholds under MICE imputation with the M-estimator, LASSO retains at least one plateau parameter ($F_a$, $T_a$, $\alpha$, or $\beta$; Section~\ref{sec:results_auc_overall} and Table~\ref{tab:LASSO}). The $z_t = 2.0$ raw-data result therefore shows that prompt-emission and absorption features are highly informative and, in this configuration, are the only features LASSO picks; it does not imply that optical plateau measurements are unnecessary at other thresholds or for other datasets. Optical plateau observables are closely linked to the afterglow emission and carry information about the burst energetics, circumburst environment, and plateau evolution \citep{dainotti2020b, Dainotti2022}, which a generic imputation procedure is not guaranteed to recover in full.

\subsection{Comparison with earlier results} \label{subsec:comparison3}

We also compare our optical classifier with earlier ML classifiers of GRB redshift \citep{morgan2012, ukwatta2016machine, 2024MNRAS.529.2676A}. All three relied exclusively on prompt $\gamma$-ray or early X-ray observables, such as $T_{90}$, PhotonIndex, and PeakFlux, and none used afterglow plateau information from either the optical or the X-ray band. \citet{morgan2012} and \citet{ukwatta2016machine} each trained a single Random Forest model on their full sample, without a held-out test set. Using 135 \textit{Swift} LGRBs, \citet{morgan2012} ranked the bursts by their classifier probability of being at high-$z$ and found that the top 40\% of this ranked list would capture $\approx 84\%$ of all high-$z$ ($z>4$) bursts. At $z_t = 4.0$, \citet{ukwatta2016machine} reported 80\% recall of true high-$z$ bursts at a 20\% false-positive rate, corresponding to an AUC of approximately 0.89. \citet{2024MNRAS.529.2676A} used prompt Fermi Gamma-ray Burst Monitor (GBM) and Konus-Wind observables with Random Forest and deep-learning models, reporting AUC values in the range $\sim 0.8$--$0.85$.

Our sample of 171 LGRBs is 27\% larger than that of \citet{morgan2012}. Our best-performing model (raw data with the M-estimator, $z_t = 2.0$) reaches a training-set AUC of 0.841. This value lies within the $\sim 0.8$--$0.85$ range of \citet{2024MNRAS.529.2676A} and is comparable to the AUC of $\approx 0.89$ of \citet{ukwatta2016machine}. It is also of the same order as the capture rate of \citet{morgan2012}, although that metric is not an AUC and is not directly comparable. Because \citet{morgan2012,ukwatta2016machine} report performance only on their training sets, this is a like-for-like comparison of training-set metrics, not of performance on held-out test data.

The X-ray classifier of MGD25 adds X-ray plateau observables, measured with the relatively uniform, space-based Swift/XRT, to the prompt-emission features used by these earlier classifiers. The present work instead separates high-$z$ from low-$z$ bursts within the optically selected population, i.e., bursts with a detected and characterized optical plateau. 

In our best-performing configuration ($z_t = 2.0$), the features retained by LASSO are prompt-emission and absorption observables of the same type used by \citet{morgan2012} and \citet{ukwatta2016machine}. In this configuration, our classifier therefore uses the same type of features as these earlier classifiers (see Section~\ref{subsec:comparison2}); its distinguishing element is the optically selected population on which it is trained.

At $z_t = 3.0$, the raw data with the M-estimator give a training-set AUC of 0.875, comparable to or above the values of these earlier studies. The corresponding training-set TPR of 0.333 reflects the smaller high-$z$ class at this threshold.

Our evaluation procedure also differs. We evaluate performance on an 80/20 train--test split at three redshift thresholds ($z_t = 2.0$, $2.5$, $3.0$) and on four preprocessing configurations (Section~\ref{sec:results}). The held-out test set gives a more conservative estimate of generalization than the full-sample training of \citet{morgan2012} and \citet{ukwatta2016machine}. Because the procedures differ, AUC values from different studies are compared approximately.

Taken together, these results show that our optical classifier reaches training-set AUC values comparable to those of earlier prompt- and X-ray-based classifiers. It does so with optical data that are much more heterogeneous and less complete than the prompt-selected samples of those studies, and with optical plateau-phase features in its candidate feature pool that those studies did not consider. The held-out test AUC and $\Delta$ for every configuration in Table~\ref{tab:AUC_for_SuperLearner} show that, for our best-performing configuration, the training-set performance is similar to the performance on the held-out test set ($1\% < \Delta < 4\%$). 
\subsection{The user-friendly web-app }
\label{subsec:webapp}

\begin{figure}[h!]
    \centering
    \fbox{\includegraphics[scale=0.2975]{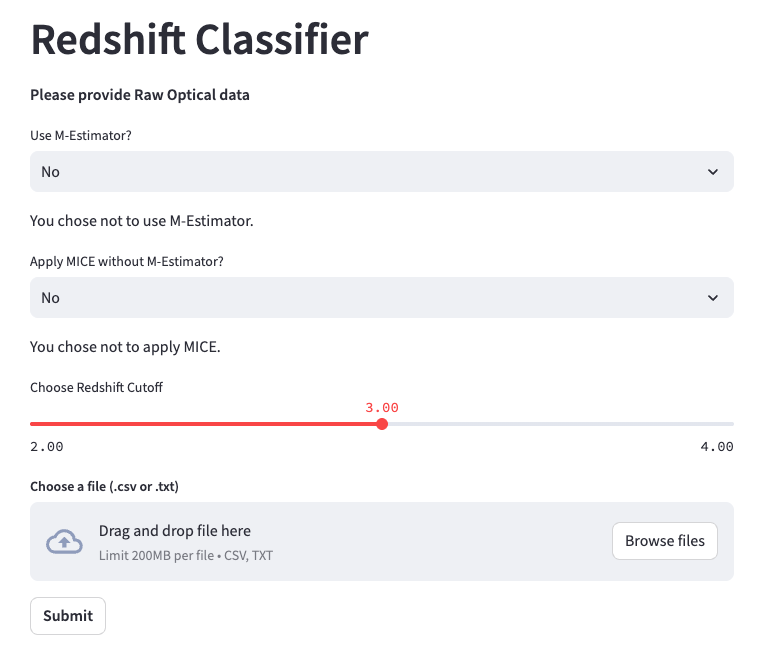}}%
    \hfill
    \fbox{\includegraphics[scale=0.3]{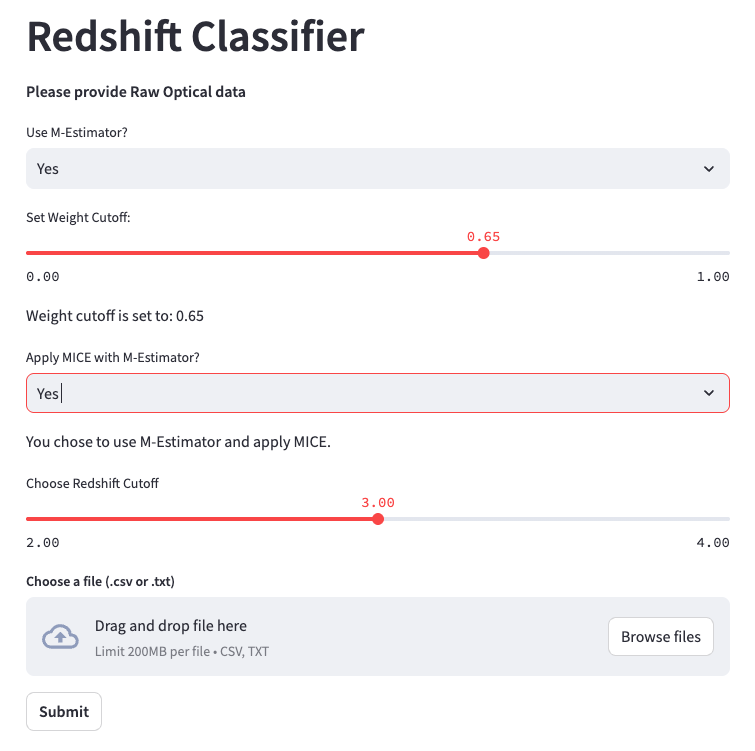}}
    \caption{Web-app interface when the user applies neither the M-estimator nor MICE imputation (top) and when the user applies both (bottom) before running the SuperLearner. The user also selects $z_{t}$ within the app.}
    \label{fig:APP}
\end{figure}

To make our work accessible, we built a web app with Streamlit, a tool for creating interactive data-science apps. The app runs locally on any machine and is available on GitHub: \href{https://github.com/Milind018/Redshift-Classifier-Optical}{Redshift-Classifier-Optical} (Figure~\ref{fig:APP}). Users select the value of $z_{t}$ and can choose to apply MICE to impute missing data and/or the M-estimator for outlier removal. They can also choose the weight threshold below which a GRB is removed as an outlier with the M-estimator. 

If neither the M-estimator nor MICE is applied, the app passes the raw data to the SuperLearner pipeline after basic filtering (removing missing values and data points with $>100\%$ relative error). When selected, the M-estimator runs first and MICE second, both before SuperLearner. 

The pipeline performs feature selection with LASSO and model selection with SuperLearner, using 100 nested 10fCVs, and produces plots of the selected features, the model coefficients, and the ROC curves and AUC scores for the training and test sets.

\section{Validation of the Shared Pipeline Using the X-ray Classifier on the Generalization Set}\label{sec:generalisation}

This section applies the X-ray classifier of MGD25 to the generalization set of \citet{Narendra_2025}. We perform this exercise because both classifiers are built with the same implementation of SuperLearner, MICE, and the M-estimator (Appendix~\ref{sec:methodology}). Applying the X-ray classifier to an independent sample is therefore an external check of this shared code. The exercise also independently assesses the X-ray classifier.

\begin{figure}[ht]
    \centering
    \includegraphics[width=0.45\textwidth]{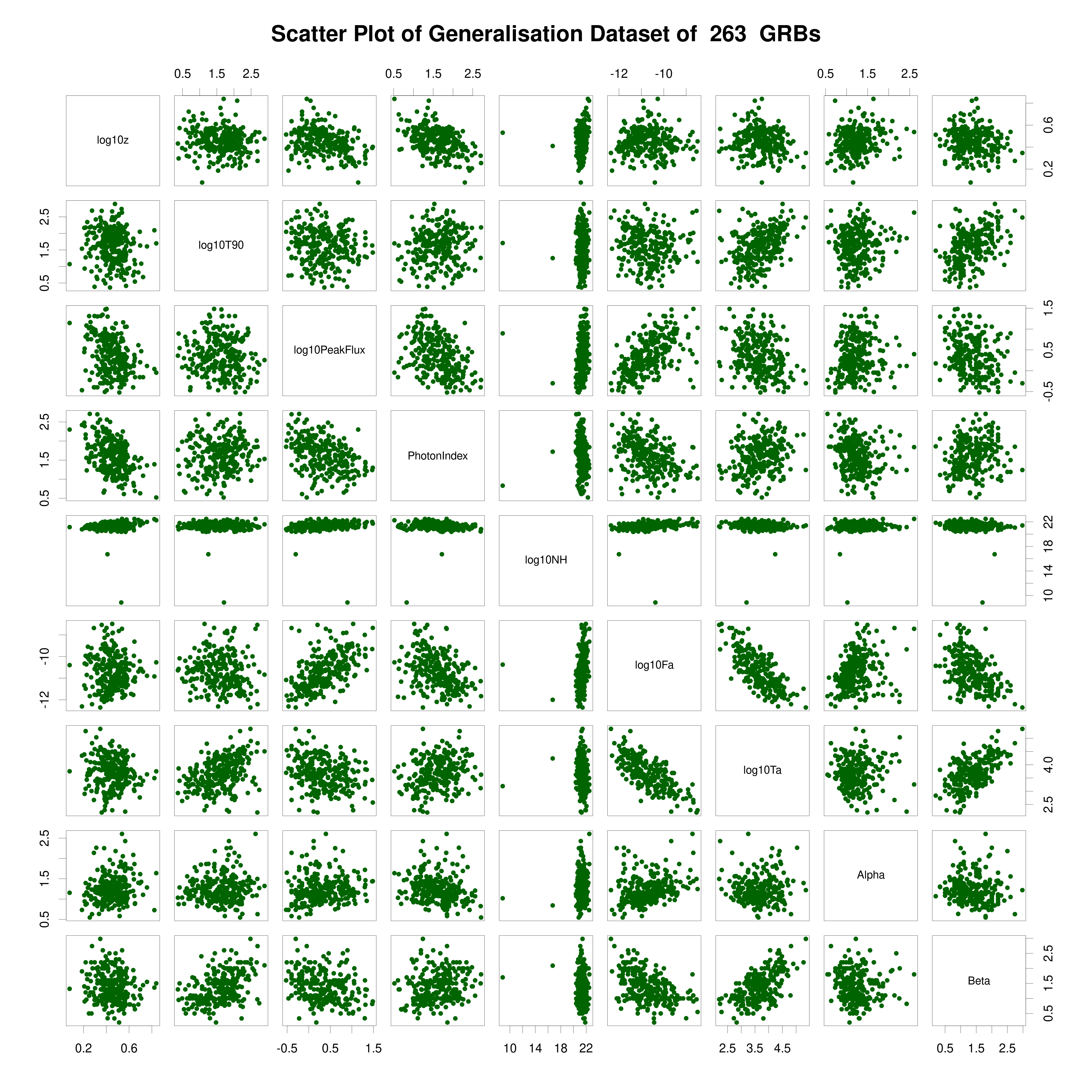}
    \caption{Scatter plot of the plateau and prompt-emission features of the generalization set used to validate the X-ray classifier of MGD25.}
    \label{fig:gen_scatter}
\end{figure}

The generalization set contains 263 LGRBs that lacked spectroscopic redshifts in that compilation but have X-ray plateau and prompt-emission features (Figure~\ref{fig:gen_scatter}) and regression-inferred redshifts. This set is not the standard validation set for the X-ray classifier, but comparing the classifier's labels with the regression-inferred redshifts tests whether it separates low- and high-redshift GRBs in the same generalization sample.

Trained with MICE and the M-estimator at $z_t = 3.5$ and scored against the regression-inferred redshifts, the X-ray SuperLearner classifier achieves an overall accuracy of $97.34\%$, a perfect specificity $(100\%)$, and a sensitivity of $12.5\%$. In the confusion matrix, 255 GRBs are correctly labeled low-$z$, with no false positives. Of the eight GRBs that are high-$z$ according to their regression-inferred redshifts, one is labeled high-$z$ (TP) and seven low-$z$ (FN).

The X-ray classifier is thus calibrated to avoid false positives. It assigns the high-$z$ label only to bursts whose features most strongly indicate a high redshift, so its high-$z$ labels are highly reliable, while most of the remaining high-$z$ bursts are labeled low-$z$.

\begin{figure}[ht]
    \centering
    \includegraphics[width=0.45\textwidth]{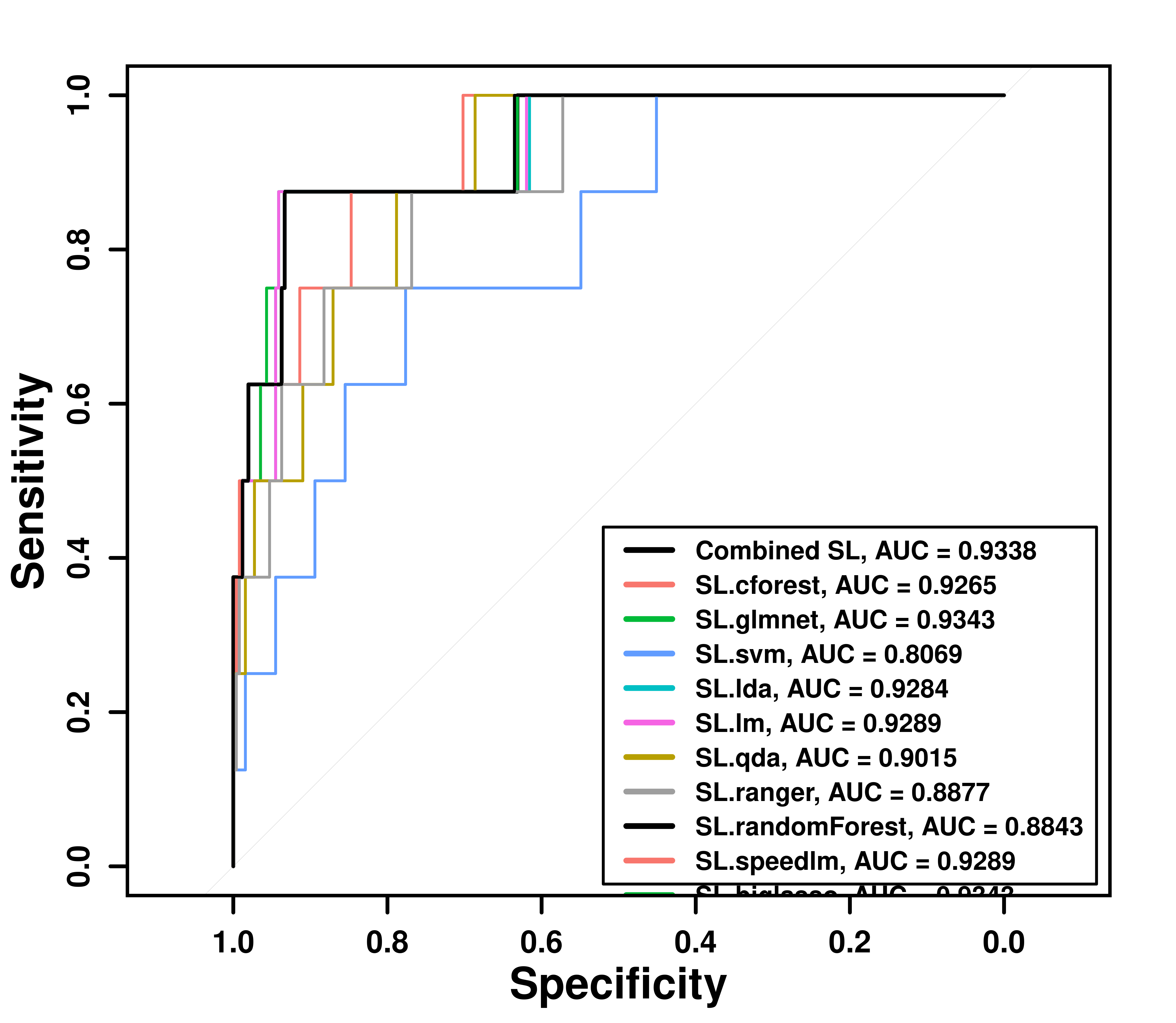}
    \caption{ROC curve for the generalization set, for the X-ray classifier of MGD25 at $z_t = 3.5$.}
    \label{fig:auc_plot}
\end{figure}

The ROC curve (Figure~\ref{fig:auc_plot}) illustrates this trade-off between specificity and sensitivity. The full SuperLearner ensemble attains an AUC of 0.934, and the best-performing individual learner (cforest) attains 0.927.
Overall, the X-ray classifier is well suited to selecting high-$z$ candidates with few false positives and to identifying targets of interest among low-$z$ bursts, which include intriguing cases of LGRBs associated with kilonovae (KNe).


\subsection{Comparison of Known Redshifts with X-ray Classifier and Estimator Results}
As a secondary objective, we test whether, for GRBs with known redshifts, the binary labels of the X-ray classifier ($z_t = 3.5$) and the continuous predictions of the regression-based estimator \citep{Narendra_2025} are consistent with the measured redshifts. 
\begin{figure}[ht]
    \centering
    \includegraphics[width=\columnwidth]{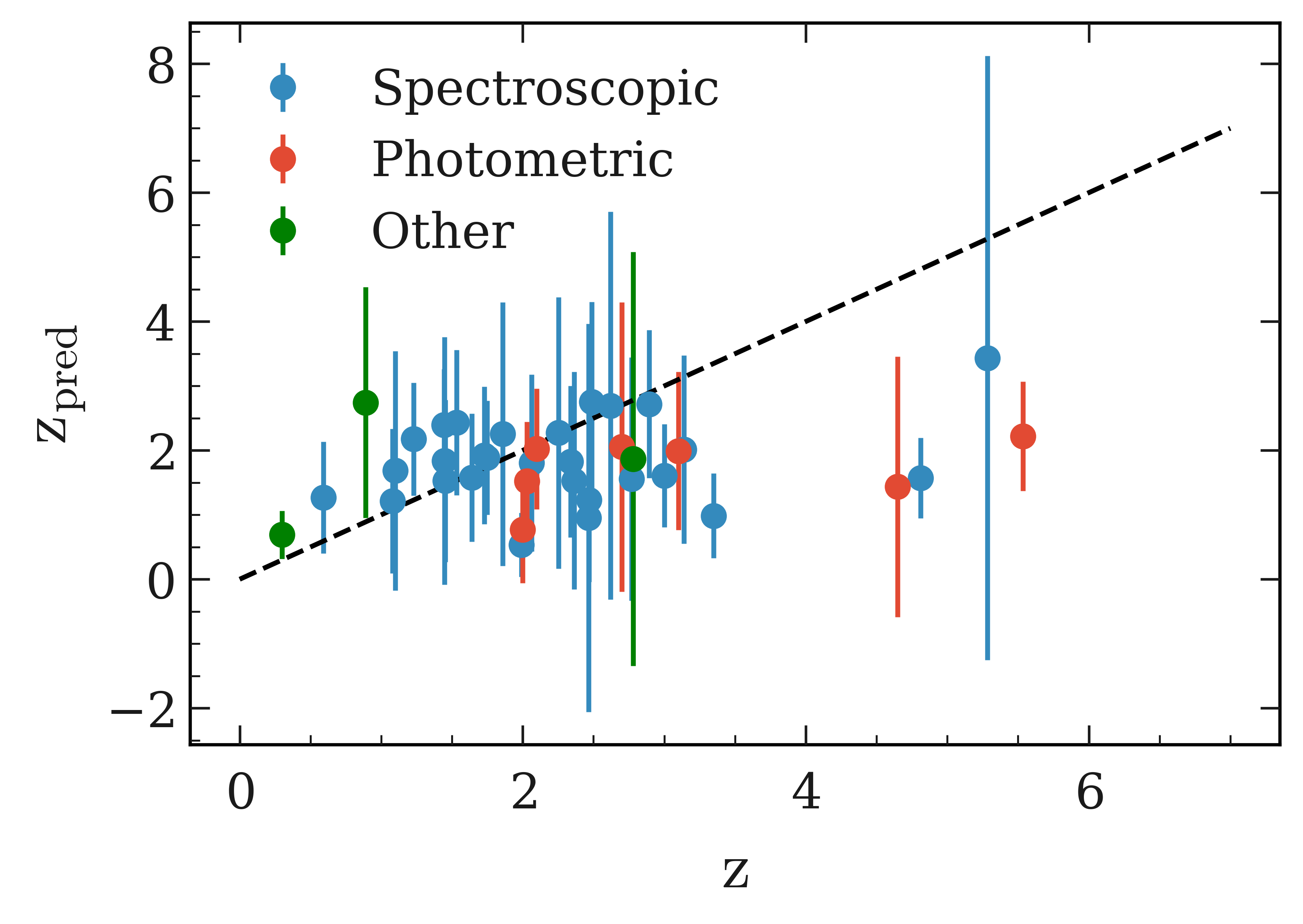}
    \includegraphics[width=\columnwidth]{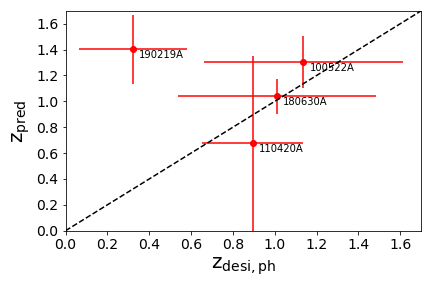}
    \caption{Upper panel: measured versus regression-predicted redshifts for the GRBs with known redshifts. The measured values were initially retrieved from GRBweb. Original references are listed in Table \ref{tab:GRBs_with_z}. Blue and red points mark spectroscopic and photometric redshifts, respectively, and green points mark GRB 130427B and GRB 150424A. The black dashed line is the 1:1 line. Lower panel: comparison of the Dark Energy Spectroscopic Instrument Legacy Survey (DESI/LS) photometric redshifts of candidate host galaxies with the regression-predicted redshifts.}
    \label{fig:zgrbweb_zpred}
\end{figure}
We then searched GRBweb \citep{grbweb_cite} for measured redshifts of GRBs in the generalization set and found 42: 29 spectroscopic, 11 photometric, and two inferred from a candidate host galaxy. The upper panel of Figure \ref{fig:zgrbweb_zpred} shows that the measured and regression-predicted redshifts closely follow the 1:1 line within the error bars.

The X-ray classifier's label agrees with the measured redshift in 38 of the 42 cases ($90.5\%$). The four disagreements (Table \ref{tab:GRBs_with_z}) are all false negatives (label $z < 3.5$, measured redshift above $3.5$), consistent with the conservative sensitivity and perfect specificity reported above. Two of these redshifts are photometric and two are spectroscopic. Excluding the two photometric cases, which are less secure at $z > 4$, gives an agreement rate of $95\%$ over the remaining 40 GRBs.
Green points in the upper panel of Figure~\ref{fig:zgrbweb_zpred} mark GRB 130427B and GRB 150424A. The redshift of GRB 130427B is tentative. The redshift of GRB 150424A is debated because two of the references listed in Table \ref{tab:GRBs_with_z} give different values for it. According to its reference, the host galaxy of GRB 161001A is a candidate, not confirmed.
A fifth GRB, 190219A, is also marked as a disagreement in Table \ref{tab:GRBs_with_z}. Its measured redshift ($0.32 \pm 0.26$) lies well below the threshold, and the X-ray classifier produced no false positives. This entry reflects a discrepancy with the continuous regression prediction, not with the binary classification, and is not counted among the four classifier disagreements.

We also searched for photometric redshifts in DESI/LS \citep{DESI}, whose magnitude limit restricts this search to GRBs with $z<1.5$. For the GRBs in the generalization set, we recorded the DESI/LS photometric redshifts of host-candidate galaxies within 5$^{\prime\prime}$ of the GRB positions. The lower panel of Figure \ref{fig:zgrbweb_zpred} compares these DESI/LS photometric redshifts with the regression-predicted redshifts. The two agree well for predicted redshifts below $1.2$.

\begin{table*}[!ht]
\centering
\begin{tabular}{c||c||c||c||c}
\hline
\hline
\multicolumn{5}{c}{\textbf{GRB Redshift Comparison Table}}\\
\hline
\hline
\textbf{GRB} & \textbf{Redshift} & \textbf{Redshift Type} & \textbf{Reference} & \textbf{Agreement}\\
\hline
\hline
050814A & $5.3 \pm 0.3$ & photometric & \citep{Jakobsson:2005jc} & No\\
\hline
060108A & 2.03 & photometric & \citep{2006GCN..4539....1M}& Yes\\
\hline
060719A & 1.532 & spectroscopic & \citep{2012ApJ...758...46K} & Yes\\
\hline
060805A & 2.363 & spectroscopic & \citep{Kruhler:2015ala}& Yes\\
\hline
061202A & 2.2543 & spectroscopic & \citep{Kruhler:2015ala}& Yes\\
\hline
070103A & 2.6208 & spectroscopic & \citep{2012ApJ...758...46K}& Yes\\
\hline
070129A & 2.3384 & spectroscopic & \citep{2012ApJ...758...46K}& Yes\\
\hline
070224A & 1.992 & spectroscopic & \citep{Kruhler:2015ala}& Yes\\
\hline
070328A & 2.06 & spectroscopic & \citep{Kruhler:2015ala}& Yes\\
\hline
080605A & 1.6403 & spectroscopic & \citep{2009ApJS..185..526F}& Yes\\
\hline
080707A & 1.23 & spectroscopic & \citep{2008GCN..7949....1F} & Yes\\
\hline
081222A & 2.77 & spectroscopic & \citep{2008GCN..8713....1C} & Yes\\
\hline
090113A & 1.7493 & spectroscopic & \citep{2012ApJ...758...46K}& Yes\\
\hline
090407A & 1.4478 & spectroscopic & \citep{2012ApJ...758...46K}& Yes\\
\hline
100302A & 4.813 & spectroscopic & \citep{2010GCN.10466....1C}& No\\
\hline
100424A & 2.4656 & spectroscopic & \citep{2013GCN.14291....1M}& Yes\\
\hline
100522A & $1.14 \pm 0.47$ & photometric & This Work & Yes \\ \hline
110420A & $0.89 \pm 0.24$ & photometric & This Work & Yes \\
\hline
110801A & 1.858 & spectroscopic & \citep{2011GCN.12234....1C}& Yes\\
\hline
111107A & 2.893 & spectroscopic & \citep{2011GCN.12537....1C}& Yes\\
\hline
111129A & 1.080 & spectroscopic & \citep{Kruhler:2015ala}& Yes\\
\hline
120119A & 1.728 & spectroscopic & \citep{2012GCN.12865....1C}& Yes\\
\hline
120224A & 1.1 & spectroscopic & \citep{Kruhler:2015ala}& Yes\\
\hline
121123A & $2.7 \pm 0.3$ & photometric & \citep{2012GCN.13992....1S}& Yes\\
\hline
121209A & $\sim 2.1$ & photometric & \citep{Kruhler:2015ala}& Yes\\
\hline
121217A & $3.1 \pm 0.1$ & photometric & \citep{Elliott:2013tfa}& Yes\\
\hline
130427B & 2.78 (tentative) & spectroscopic & \citep{2013GCN.14493....1F}& Yes\\
\hline
140114A & 3 & spectroscopic & \citep{Kruhler:2015ala}& Yes\\
\hline
140304A & 5.283 & spectroscopic & \citep{2014GCN.15924....1D}& No\\
\hline
140331A & 4.65 & photometric & \citep{2014GCN.16050....1L}& No\\
\hline
140703A & 3.14 & spectroscopic & \citep{2014GCN.16505....1C}& Yes\\
\hline
141026A & 3.35 & spectroscopic & \citep{2014GCN.16968....1D}& Yes\\
\hline
141221A & 1.452 & spectroscopic & \citep{2014GCN.17228....1P}& Yes\\
\hline
150323A & 0.593 & spectroscopic & \citep{2015GCN.17616....1P}& Yes\\
\hline
150424A & $0.3$ or $1^{+0.3}_{-0.2}$ & candidate host & \citep{2015GCN.17758....1C,Knust:2017ysj}& Yes\\
\hline
161001A & 0.891 & candidate host & \citep{Selsing:2018dwd} & Yes\\
\hline
180115A & 2.487 & spectroscopic & \citep{2018GCN.22346....1D}& Yes\\
\hline
180314A & 1.445 & spectroscopic & \citep{2018GCN.22484....1S}& Yes\\
\hline
180630A & $1.01 \pm 0.47$ & photometric & This Work & Yes \\\hline
190219A & $0.32 \pm 0.26$ & photometric & This Work & No \\\hline
190719C & 2.469 & spectroscopic & \citep{2019GCN.25252....1R}& Yes\\
\hline
210112A & $\sim 2$ & photometric & \citep{2021GCN.29296....1K} & Yes\\
\hline
\hline
\end{tabular}
\caption{The 42 GRBs in the generalization set with known redshifts, used to test the consistency and reliability of the regression estimator and the X-ray classifier. Column 1: GRB name. Column 2: measured redshift. Column 3: method used to measure the redshift. Column 4: literature reference. Column 5: whether the binary classification by the X-ray classifier of MGD25 at $z_t = 3.5$ agrees with the measured redshift. For GRB 190219A only, the entry instead refers to a discrepancy between the measured redshift and the continuous prediction of the regression estimator of \citet{Narendra_2025}.}
\label{tab:GRBs_with_z}
\end{table*}

\section{Summary and Conclusion}
\label{sec:conclusion}

We introduce an ensemble ML framework that classifies LGRBs as high- or low-$z$, using prompt, plateau, and afterglow emission features of 171 LGRBs observed by \textit{Swift} UVOT and by 455 ground-based telescopes and detectors. The aim is to flag newly observed high-$z$ GRBs reliably, enabling timely follow-up and thus increasing the number of high-$z$ GRBs with measured redshifts.

The classifier training pipeline comprises several key steps, illustrated in Figure \ref{Fig:flowchart} and summarized below:

\begin{enumerate}
    \item Initial preprocessing that removes physically implausible or poorly constrained values, discarding measurements with relative errors exceeding unity. 
    \item Outliers were identified with the M-estimator as GRBs that receive low weights in a robust regression of $\log(1+z)$ on the features, and were removed.
    \item Missing data were addressed through the MICE procedure.
    \item LASSO was then applied to select the most informative features and reduce dimensionality.
    \item Model evaluation and selection were carried out via SuperLearner, implementing a nested 10fCV to identify the best-performing models.
    \item The preferred configuration of the final SuperLearner ensemble was selected to maximize sensitivity while maintaining a high AUC and minimizing overfitting, measured by $\Delta$, the fractional difference between the training and test AUC.
\end{enumerate}

As emphasized in Section \ref{sec:intro}, the central novelty of this study is a redshift classifier trained on an optical-plateau-selected sample. Which of the optical plateau parameters survive feature selection depends on the redshift threshold and the missing-data treatment: at $z_t = 2.0$ and $3.0$ on the raw data, none are retained; at $z_t = 2.5$ on the raw data the spectral index $\beta$ is retained; and under MICE imputation with the M-estimator, one or more plateau parameters are retained at every threshold. The best classification performance for our dataset is obtained at $z_t = 2.0$ on the raw dataset with the M-estimator, with a training-set AUC of 0.841, a high training-set TPR of 0.741, and $\Delta$ between $1\%$ and $4\%$. This $\Delta$ is the lowest of the four datasets at $z_t = 2.0$ and lower than that of the same dataset at $z_t = 2.5$ and $3.0$, so this configuration transfers reliably to the held-out test set.
Beyond these performance metrics, comparing the optical and X-ray features directly establishes two results. First, the optical plateau flux $F_{a,\mathrm{opt}}$ is systematically offset from its X-ray counterpart by $\sim 1.6$ dex, whereas the optical and X-ray values of $T_a$, $\alpha$, and $\beta$ do not differ from a statistical point of view. This offset is most likely caused by dust extinction and by sparse or uneven observing cadence, which suppress the measured optical afterglow flux. The offset provides a physically motivated explanation for why LASSO selects prompt-emission observables in preference to $F_{a,\mathrm{opt}}$ at $z_t = 2.0$. Second, $\log(\mathrm{NH})$ and $\log(\mathrm{PeakFlux})$ are the only features retained by LASSO in essentially every configuration of both classifiers: $\log(\mathrm{NH})$ in all twelve optical and all sixteen X-ray cases, and $\log(\mathrm{PeakFlux})$ in eleven of twelve and all sixteen, respectively. We attribute this stability to their being measured directly from high-energy prompt and afterglow data rather than from band-specific afterglow follow-up, so they are unaffected by the wavelength-dependent selection effects (such as extinction and uneven cadence in the optical band) that differ between the optical and X-ray samples.
Taken together, these results show that the optical and X-ray classifiers are affected by different sources of systematic error and missing data. They also respond differently to MICE imputation: the optical classifier performs best on the raw data, whereas the X-ray classifier is essentially unchanged at $z_t = 2.0$ and improves mildly with imputation at higher thresholds.
 
The optical classifier complements the X-ray model of MGD25 rather than competing with it. At $z_t = 2.0$ on the raw data with the M-estimator, the two classifiers have effectively equal training-set sensitivities, but the optical classifier's performance transfers far better to its test set. The X-ray classifier reaches its smallest $\Delta$ at $z_t = 3.5$, whereas the current optical sample does not yet contain enough bursts above $z = 3.5$ for that threshold. The two models are therefore suited to different redshift thresholds. The X-ray classifier was also validated externally on the generalization set (Section \ref{sec:generalisation}). For the 42 GRBs with known redshifts, its binary labels agree with the measured redshifts in 38 cases, and the regression-based estimates of \citet{Narendra_2025} closely follow the measured redshifts along the 1:1 line (Figure~\ref{fig:zgrbweb_zpred}). Because the two classifiers share the same implementation, this validation lends additional confidence to the shared pipeline used throughout this work. Our training-set AUC values are comparable to those reported by earlier ML classifiers based on prompt-emission and X-ray data \citep{morgan2012,ukwatta2016machine,2024MNRAS.529.2676A}. Binary classifiers built with the shared pipeline of this work and MGD25 also offer a complementary, independent check on reliable ML-based redshift estimators \citep{Narendra_2025,Dainotti_2024_ApJS,2024MNRAS.529.2676A}.

We also introduce a public web application that applies our ML models to GRB redshift classification. Users can tailor the classifier to specific follow-up programs through the options described in Section~\ref{subsec:webapp}.

This work demonstrates the significant potential of ensemble ML techniques in astrophysical research. By enabling more accurate identification of high-$z$ and low-$z$ LGRBs (identifying low-$z$ LGRBs is particularly important for finding the peculiar KNe associated with LGRBs, \citet{rastinejad2022grb211211a}), this work lays the groundwork for expanding the sample of LGRBs available for detailed photometric and spectroscopic investigation. A larger high-$z$ LGRB dataset will, in turn, improve constraints on the GRB luminosity function and redshift evolution, and will allow us to better address why the density rate of low-$z$ GRBs deviates from the general star formation rate evolution \citep{Dainotti2024ApJ...967L..30D,2024ApJ...963L..12P,bal2025probingevolutionlonggrb,Khatiya_2025}. As both GRB datasets and ML algorithms continue to evolve, we expect significant gains in classification precision and recall in future research. Because the framework is general, it is also suitable for applications beyond GRBs, such as the redshift classification of AGN, possibly including BL Lacs, whose optical spectra lack the absorption or emission lines needed to measure a spectroscopic redshift \citep{narendra2022predicting, dainotti2021predicting, gibson2022}. Our study advances the goal of utilizing GRBs as astrophysical probes of the early Universe.

\section*{Data and Code Availability}

The optical plateau and prompt-emission parameters used in this work are drawn from the compilations of \citet{Dainotti2022} and \citet{Dainotti2024ApJ...967L..30D}, and the generalization set from \citet{Narendra_2025}. The analysis scripts and the web application implementing the optical classifier are publicly available at \url{https://github.com/Milind018/Redshift-Classifier-Optical} under a GNU GPL v3.0 license, together with the input data and the package requirements needed to reproduce the environment. The optical package builds on the X-ray classifier code of MGD25.

\begin{acknowledgments}
The authors thank the UK \textit{Swift} Science Data Center at the University of Leicester for providing access to the \textit{Swift} BAT+XRT data used in this study. 
MGD acknowledges the support of the DoS and of JSPS Grant-in-Aid for Scientific Research (KAKENHI) (A), Grant Number JP25H00675. \newline
AP is grateful for support from the Polish National Science Centre \newline
grant 2023/50/A/ST9/00579. NF  acknowledges financial support from UNAM-DGAPA-PAPIIT  through grant  IN112525.  We acknowledge Aditya Narendra for his early help in setting up the analysis and guiding MS in writing the paper. We thank Simanta Deka for his help with the computations. We also thank the wider astrophysics community for maintaining and supporting publicly available datasets, which make independent research like this possible. In addition, we recognize the open-source software ecosystem that enabled our analysis and visualization workflows.
\end{acknowledgments}

\software{SuperLearner \citep{vanderLaanPolleyHubbard+2007}, mice \citep{van2011mice}, MASS \citep{MASS}, glmnet, unbalanced \citep{chawla2002smote}, R \citep{R}}

\section*{AI Disclosure Statement}
Large Language Models (LLMs), including Claude Opus 5.5 (Anthropic), were used to edit the text of this manuscript. All AI-assisted output was reviewed and verified by the authors, who take full responsibility for the content of this manuscript.

\appendix

\section{Detailed Methodology and Technical Descriptions}
\label{sec:methodology}

\subsection{Outlier Removal}
\label{sec:Mestimator}

Our dataset contains outliers, i.e., GRBs that deviate strongly from the trend followed by the rest of the sample and could therefore bias the predictions of our ML model.
We remove them with the M-estimator, a generalized maximum likelihood method \citep{huber1964,huber1996robust,DEMENEZES2021107254} and
a reliable alternative to the ordinary least squares (OLS) method.
OLS minimizes the sum of squared residuals, so a few GRBs with large residuals can dominate the fit. The M-estimator instead minimizes the Huber function of the residuals \citep{huber1964, huber2009robust}, which grows only linearly for large residuals and therefore reduces their influence.
The M-estimator fits a generalized linear model \citep[GLM,][]{68aee965-a8a0-3e72-9f89-8d89ae91a62b} that includes squared feature terms to capture nonlinear correlations with $z$.
The M-estimator formula is:

\begin{align}
\log(1+z) = & \; \bigg( (\log{\rm{PeakFlux}})^2 + (\log{\rm{NH}})^2 \notag \\
& + (\rm{PhotonIndex})^2 + \beta^2 \notag \\
& + \log{\rm{PeakFlux}} + \alpha \bigg)^2 \notag \\
& +\log{F_a} + \log{T_a} + \beta \notag \\
& + \rm{PhotonIndex} + \log{\rm{NH}} \notag \\
& + (\log{F_a})^2 + (\log{T_a})^2 + \beta^2.
\label{Eq:1}
\end{align}

This empirical formula is Eq. 1 of \citet{Dainotti2024ApJ...967L..30D}, which obtained it by exploring 161,000 combinations of the variables.
Because all its smoothing functions are set to 1, it is a GLM rather than a Generalized Additive Model \citep{hastie1987generalized}, in which each term would enter through a fitted smoothing function. 

We implemented the M-estimator method using the MASS package \citep{MASS} of R \citep{R}.

\begin{figure*}[htbp]
\centering
    \includegraphics[width = 0.6\textwidth]{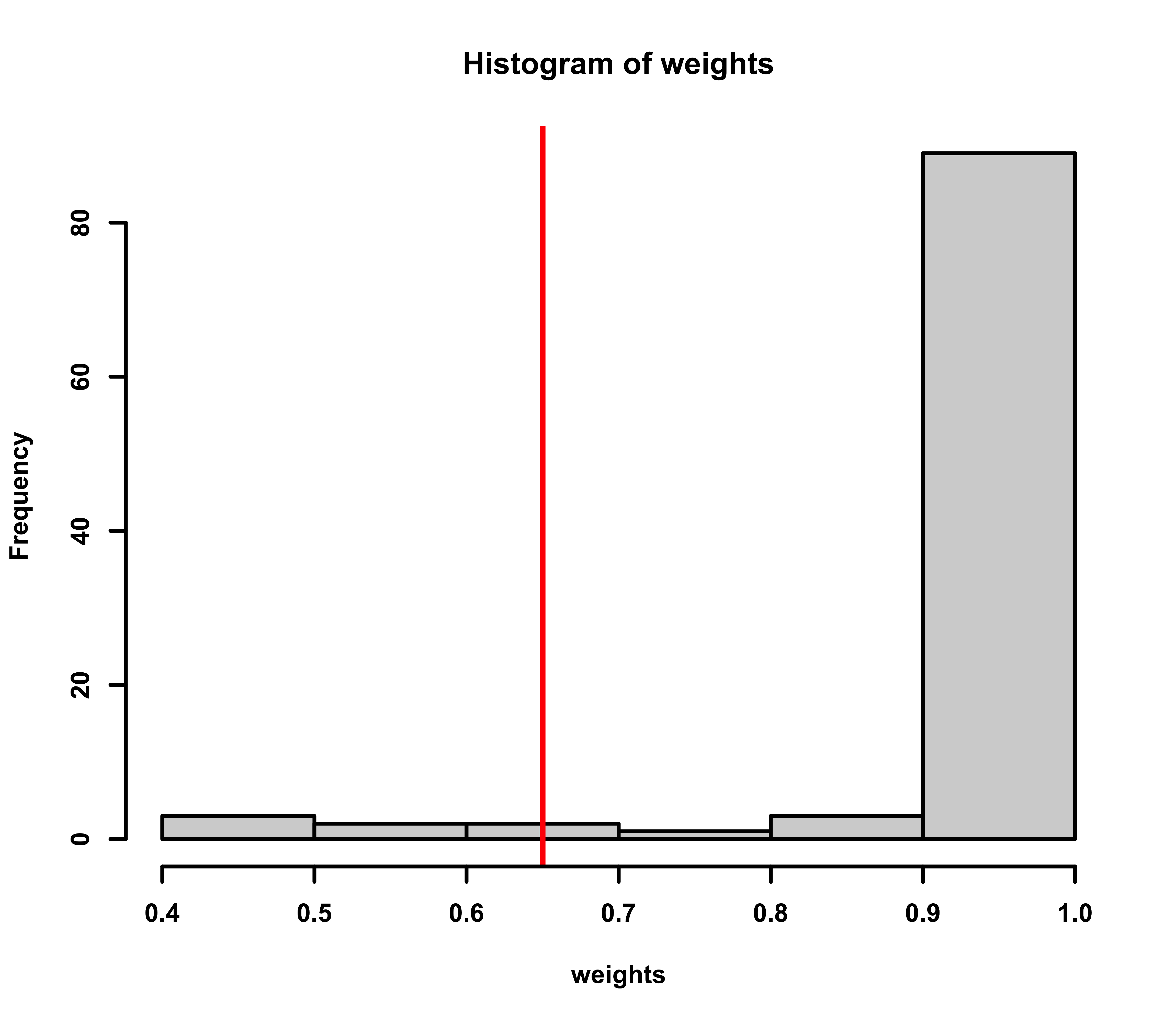}
\caption
{Distribution of the weights assigned by the M-estimator to each GRB of the raw data. The red vertical line marks the outlier cutoff of 0.65.}
\label{Fig:M-estimator weights}
\end{figure*}

\begin{figure*}[htbp]
\centering
    \includegraphics[width = 1\textwidth]{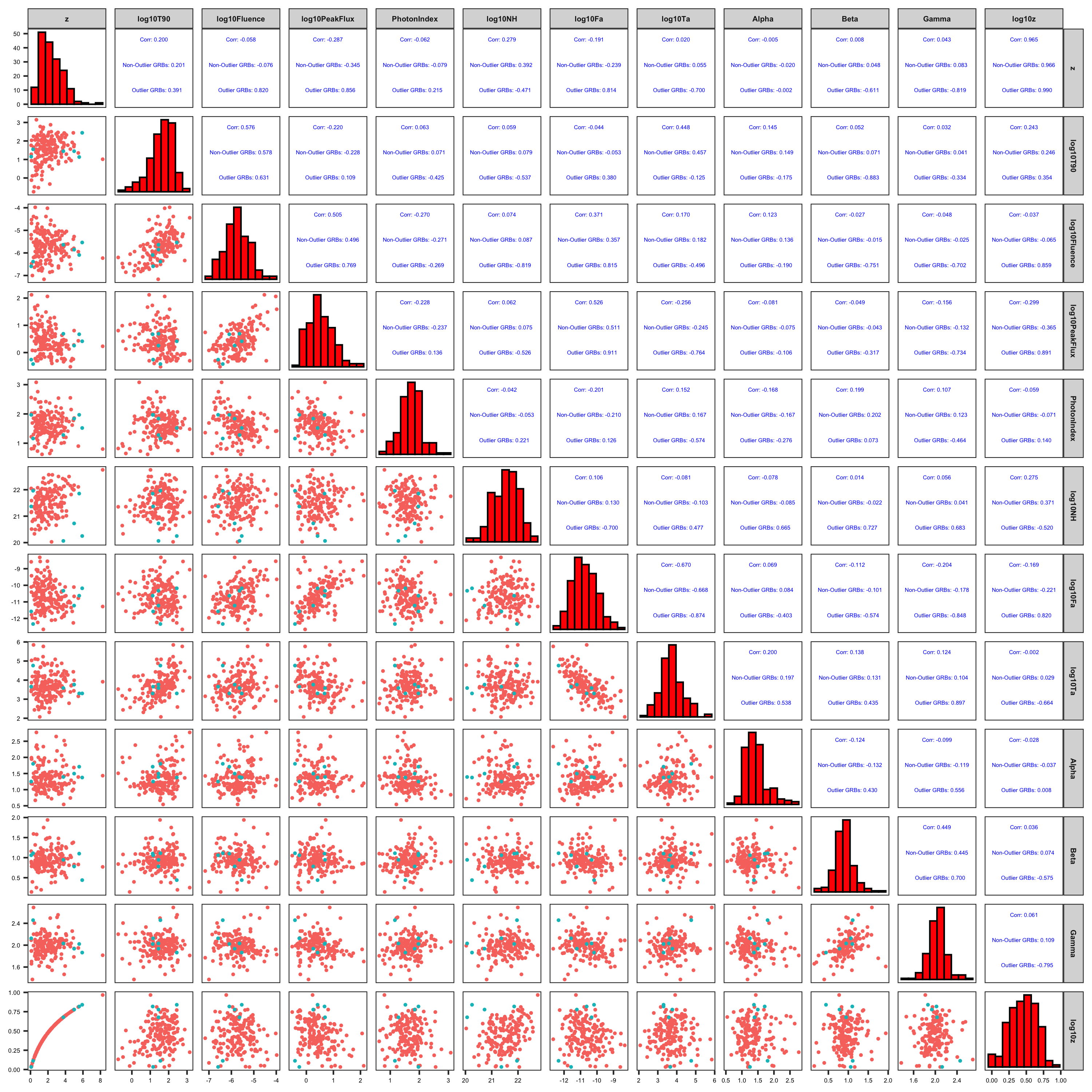}
\caption{Scatter matrix plot of the raw data, with M-estimator outliers in green and the remaining GRBs in red.}
\label{Fig:M-estimator Scatter matrix plot-raw}
\end{figure*}

Figure \ref{Fig:M-estimator weights} shows the distribution of the weights assigned by the M-estimator to the GRBs of the raw (complete-case) sample; a GRB with a large residual from the fit of the formula above receives a low weight.
GRBs with weights below the cutoff of 0.65 (red line) are considered outliers.
We select this cutoff so that only a small fraction of each sample is removed as outliers.
The M-estimator is applied after the relative-error cut described in Section~\ref{sec:sample}. In the complete-case sample of 97 GRBs, the 0.65 cutoff removes 7 GRBs (7.2\%: GRB170714A, GRB181010A, GRB070208A, GRB070318A, GRB070419A, GRB071025A, and GRB091024), leaving 90. MICE is then applied to these 90 GRBs, yielding the 147 GRBs of the MICE-imputed dataset with the M-estimator (Figure~\ref{Fig:flowchart}).
Figure \ref{Fig:M-estimator Scatter matrix plot-raw} shows the scatter matrix plot of the raw (complete-case) sample, with the M-estimator outliers highlighted in green.
The upper triangle shows the pairwise feature correlations for outlier and non-outlier GRBs.
We also computed the variance inflation factor (VIF) of each predictor, which measures how much the variance of its fitted coefficient is inflated by correlations with the other predictors (VIF $=1$ means no such correlation). The VIF values range from 1.0 to 2.59, below the critical threshold of 10, so collinearity is minimal and does not compromise the M-estimator fit or outlier identification \citep{Brown2009}.

\subsection{Data Imputation}
\label{sec:mice}

\begin{figure*}[htbp]
\centering
    \includegraphics[width = 1\textwidth]{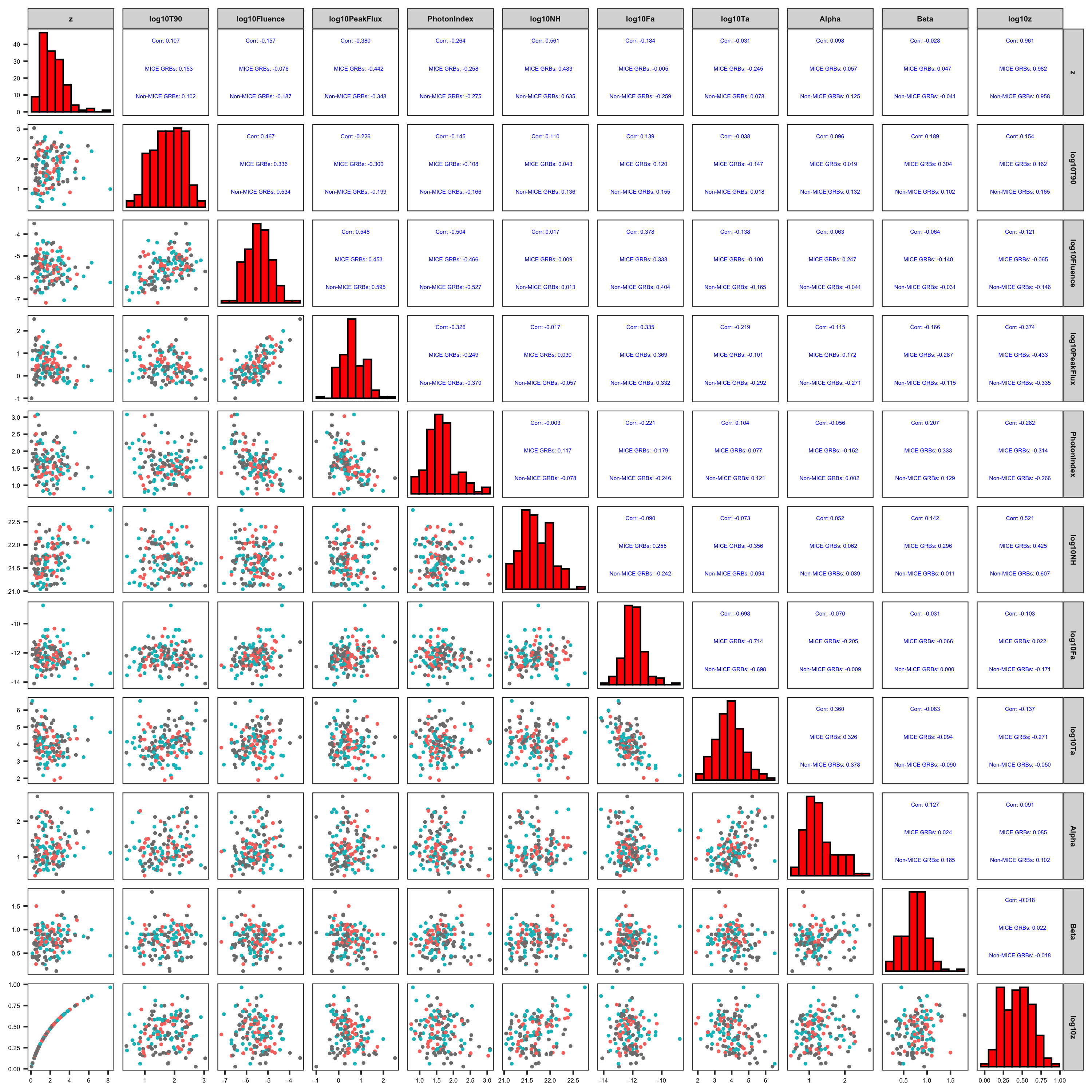}
\caption{Scatter matrix plot of the MICE-imputed dataset, with imputed values in red and observed values in cyan.}
\label{Fig:MICE scatter plot}
\end{figure*}

\begin{figure*}[htbp]
\centering
\includegraphics[width = 1\textwidth]{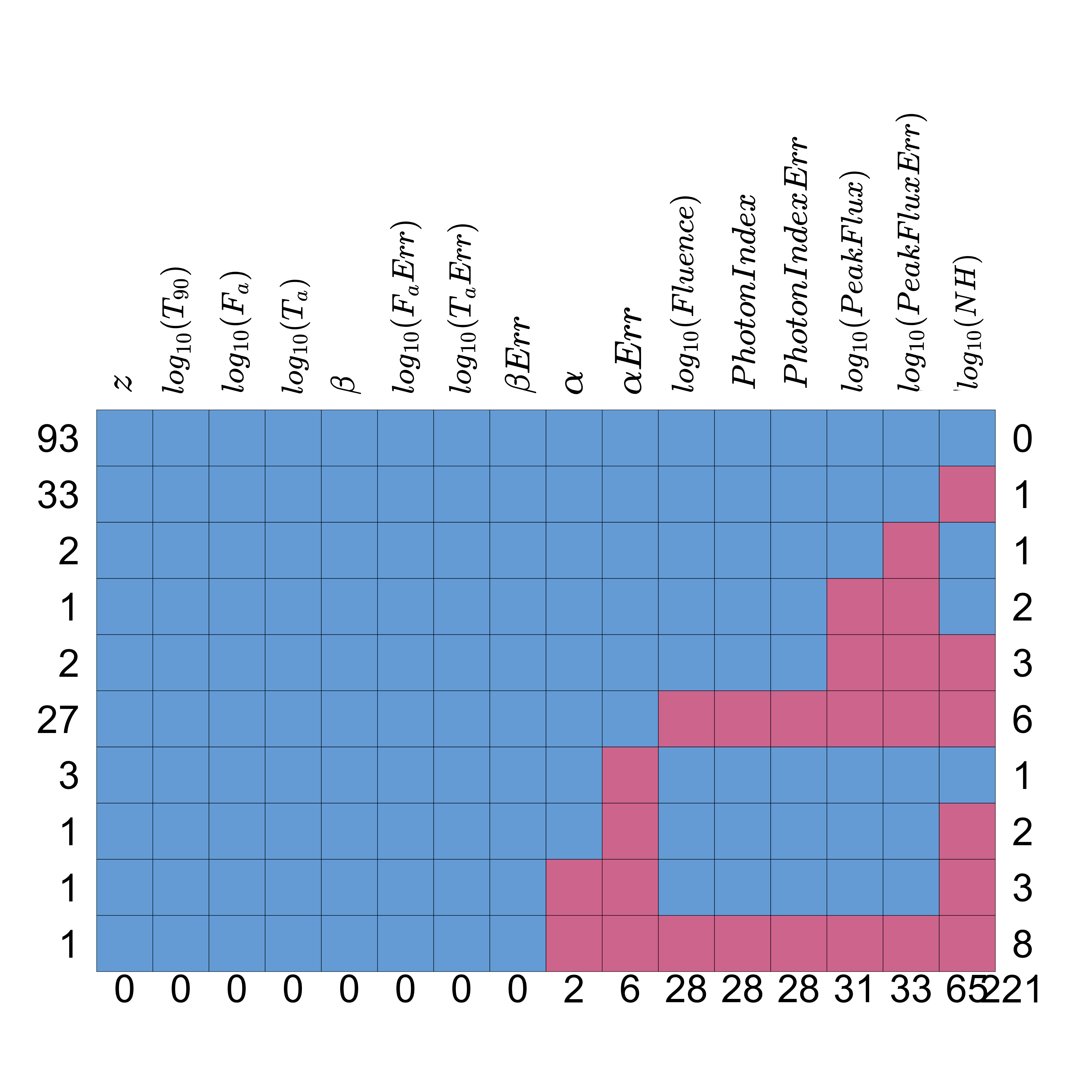}
\caption{The distribution of missing data in our sample. Each row is one pattern of missing data; pink boxes mark missing and blue boxes measured values of the variables named on the top axis. The left axis gives the number of GRBs sharing each pattern, the right axis the number of features missing in it, and the bottom axis the total number of GRBs in which each feature is missing. For instance, 93 GRBs have no missing entry in any column, 33 GRBs miss only $\log(\rm{NH})$, 2 GRBs miss $\log(PeakFlux_{\rm{err}})$, and so on. The count of 93 includes the measurement-uncertainty columns.}
\label{Fig:mices}
\end{figure*}

\begin{figure*}[htbp]
\centering
    \includegraphics[width = 1\textwidth]{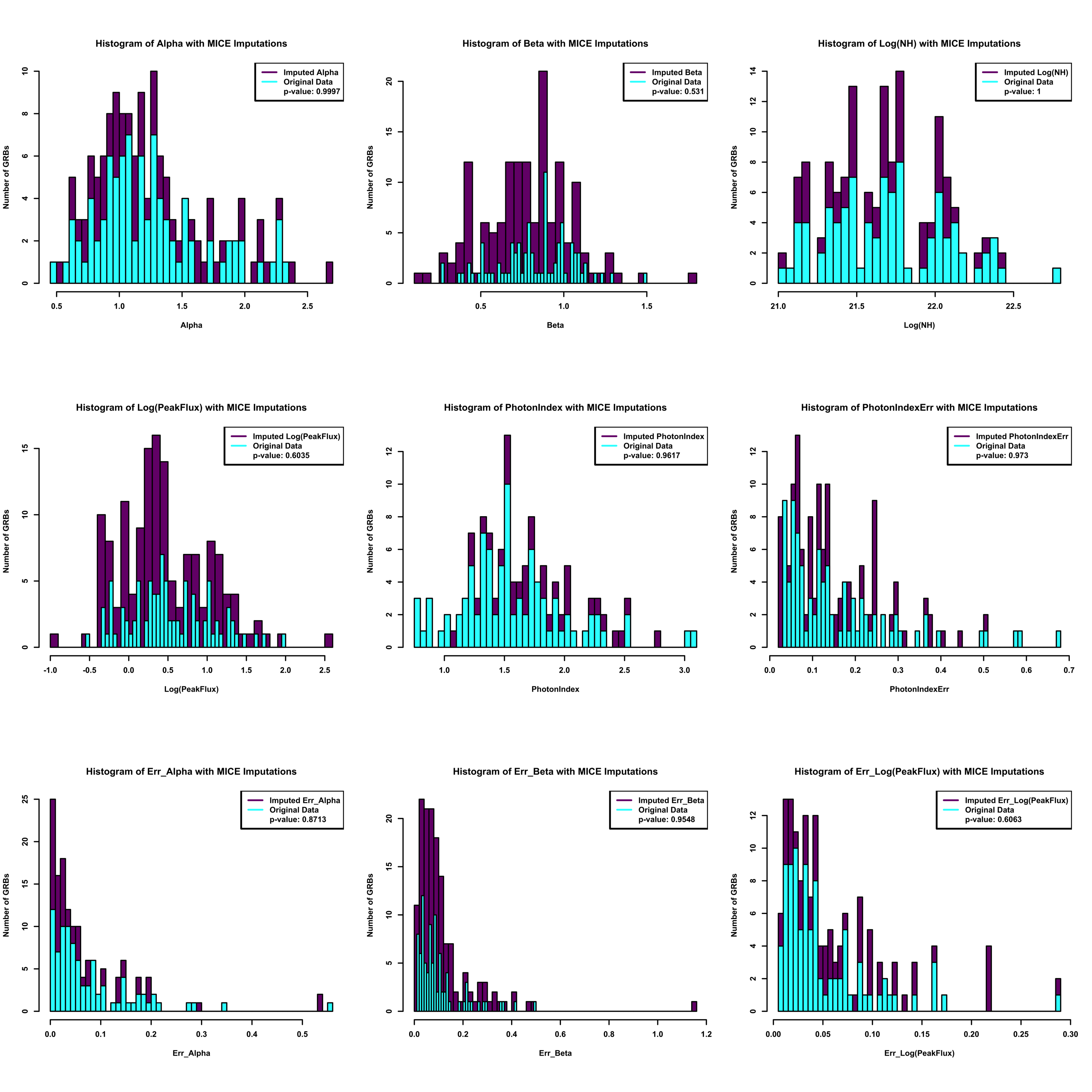}    
    \caption{Panels (a) to (i) show the distributions of $\alpha$, $\beta$, $\log({\rm{NH})}$, $\log({\rm{PeakFlux})}$, PhotonIndex, PhotonIndexErr, Err $\alpha$, Err $\beta$, and Err $\log({\rm{PeakFlux})}$, with MICE-imputed data in magenta, the original data in blue, and the KS-test $p$-value.}
    \label{fig:combined_hist}
\end{figure*}

MICE \citep{schafer2002missing, van2011mice} is an iterative method for imputing missing values in multivariate datasets, and requires that the data be MAR \citep{rubin1976inference}; that is, that the probability of a value being missing depends only on the observed data and not on the unobserved value itself.
We adopt this assumption because the orbital gaps appear randomly, and consequently the time and flux of the plateau can be randomly affected by them. The effect of imputation on classifier performance is discussed in Sections \ref{sec:results} and \ref{subsec:comparison2}.
We implement MICE in \textsc{R} using PMM with the \textit{midastouch} option \citep{little2019statistical}. In PMM, a regression on the other features predicts a value for each GRB, and each missing value is replaced by the observed value of another GRB (the ``donor'') whose predicted value is close to that of the GRB with the missing entry. In the \textit{midastouch} variant, the donor is drawn at random, and GRBs with closer predicted values have a higher probability of being drawn.
Initially, each missing value is filled with the mean of its feature.
Then, in each iteration, each incomplete feature is regressed on the other features, and its missing entries are updated with PMM. 
We repeat this cycle for 20 iterations, by which point the imputed values have stabilized.
The relative-error cut ($\Delta x/x > 1$ for any feature $x$) is part of the initial data filtering and is therefore applied before both the M-estimator and MICE (Figure~\ref{Fig:flowchart}).
Figure \ref{Fig:MICE scatter plot} shows the scatter matrix plot of the MICE-imputed sample, with imputed values in red. 
Figure \ref{Fig:mices} shows the pattern of missing data in our sample.
To check the imputation, we compared the imputed and observed distributions of each variable using a two-sided KS test \citep{doi:10.1080/01621459.1968.11009335}. The $p$-values average 0.804 across all variables.
The KS test finds no significant difference between the imputed and observed distributions. Because PMM draws imputed values from the pool of observed values, this agreement is expected by construction; the KS test therefore checks that the imputation does not distort the distribution of each feature, rather than validating individual imputed values. Figure \ref{fig:combined_hist} shows the distributions of the features before and after MICE, with their KS-test $p$-values.
As for the M-estimator, every subsequent step is carried out both with and without MICE (Figure~\ref{Fig:flowchart}).
Each of the four resulting datasets was then randomly split into an 80\% training set and a 20\% test set. 
MGD25 used a comparable MICE procedure; the differences between the imputed datasets of the two studies are described in Section \ref{subsec:comparison}.
To address class imbalance (the high-$z$ class is the smaller one), we initially applied SMOTE \citep{chawla2002smote}, which creates synthetic high-$z$ GRBs by interpolating between the features of existing ones. Compared with duplicating existing high-$z$ GRBs, this interpolation increases the number of high-$z$ examples while reducing overfitting. The \texttt{ubSMOTE} method of the \texttt{unbalanced} R package added 124, 92, and 58 synthetic GRBs to the 147 GRBs of the MICE-imputed dataset with the M-estimator at $z_{t} = 2.0$, $z_{t} = 2.5$, and $z_{t} = 3.0$, respectively; these numbers are large relative to the original optical sample. Training on these SMOTE-augmented datasets did not improve classifier performance, and the synthetic GRBs risked distorting the intrinsic distributions of the features. We therefore did not adopt SMOTE, and all results in this work come from the analyses without SMOTE (Section~\ref{sec:results}).

\subsection{Feature Selection}
\label{sec:LASSO}

\begin{figure*}[htbp]
\centering
    \includegraphics[width = 1\textwidth]{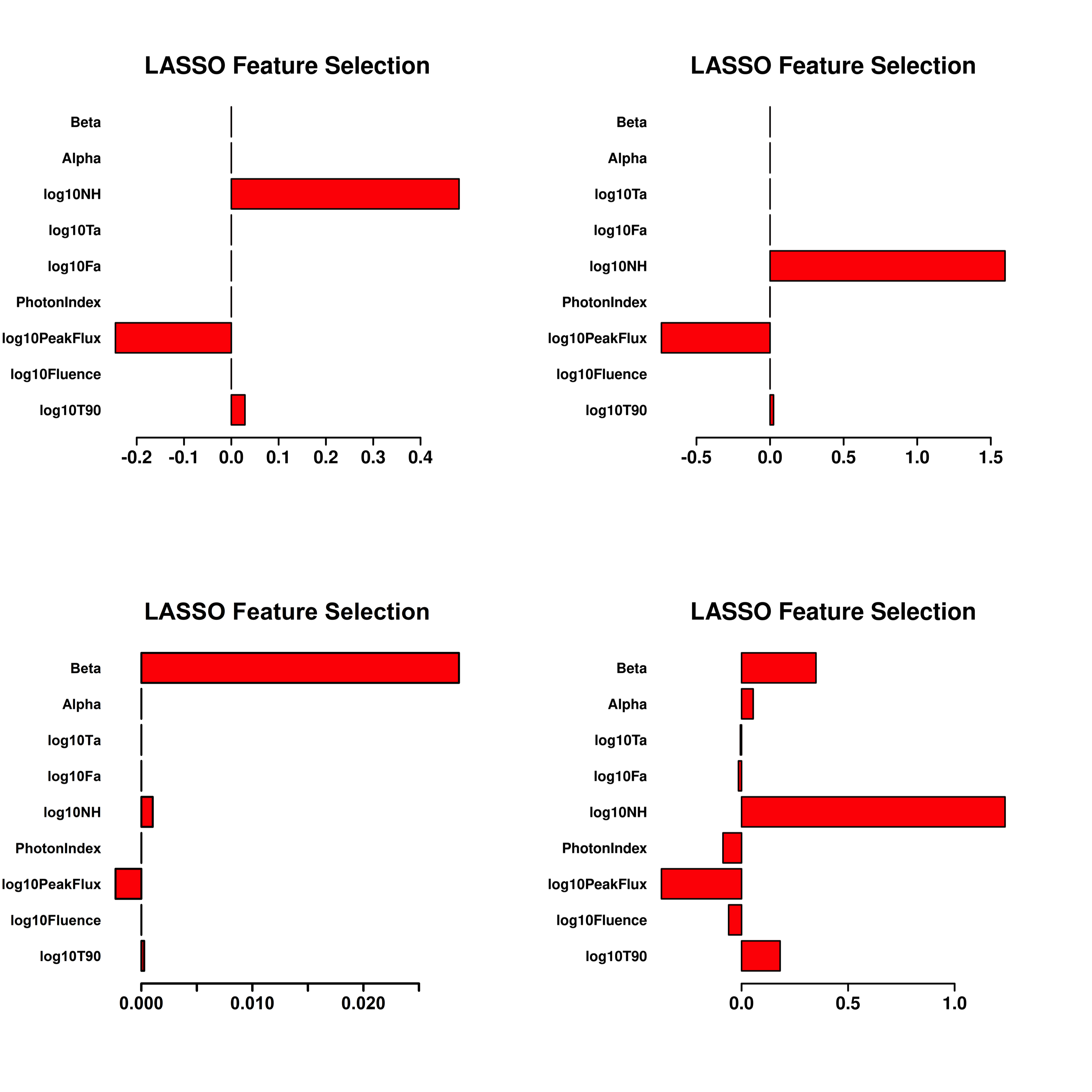}
\caption{Weight assigned to each feature by LASSO for $z_{t}$ = 3.0 with the raw dataset without the M-estimator (top left), the raw dataset with the M-estimator (top right), the MICE-imputed dataset without the M-estimator (bottom left), and the MICE-imputed dataset with the M-estimator (bottom right). Features with weights $>$ 2\% are selected.}
\label{Fig:LASSO}
\end{figure*}

To select the features that best separate high-$z$ from low-$z$ GRBs, we applied LASSO \citep{TibshiraniLasso}, which predicts the response variable, $z$, by minimizing the residual sum of squares while constraining the sum of the coefficients' absolute values. 
To check that the selection is stable, we repeated the LASSO fit 100, 200, 300, 500, and 1000 times and averaged the weight that LASSO assigns to each feature. 
Features with an average weight above 2\% were retained as the most informative. 
Lowering this threshold to 1\% left both the selected features and the prediction accuracy unchanged, owing to the limited size of the sample. 
Figure \ref{Fig:LASSO} shows the LASSO weights at $z_{t}=3.0$ for the four datasets. 

\subsection{Algorithms}
\label{sec:algorithms}

The SuperLearner ensemble combines algorithms from three broad categories:

\begin{itemize}
\item \textbf{Parametric methods:} lm (linear regression), glm (generalized linear model), glmnet (GLM with LASSO/elastic-net regularization), speedglm and speedlm (fast implementations of glm and lm), bayesglm (Bayesian GLM), and lda and qda (linear and quadratic discriminant analysis).
\item \textbf{Semi-parametric methods:} biglasso (LASSO regression for large datasets) and earth (Multivariate Adaptive Regression Splines, MARS).
\item \textbf{Non-parametric methods:} randomForest, cforest, and ranger (random forests of decision trees; cforest uses conditional-inference trees and ranger is a fast implementation), caret.rpart (a single decision tree), xgboost (gradient-boosted decision trees), kernelKnn (kernel-weighted $k$-nearest neighbors), and svm and ksvm (support vector machines).
\end{itemize}

These algorithms span model assumptions from linear models to tree ensembles; their full descriptions are given in MGD25, where the same SuperLearner framework was developed and validated.

\section{Redshift threshold of 2.0}
\label{sec:z2.0}

For each of the twelve configurations (three values of $z_t$ times four datasets), this and the following two appendices report the AUC of the full SuperLearner ensemble (the ``combined'' AUC) and of the best- and least-performing individual algorithms (base learners). All AUC values quoted here are training-set quantities; the test values and the train-to-test differences $\Delta$ are given in Table \ref{tab:AUC_for_SuperLearner}. The AUC of an individual base learner varies more between configurations than the ensemble AUC, because each learner is fitted to at most a few tens of high-$z$ bursts. For example, `xgboost' returns an AUC of 0.1909 on the raw data without the M-estimator at $z_t = 2.0$, far below the value of 0.5 of a classifier that assigns classes at random, and 0.9592 on the same preprocessing at $z_t = 3.0$, where only thirteen high-$z$ bursts are available for training. Neither value is representative of that learner's discriminatory power. We therefore interpret only the ensemble AUC, and give the individual values to indicate which learners were removed and why. Because the test-set AUC varies by up to about $0.065$ (and the training-set AUC by less than $0.05$) between repeated runs of the fitting procedure (Section \ref{sec:results}), we treat changes in the combined AUC smaller than $0.065$ as within the run-to-run variation.

\subsection{Raw dataset without M-estimator}
\label{sec:raw_z2.0}
The combined AUC is 0.7622; the best-performing individual algorithm is `lm' (AUC = 0.7785) and the least-performing is `xgboost' (AUC = 0.1909; Figure \ref{fig:combined_raw_z2.0} top left). Removing the least-performing algorithms (`cforest', `ksvm', `qda', `xgboost', and `earth') increased the combined AUC marginally from 0.7622 to 0.7751 (Figure \ref{fig:combined_raw_z2.0} top right).

\subsection{Raw dataset with M-estimator}
\label{sec:raw_Mestimator_z2.0}
With the outliers removed (Appendix~\ref{sec:Mestimator}), the combined AUC is 0.8261; the best-performing individual algorithm is `lda' (AUC = 0.8434) and the least-performing is `ksvm' (AUC = 0.7482; Figure \ref{fig:combined_raw_z2.0} second row left). Removing `ksvm' increased the combined AUC from 0.8261 to 0.8411 (Figure \ref{fig:combined_raw_z2.0} second row right).

\subsection{MICE-imputed dataset without M-estimator}
\label{sec:BALANCED_Mestimator_z2.0}
The combined AUC is 0.6967; the best-performing individual algorithms are `lm' and `speedlm' (both AUC = 0.714) and the least-performing is `ranger' (AUC = 0.6698; Figure \ref{fig:combined_raw_z2.0} third row left). Removing the least-performing algorithms (`randomForest', `ranger') increased the combined AUC from 0.6967 to 0.7061 (Figure \ref{fig:combined_raw_z2.0} third row right).

\subsection{MICE-imputed dataset with M-estimator}
\label{sec:MICE_Mestimator_z2.0}
The combined AUC is 0.8021; the best-performing individual algorithm is `qda' (AUC = 0.8042) and the least-performing is `kernelKnn' (AUC = 0.7078; Figure \ref{fig:combined_raw_z2.0} bottom left). Removing `kernelKnn' changed the combined AUC from 0.8021 to 0.7817 (Figure \ref{fig:combined_raw_z2.0} bottom right), a difference within the run-to-run variation of the AUC described in Section \ref{sec:results}.

\section{Redshift threshold of 2.5}
\label{sec:z2.5}
\subsection{Raw dataset without M-estimator}
\label{sec:raw_z2.5}
The combined AUC is 0.804; the best-performing individual algorithm is `speedlm' (AUC = 0.8134) and the least-performing is `caret.rpart' (AUC = 0.6221; Figure \ref{fig:combined_raw_z2.5} top left). Removing the least-performing algorithms (`caret.rpart', `earth', `kernelKnn') increased the combined AUC slightly from 0.804 to 0.8071 (Figure \ref{fig:combined_raw_z2.5} top right).

\subsection{Raw dataset with M-estimator}
\label{sec:raw_Mestimator_z2.5}
The combined AUC is 0.8727; the best-performing individual algorithm is `glmnet' (AUC = 0.8691) and the least-performing is `svm' (AUC = 0.7082; Figure \ref{fig:combined_raw_z2.5} second row left). Removing `svm' increased the combined AUC slightly from 0.8727 to 0.8773 (Figure \ref{fig:combined_raw_z2.5} second row right).

\subsection{MICE-imputed dataset without M-estimator}
\label{sec:BALANCED_Mestimator_z2.5}
The combined AUC is 0.7001; the best-performing individual algorithm is `qda' (AUC = 0.7129) and the least-performing is `xgboost' (AUC = 0.6525; Figure \ref{fig:combined_raw_z2.5} third row left). 
Removing `xgboost' left the combined training-set AUC essentially unchanged at 0.698 (Figure \ref{fig:combined_raw_z2.5} third row right), while the corresponding test AUC rose from 0.581 to 0.652 (Table \ref{tab:AUC_for_SuperLearner}).

\subsection{MICE-imputed dataset with M-estimator}
\label{sec:MICE_Mestimator_z2.5}
The combined AUC is 0.8343; the best-performing individual algorithm is `qda' (AUC = 0.8316) and the least-performing is `ksvm' (AUC = 0.73; Figure \ref{fig:combined_raw_z2.5} bottom left). 
Removing the least-performing algorithms (`cforest' and `ksvm') gave a combined AUC of 0.8286 (Figure \ref{fig:combined_raw_z2.5} bottom right), a change within the run-to-run variation of the AUC described in Section \ref{sec:results}.

\section{Redshift threshold of 3.0}
\label{sec:z3.0}
\subsection{Raw dataset without M-estimator}
\label{sec:raw_z3.0}
The combined AUC is 0.8; the best-performing individual algorithm is `xgboost' (AUC = 0.9592) and the least-performing is `ksvm' (AUC = 0.5893; Figure \ref{fig:combined_raw_z3.0} top left). As noted at the start of Appendix \ref{sec:z2.0}, this `xgboost' value is not representative: a flexible boosted-tree learner can separate the thirteen high-$z$ training bursts almost perfectly without this separation carrying over to unseen bursts. The SuperLearner correctly does not give this learner a dominant weight in the ensemble.
Removing the least-performing algorithms (`kernelKnn' and `ksvm') increased the combined AUC from 0.8 to 0.8095 (Figure \ref{fig:combined_raw_z3.0} top right).

\subsection{Raw dataset with M-estimator}
\label{sec:raw_Mestimator_z3.0}
The combined AUC is 0.8764; the best-performing individual algorithm is `lm' (AUC = 0.8944) and the least-performing is `lda' (AUC = 0.8229; Figure \ref{fig:combined_raw_z3.0} second row left). 
Removing the least-performing algorithms (`glm' and `lda') changed the combined AUC from 0.8764 to 0.875 (Figure \ref{fig:combined_raw_z3.0} second row right), a difference well below the run-to-run variation of the AUC described in Section \ref{sec:results}.

\subsection{MICE-imputed dataset without M-estimator}
\label{sec:BALANCED_Mestimator_z3.0}
The combined AUC is 0.7354; the best-performing individual algorithm is `ranger' (AUC = 0.7434) and the least-performing is `svm' (AUC = 0.5826; Figure \ref{fig:combined_raw_z3.0} third row left). 
Removing the least-performing algorithms (`biglasso' and `svm') increased the combined AUC from 0.7354 to 0.7521 (Figure \ref{fig:combined_raw_z3.0} third row right).

\subsection{MICE-imputed dataset with M-estimator}
\label{sec:MICE_Mestimator_z3.0}
The combined AUC is 0.8197; the best-performing individual algorithm is `ksvm' (AUC = 0.832) and the least-performing is `qda' (AUC = 0.7066; Figure \ref{fig:combined_raw_z3.0} bottom left).  
Removing the least-performing algorithms (`svm', `qda', and `kernelKnn') gave a combined AUC of 0.8137 (Figure \ref{fig:combined_raw_z3.0} bottom right), a change within the run-to-run variation of the AUC described in Section \ref{sec:results}.

\bibliography{sample701}{}

\begin{thebibliography}{}
\expandafter\ifx\csname natexlab\endcsname\relax\def\natexlab#1{#1}\fi
\providecommand{\url}[1]{\href{#1}{#1}}
\providecommand{\dodoi}[1]{doi:~\href{http://doi.org/#1}{\nolinkurl{#1}}}
\providecommand{\doeprint}[1]{\href{http://ascl.net/#1}{\nolinkurl{http://ascl.net/#1}}}
\providecommand{\doarXiv}[1]{\href{https://arxiv.org/abs/#1}{\nolinkurl{https://arxiv.org/abs/#1}}}

\bibitem[{T. {Aldowma} \& S. {Razzaque}(2024){Aldowma} \&
  {Razzaque}}]{2024MNRAS.529.2676A}
{Aldowma}, T., \& {Razzaque}, S. 2024, \bibinfo{title}{{Deep Neural Networks
  for estimation of gamma-ray burst redshifts},} MNRAS, 529, 2676,
  \dodoi{10.1093/mnras/stae535}

\bibitem[{D.~S. Bal {et~al.}(2025)Bal, Narendra, Dainotti, Khatiya, Lenart, \&
  Hartmann}]{bal2025probingevolutionlonggrb}
Bal, D.~S., Narendra, A., Dainotti, M.~G., {et~al.} 2025, Probing evolution of
  Long GRB properties through their cosmic formation history aided by Machine
  Learning predicted redshifts, \doarXiv{2510.07306}

\bibitem[{S.~D. {Barthelmy} {et~al.}(2005){Barthelmy}, {Barbier}, {Cummings},
  {Fenimore}, {Gehrels}, {Hullinger}, {Krimm}, {Markwardt}, {Palmer},
  {Parsons}, {Sato}, {Suzuki}, {Takahashi}, {Tashiro}, \&
  {Tueller}}]{barthelmy2005burst}
{Barthelmy}, S.~D., {Barbier}, L.~M., {Cummings}, J.~R., {et~al.} 2005,
  \bibinfo{title}{{The Burst Alert Telescope (BAT) on the SWIFT Midex
  Mission},} SSRv, 120, 143, \dodoi{10.1007/s11214-005-5096-3}

\bibitem[{E. {Berger}(2014){Berger}}]{Berger2014}
{Berger}, E. 2014, \bibinfo{title}{{Short-Duration Gamma-Ray Bursts},} ARA\&A,
  52, 43, \dodoi{10.1146/annurev-astro-081913-035926}

\bibitem[{G. {Beskin} {et~al.}(2010){Beskin}, {Karpov}, {Bondar}, {Greco},
  {Guarnieri}, {Bartolini}, \& {Piccioni}}]{Beskin2010ApJ}
{Beskin}, G., {Karpov}, S., {Bondar}, S., {et~al.} 2010, \bibinfo{title}{{Fast
  Optical Variability of a Naked-eye Burst{\textemdash}Manifestation of the
  Periodic Activity of an Internal Engine},} ApJL, 719, L10,
  \dodoi{10.1088/2041-8205/719/1/L10}

\bibitem[{S. Bhardwaj {et~al.}(2023)Bhardwaj, Dainotti, Venkatesh, Narendra,
  Kalsi, Rinaldi, \& Pollo}]{10.1093/mnras/stad2593}
Bhardwaj, S., Dainotti, M.~G., Venkatesh, S., {et~al.} 2023,
  \bibinfo{title}{{GRB optical and X-ray plateau properties classifier using
  unsupervised machine learning},} MNRAS, 525, 5204,
  \dodoi{10.1093/mnras/stad2593}

\bibitem[{L. Breiman(2001)Breiman}]{breiman2001randomforest}
Breiman, L. 2001, \bibinfo{title}{Random Forests,} Machine Learning, 45, 5,
  \dodoi{10.1023/A:1010933404324}

\bibitem[{S.~D. Brown {et~al.}(2009)Brown, Tauler, \& Walczak}]{Brown2009}
Brown, S.~D., Tauler, R., \& Walczak, B., eds. 2009, Comprehensive
  Chemometrics: Chemical and Biochemical Data Analysis (Elsevier),
  \dodoi{10.1016/B978-0-444-64163-0.X0001-7}

\bibitem[{D.~N. {Burrows} {et~al.}(2005){Burrows}, {Hill}, {Nousek}, {Kennea},
  {Wells}, {Osborne}, {Abbey}, {Beardmore}, {Mukerjee}, {Short}, {Chincarini},
  {Campana}, {Citterio}, {Moretti}, {Pagani}, {Tagliaferri}, {Giommi},
  {Capalbi}, {Tamburelli}, {Angelini}, {Cusumano}, {Br{\"a}uninger}, {Burkert},
  \& {Hartner}}]{2005SSRv..120..165B}
{Burrows}, D.~N., {Hill}, J.~E., {Nousek}, J.~A., {et~al.} 2005,
  \bibinfo{title}{{The Swift X-Ray Telescope},} SSRv, 120, 165,
  \dodoi{10.1007/s11214-005-5097-2}

\bibitem[{A. {Cabrera Lavers} {et~al.}(2011){Cabrera Lavers}, {de Ugarte
  Postigo}, {Castro-Tirado}, {Gorosabel}, {Thoene}, \&
  {Dominguez}}]{2011GCN.12234....1C}
{Cabrera Lavers}, A., {de Ugarte Postigo}, A., {Castro-Tirado}, A.~J., {et~al.}
  2011, \bibinfo{title}{{GRB 110801A: afterglow redshift from 10.4m GTC.},} GRB
  Coordinates Network, 12234, 1

\bibitem[{S. Campana {et~al.}(2012)Campana, Salvaterra, Melandri, Vergani,
  Covino, D’Avanzo, Fugazza, Ghisellini, Sbarufatti, \&
  Tagliaferri}]{Campana_2012}
Campana, S., Salvaterra, R., Melandri, A., {et~al.} 2012, \bibinfo{title}{The
  X-ray absorbing column density of a complete sample of
  bright<i>Swift</i>gamma-ray bursts: <i>Column densities of bright Swift
  GRBs</i>,} MNRAS, 421, 1697–1702, \dodoi{10.1111/j.1365-2966.2012.20428.x}

\bibitem[{A.~J. {Castro-Tirado} {et~al.}(2015){Castro-Tirado},
  {Sanchez-Ramirez}, {Lombardi}, \& {Rivero}}]{2015GCN.17758....1C}
{Castro-Tirado}, A.~J., {Sanchez-Ramirez}, R., {Lombardi}, G., \& {Rivero},
  M.~A. 2015, \bibinfo{title}{{GRB 150424A: 10.4m GTC spectroscopy.},} GRB
  Coordinates Network, 17758, 1

\bibitem[{A.~J. {Castro-Tirado} {et~al.}(2014){Castro-Tirado}, {Cunniffe},
  {Sanchez-Ramirez}, {Gorosabel}, {Jelinek}, {Oates}, {Jeong}, {Tello}, \&
  {Pandey}}]{2014GCN.16505....1C}
{Castro-Tirado}, A.~J., {Cunniffe}, R., {Sanchez-Ramirez}, R., {et~al.} 2014,
  \bibinfo{title}{{GRB140703A: 10.4m GTC redshift.},} GRB Coordinates Network,
  16505, 1

\bibitem[{N.~V. Chawla {et~al.}(2002)Chawla, Bowyer, Hall, \&
  Kegelmeyer}]{chawla2002smote}
Chawla, N.~V., Bowyer, K.~W., Hall, L.~O., \& Kegelmeyer, W.~P. 2002,
  \bibinfo{title}{SMOTE: synthetic minority over-sampling technique,} Journal
  of artificial intelligence research, 16, 321

\bibitem[{R. {Chornock} {et~al.}(2011){Chornock}, {Berger}, \&
  {Fox}}]{2011GCN.12537....1C}
{Chornock}, R., {Berger}, E., \& {Fox}, D. 2011, \bibinfo{title}{{GRB 111107A:
  gemini-south redshift.},} GRB Coordinates Network, 12537, 1

\bibitem[{R. {Chornock} {et~al.}(2010){Chornock}, {Cucchiara}, {Fox}, \&
  {Berger}}]{2010GCN.10466....1C}
{Chornock}, R., {Cucchiara}, A., {Fox}, D., \& {Berger}, E. 2010,
  \bibinfo{title}{{GRB 100302A: gemini-north redshift.},} GRB Coordinates
  Network, 10466, 1

\bibitem[{E. {Costa} {et~al.}(1997){Costa}, {Frontera}, {Heise}, {Feroci},
  {in't Zand}, {Fiore}, {Cinti}, {Dal Fiume}, {Nicastro}, {Orlandini},
  {Palazzi}, {Rapisarda\#}, {Zavattini}, {Jager}, {Parmar}, {Owens}, {Molendi},
  {Cusumano}, {Maccarone}, {Giarrusso}, {Coletta}, {Antonelli}, {Giommi},
  {Muller}, {Piro}, \& {Butler}}]{costa1997}
{Costa}, E., {Frontera}, F., {Heise}, J., {et~al.} 1997,
  \bibinfo{title}{{Discovery of an X-ray afterglow associated with the
  {\ensuremath{\gamma}}-ray burst of 28 February 1997},} Nature, 387, 783,
  \dodoi{10.1038/42885}

\bibitem[{S. Covino {et~al.}(2013)Covino, Melandri, Salvaterra, Campana,
  Vergani, Bernardini, D’Avanzo, D’Elia, Fugazza, Ghirlanda, Ghisellini,
  Gomboc, Jin, Krühler, Malesani, Nava, Sbarufatti, \&
  Tagliaferri}]{Covino_2013}
Covino, S., Melandri, A., Salvaterra, R., {et~al.} 2013, \bibinfo{title}{Dust
  extinctions for an unbiased sample of gamma-ray burst afterglows,} MNRAS,
  432, 1231–1244, \dodoi{10.1093/mnras/stt540}

\bibitem[{D.~M. Coward(2009)Coward}]{Coward_2009}
Coward, D.~M. 2009, \bibinfo{title}{Gamma-ray burst optical afterglow and
  redshift selection effects: the learning curve effect at work,} MNRASL, 393,
  L65–L69, \dodoi{10.1111/j.1745-3933.2008.00601.x}

\bibitem[{D.~M. Coward {et~al.}(2013)Coward, Howell, Branchesi, Stratta,
  Guetta, Gendre, \& Macpherson}]{Coward_2013}
Coward, D.~M., Howell, E.~J., Branchesi, M., {et~al.} 2013, \bibinfo{title}{The
  Swift gamma-ray burst redshift distribution: selection biases and optical
  brightness evolution at high z?} MNRAS, 432, 2141–2149,
  \dodoi{10.1093/mnras/stt537}

\bibitem[{A. {Cucchiara} {et~al.}(2008){Cucchiara}, {Fox}, {Cenko}, \&
  {Berger}}]{2008GCN..8713....1C}
{Cucchiara}, A., {Fox}, D.~B., {Cenko}, S.~B., \& {Berger}, E. 2008,
  \bibinfo{title}{{GRB 081222: gemini-south absorption redshift.},} GRB
  Coordinates Network, 8713, 1

\bibitem[{A. {Cucchiara} \& J.~X. {Prochaska}(2012){Cucchiara} \&
  {Prochaska}}]{2012GCN.12865....1C}
{Cucchiara}, A., \& {Prochaska}, J.~X. 2012, \bibinfo{title}{{GRB 120119A:
  gemini-s redshift.},} GRB Coordinates Network, 12865, 1

\bibitem[{M.~G. {Dainotti} {et~al.}(2008){Dainotti}, {Cardone}, \&
  {Capozziello}}]{Dainotti2008}
{Dainotti}, M.~G., {Cardone}, V.~F., \& {Capozziello}, S. 2008,
  \bibinfo{title}{{A time-luminosity correlation for {\ensuremath{\gamma}}-ray
  bursts in the X-rays},} MNRAS, 391, L79,
  \dodoi{10.1111/j.1745-3933.2008.00560.x}

\bibitem[{M.~G. {Dainotti} {et~al.}(2020){Dainotti}, {Livermore}, {Kann}, {Li},
  {Oates}, {Yi}, {Zhang}, {Gendre}, {Cenko}, \& {Fraija}}]{dainotti2020b}
{Dainotti}, M.~G., {Livermore}, S., {Kann}, D.~A., {et~al.} 2020,
  \bibinfo{title}{{The Optical Luminosity-Time Correlation for More than 100
  Gamma-Ray Burst Afterglows},} ApJL, 905, L26,
  \dodoi{10.3847/2041-8213/abcda9}

\bibitem[{M.~G. {Dainotti} {et~al.}(2021{\natexlab{a}}){Dainotti}, {Omodei},
  {Srinivasaragavan}, {Vianello}, {Willingale}, {O'Brien}, {Nagataki},
  {Petrosian}, {Nuygen}, {Hernandez}, {Axelsson}, {Bissaldi}, \&
  {Longo}}]{2021ApJS..255...13D}
{Dainotti}, M.~G., {Omodei}, N., {Srinivasaragavan}, G.~P., {et~al.}
  2021{\natexlab{a}}, \bibinfo{title}{{On the Existence of the Plateau Emission
  in High-energy Gamma-Ray Burst Light Curves Observed by Fermi-LAT},} ApJS,
  255, 13, \dodoi{10.3847/1538-4365/abfe17}

\bibitem[{M.~G. {Dainotti} {et~al.}(2021{\natexlab{b}}){Dainotti}, {Bogdan},
  {Narendra}, {Gibson}, {Miasojedow}, {Liodakis}, {Pollo}, {Nelson}, {Wozniak},
  {Nguyen}, \& {Larrson}}]{dainotti2021predicting}
{Dainotti}, M.~G., {Bogdan}, M., {Narendra}, A., {et~al.} 2021{\natexlab{b}},
  \bibinfo{title}{Predicting the Redshift of $\Gamma$-Ray-loud {AGNs} Using
  Supervised Machine Learning,} ApJ, 920, 118, \dodoi{10.3847/1538-4357/ac1748}

\bibitem[{M.~G. {Dainotti} {et~al.}(2022){Dainotti}, {Young}, {Li}, {Levine},
  {Kalinowski}, {Kann}, {Tran}, {Zambrano-Tapia}, {Zambrano-Tapia}, {Cenko},
  {Fuentes}, {S{\'a}nchez-V{\'a}zquez}, {Oates}, {Fraija}, {Becerra}, {Watson},
  {Butler}, {Gonz{\'a}lez}, {Kutyrev}, {Lee}, {Prochaska}, {Ramirez-Ruiz},
  {Richer}, \& {Zola}}]{Dainotti2022}
{Dainotti}, M.~G., {Young}, S., {Li}, L., {et~al.} 2022, \bibinfo{title}{{The
  Optical Two- and Three-dimensional Fundamental Plane Correlations for Nearly
  180 Gamma-Ray Burst Afterglows with Swift/UVOT, RATIR, and the Subaru
  Telescope},} The Astrophysical Journal Supplement Series, 261, 25,
  \dodoi{10.3847/1538-4365/ac7c64}

\bibitem[{M.~G. Dainotti {et~al.}(2024{\natexlab{a}})Dainotti, Taira, Wang,
  Lehman, Narendra, Pollo, Madejski, Petrosian, Bogdan, Dey, \&
  Bhardwaj}]{Dainotti_2024_ApJS}
Dainotti, M.~G., Taira, E., Wang, E., {et~al.} 2024{\natexlab{a}},
  \bibinfo{title}{Inferring the Redshift of More than 150 GRBs with a
  Machine-learning Ensemble Model,} The Astrophysical Journal Supplement
  Series, 271, 22, \dodoi{10.3847/1538-4365/ad1aaf}

\bibitem[{M.~G. Dainotti {et~al.}(2024{\natexlab{b}})Dainotti, Narendra, Pollo,
  Petrosian, Bogdan, Iwasaki, Prochaska, Rinaldi, \&
  Zhou}]{Dainotti2024ApJ...967L..30D}
Dainotti, M.~G., Narendra, A., Pollo, A., {et~al.} 2024{\natexlab{b}},
  \bibinfo{title}{Gamma-Ray Bursts as Distance Indicators by a Statistical
  Learning Approach,} \apjl, 967, L30, \dodoi{10.3847/2041-8213/ad4970}

\bibitem[{M.~G. Dainotti {et~al.}(2025)Dainotti, Bhardwaj, Cook, Ange,
  Lamichhane, Bogdan, McGee, Nadolsky, Sarkar, Pollo, \&
  Nagataki}]{dainotti2025grbredshiftclassifierfollowup}
Dainotti, M.~G., Bhardwaj, S., Cook, C., {et~al.} 2025, \bibinfo{title}{GRB
  Redshift Classifier to Follow up High-redshift GRBs Using Supervised Machine
  Learning,} The Astrophysical Journal Supplement Series, 277, 31,
  \dodoi{10.3847/1538-4365/adafa9}

\bibitem[{D. {de Menezes} {et~al.}(2021){de Menezes}, Prata, Secchi, \&
  Pinto}]{DEMENEZES2021107254}
{de Menezes}, D., Prata, D., Secchi, A., \& Pinto, J. 2021, \bibinfo{title}{A
  review on robust M-estimators for regression analysis,} Computers \& Chemical
  Engineering, 147, 107254,
  \dodoi{https://doi.org/10.1016/j.compchemeng.2021.107254}

\bibitem[{A. {de Ugarte Postigo} {et~al.}(2018){de Ugarte Postigo}, {Cano},
  {Izzo}, {Thoene}, {Kann}, {Castro-Rodriguez}, \&
  {Valladares}}]{2018GCN.22346....1D}
{de Ugarte Postigo}, A., {Cano}, Z., {Izzo}, L., {et~al.} 2018,
  \bibinfo{title}{{GRB 180115A: Redshift from OSIRIS/GTC.},} GRB Coordinates
  Network, 22346, 1

\bibitem[{A. {de Ugarte Postigo} {et~al.}(2014{\natexlab{a}}){de Ugarte
  Postigo}, {Thoene}, {Tanvir}, {Gorosabel}, {Fynbo}, {Lombardi},
  {Reverte-Paya}, \& {Perez}}]{2014GCN.16968....1D}
{de Ugarte Postigo}, A., {Thoene}, C.~C., {Tanvir}, N.~R., {et~al.}
  2014{\natexlab{a}}, \bibinfo{title}{{GRB 141026A: GTC spectroscopy.},} GRB
  Coordinates Network, 16968, 1

\bibitem[{A. {de Ugarte Postigo} {et~al.}(2014{\natexlab{b}}){de Ugarte
  Postigo}, {Xu}, {Gorosabel}, {Malesani}, {Milvang-Jensen}, {Andersen},
  {Jakobsson}, {Fynbo}, \& {Somero}}]{2014GCN.15924....1D}
{de Ugarte Postigo}, A., {Xu}, D., {Gorosabel}, J., {et~al.}
  2014{\natexlab{b}}, \bibinfo{title}{{GRB 140304A: redshift from the NOT.},}
  GRB Coordinates Network, 15924, 1

\bibitem[{A. {Dey} {et~al.}(2019){Dey}, {Schlegel}, {Lang}, {Blum}, {Burleigh},
  {Fan}, {Findlay}, {Finkbeiner}, {Herrera}, {Juneau}, {Landriau}, {Levi},
  {McGreer}, {Meisner}, {Myers}, {Moustakas}, {Nugent}, {Patej}, {Schlafly},
  {Walker}, {Valdes}, {Weaver}, {Y{\`e}che}, {Zou}, {Zhou}, {Abareshi},
  {Abbott}, {Abolfathi}, {Aguilera}, {Alam}, {Allen}, {Alvarez}, {Annis},
  {Ansarinejad}, {Aubert}, {Beechert}, {Bell}, {BenZvi}, {Beutler}, {Bielby},
  {Bolton}, {Brice{\~n}o}, {Buckley-Geer}, {Butler}, {Calamida}, {Carlberg},
  {Carter}, {Casas}, {Castander}, {Choi}, {Comparat}, {Cukanovaite}, {Delubac},
  {DeVries}, {Dey}, {Dhungana}, {Dickinson}, {Ding}, {Donaldson}, {Duan},
  {Duckworth}, {Eftekharzadeh}, {Eisenstein}, {Etourneau}, {Fagrelius},
  {Farihi}, {Fitzpatrick}, {Font-Ribera}, {Fulmer}, {G{\"a}nsicke},
  {Gaztanaga}, {George}, {Gerdes}, {Gontcho}, {Gorgoni}, {Green}, {Guy},
  {Harmer}, {Hernandez}, {Honscheid}, {Huang}, {James}, {Jannuzi}, {Jiang},
  {Joyce}, {Karcher}, {Karkar}, {Kehoe}, {Kneib}, {Kueter-Young}, {Lan},
  {Lauer}, {Le Guillou}, {Le Van Suu}, {Lee}, {Lesser}, {Perreault Levasseur},
  {Li}, {Mann}, {Marshall}, {Mart{\'\i}nez-V{\'a}zquez}, {Martini}, {du Mas des
  Bourboux}, {McManus}, {Meier}, {M{\'e}nard}, {Metcalfe},
  {Mu{\~n}oz-Guti{\'e}rrez}, {Najita}, {Napier}, {Narayan}, {Newman}, {Nie},
  {Nord}, {Norman}, {Olsen}, {Paat}, {Palanque-Delabrouille}, {Peng},
  {Poppett}, {Poremba}, {Prakash}, {Rabinowitz}, {Raichoor}, {Rezaie},
  {Robertson}, {Roe}, {Ross}, {Ross}, {Rudnick}, {Safonova}, {Saha},
  {S{\'a}nchez}, {Savary}, {Schweiker}, {Scott}, {Seo}, {Shan}, {Silva},
  {Slepian}, {Soto}, {Sprayberry}, {Staten}, {Stillman}, {Stupak}, {Summers},
  {Sien Tie}, {Tirado}, {Vargas-Maga{\~n}a}, {Vivas}, {Wechsler}, {Williams},
  {Yang}, {Yang}, {Yapici}, {Zaritsky}, {Zenteno}, {Zhang}, {Zhang}, {Zhou}, \&
  {Zhou}}]{DESI}
{Dey}, A., {Schlegel}, D.~J., {Lang}, D., {et~al.} 2019,
  \bibinfo{title}{{Overview of the DESI Legacy Imaging Surveys},} \aj, 157,
  168, \dodoi{10.3847/1538-3881/ab089d}

\bibitem[{S.~G. {Djorgovski} {et~al.}(2001){Djorgovski}, {Frail}, {Kulkarni},
  {Bloom}, {Odewahn}, \& {Diercks}}]{Djorgovski2001}
{Djorgovski}, S.~G., {Frail}, D.~A., {Kulkarni}, S.~R., {et~al.} 2001,
  \bibinfo{title}{{The Afterglow and the Host Galaxy of the Dark Burst GRB
  970828},} ApJ, 562, 654, \dodoi{10.1086/323845}

\bibitem[{D. {Eichler} {et~al.}(1989){Eichler}, {Livio}, {Piran}, \&
  {Schramm}}]{Eichler1989}
{Eichler}, D., {Livio}, M., {Piran}, T., \& {Schramm}, D.~N. 1989,
  \bibinfo{title}{{Nucleosynthesis, neutrino bursts and
  {\ensuremath{\gamma}}-rays from coalescing neutron stars},} \nat, 340, 126,
  \dodoi{10.1038/340126a0}

\bibitem[{J. Elliott {et~al.}(2014)Elliott {et~al.}}]{Elliott:2013tfa}
Elliott, J., {et~al.} 2014, \bibinfo{title}{{Prompt emission of GRB 121217A
  from gamma-rays to the near-infrared},} Astron. Astrophys., 562, A100,
  \dodoi{10.1051/0004-6361/201322600}

\bibitem[{P.~A. {Evans} {et~al.}(2009){Evans}, {Beardmore}, {Page}, {Osborne},
  {O'Brien}, {Willingale}, {Starling}, {Burrows}, {Godet}, {Vetere}, {Racusin},
  {Goad}, {Wiersema}, {Angelini}, {Capalbi}, {Chincarini}, {Gehrels}, {Kennea},
  {Margutti}, {Morris}, {Mountford}, {Pagani}, {Perri}, {Romano}, \&
  {Tanvir}}]{Evans2009}
{Evans}, P.~A., {Beardmore}, A.~P., {Page}, K.~L., {et~al.} 2009,
  \bibinfo{title}{{Methods and results of an automatic analysis of a complete
  sample of Swift-XRT observations of GRBs},} MNRAS, 397, 1177,
  \dodoi{10.1111/j.1365-2966.2009.14913.x}

\bibitem[{H. {Flores} {et~al.}(2013){Flores}, {Covino}, {de Ugarte Postigo},
  {Fynbo}, {Kruehler}, {Xu}, \& {Tanvir}}]{2013GCN.14493....1F}
{Flores}, H., {Covino}, S., {de Ugarte Postigo}, A., {et~al.} 2013,
  \bibinfo{title}{{GRB 130427B: tentative VLT/X-shooter redshift.},} GRB
  Coordinates Network, 14493, 1

\bibitem[{J.~P.~U. {Fynbo} {et~al.}(2008){Fynbo}, {Malesani}, \&
  {Milvang-Jensen}}]{2008GCN..7949....1F}
{Fynbo}, J.~P.~U., {Malesani}, D., \& {Milvang-Jensen}, B. 2008,
  \bibinfo{title}{{GRB 080707: VLT redshift.},} GRB Coordinates Network, 7949,
  1

\bibitem[{J.~P.~U. {Fynbo} {et~al.}(2001){Fynbo}, {Jensen}, {Gorosabel},
  {Hjorth}, {Pedersen}, {M{\o}ller}, {Abbott}, {Castro-Tirado}, {Delgado},
  {Greiner}, {Henden}, {Magazz{\`u}}, {Masetti}, {Merlino}, {Masegosa},
  {{\O}stensen}, {Palazzi}, {Pian}, {Schwarz}, {Cline}, {Guidorzi}, {Goldsten},
  {Hurley}, {Mazets}, {McClanahan}, {Montanari}, {Starr}, \&
  {Trombka}}]{Fynbo2001}
{Fynbo}, J.~P.~U., {Jensen}, B.~L., {Gorosabel}, J., {et~al.} 2001,
  \bibinfo{title}{{Detection of the optical afterglow of GRB 000630:
  Implications for dark bursts},} A\&A, 369, 373,
  \dodoi{10.1051/0004-6361:20010112}

\bibitem[{J.~P.~U. {Fynbo} {et~al.}(2009){Fynbo}, {Jakobsson}, {Prochaska},
  {Malesani}, {Ledoux}, {de Ugarte Postigo}, {Nardini}, {Vreeswijk},
  {Wiersema}, {Hjorth}, {Sollerman}, {Chen}, {Th{\"o}ne}, {Bj{\"o}rnsson},
  {Bloom}, {Castro-Tirado}, {Christensen}, {De Cia}, {Fruchter}, {Gorosabel},
  {Graham}, {Jaunsen}, {Jensen}, {Kann}, {Kouveliotou}, {Levan}, {Maund},
  {Masetti}, {Milvang-Jensen}, {Palazzi}, {Perley}, {Pian}, {Rol}, {Schady},
  {Starling}, {Tanvir}, {Watson}, {Xu}, {Augusteijn}, {Grundahl}, {Telting}, \&
  {Quirion}}]{2009ApJS..185..526F}
{Fynbo}, J.~P.~U., {Jakobsson}, P., {Prochaska}, J.~X., {et~al.} 2009,
  \bibinfo{title}{{Low-resolution Spectroscopy of Gamma-ray Burst Optical
  Afterglows: Biases in the Swift Sample and Characterization of the
  Absorbers},} \apjs, 185, 526, \dodoi{10.1088/0067-0049/185/2/526}

\bibitem[{N. {Gehrels} {et~al.}(2004){Gehrels}, {Chincarini}, {Giommi},
  {Mason}, {Nousek}, {Wells}, {White}, {Barthelmy}, {Burrows}, {Cominsky},
  {Hurley}, {Marshall}, {M{\'e}sz{\'a}ros}, {Roming}, {Angelini}, {Barbier},
  {Belloni}, {Campana}, {Caraveo}, {Chester}, {Citterio}, {Cline}, {Cropper},
  {Cummings}, {Dean}, {Feigelson}, {Fenimore}, {Frail}, {Fruchter}, {Garmire},
  {Gendreau}, {Ghisellini}, {Greiner}, {Hill}, {Hunsberger}, {Krimm},
  {Kulkarni}, {Kumar}, {Lebrun}, {Lloyd-Ronning}, {Markwardt}, {Mattson},
  {Mushotzky}, {Norris}, {Osborne}, {Paczynski}, {Palmer}, {Park}, {Parsons},
  {Paul}, {Rees}, {Reynolds}, {Rhoads}, {Sasseen}, {Schaefer}, {Short},
  {Smale}, {Smith}, {Stella}, {Tagliaferri}, {Takahashi}, {Tashiro},
  {Townsley}, {Tueller}, {Turner}, {Vietri}, {Voges}, {Ward}, {Willingale},
  {Zerbi}, \& {Zhang}}]{Gehrels2004ApJ...611.1005G}
{Gehrels}, N., {Chincarini}, G., {Giommi}, P., {et~al.} 2004,
  \bibinfo{title}{{The Swift Gamma-Ray Burst Mission},} ApJ, 611, 1005,
  \dodoi{10.1086/422091}

\bibitem[{S.~J. Gibson {et~al.}(2022)Gibson, Narendra, Dainotti, Bogdan, Pollo,
  Poliszczuk, Rinaldi, \& Liodakis}]{gibson2022}
Gibson, S.~J., Narendra, A., Dainotti, M.~G., {et~al.} 2022,
  \bibinfo{title}{Using Multivariate Imputation by Chained Equations to Predict
  Redshifts of Active Galactic Nuclei,} Frontiers in Astronomy and Space
  Sciences, 9, \dodoi{10.3389/fspas.2022.836215}

\bibitem[{E.~S. {Gorbovskoy} {et~al.}(2012){Gorbovskoy}, {Lipunova}, {Lipunov},
  {Kornilov}, {Belinski}, {Shatskiy}, {Tyurina}, {Kuvshinov}, {Balanutsa},
  {Chazov}, {Kuznetsov}, {Zimnukhov}, {Kornilov}, {Sankovich}, {Krylov},
  {Ivanov}, {Chvalaev}, {Poleschuk}, {Konstantinov}, {Gress}, {Yazev},
  {Budnev}, {Krushinski}, {Zalozhnich}, {Popov}, {Tlatov}, {Parhomenko},
  {Dormidontov}, {Senik}, {Yurkov}, {Sergienko}, {Varda}, {Kudelina},
  {Castro-Tirado}, {Gorosabel}, {S{\'a}nchez-Ram{\'\i}rez}, {Jelinek}, \&
  {Tello}}]{2012MNRAS.421.1874G}
{Gorbovskoy}, E.~S., {Lipunova}, G.~V., {Lipunov}, V.~M., {et~al.} 2012,
  \bibinfo{title}{{Prompt, early and afterglow optical observations of five
  {\ensuremath{\gamma}}-ray bursts: GRB 100901A, GRB 100902A, GRB 100905A, GRB
  100906A and GRB 101020A},} MNRAS, 421, 1874,
  \dodoi{10.1111/j.1365-2966.2012.20195.x}

\bibitem[{J. Greiner {et~al.}(2011)Greiner, Kr{\"u}hler, Klose,
  {et~al.}}]{Greiner2011}
Greiner, J., Kr{\"u}hler, T., Klose, S., {et~al.} 2011, \bibinfo{title}{The
  Nature of Dark Gamma-Ray Bursts,} A\&A, 526, A30,
  \dodoi{10.1051/0004-6361/201015458}

\bibitem[{D. Grupe {et~al.}(2006)Grupe, Burrows, Patel, Kouveliotou, Zhang,
  Meszaros, Wijers, \& Gehrels}]{Grupe_2006}
Grupe, D., Burrows, D.~N., Patel, S.~K., {et~al.} 2006, \bibinfo{title}{Jet
  Breaks in Short Gamma‐Ray Bursts. I. The Uncollimated Afterglow of GRB
  050724,} The Astrophysical Journal, 653, 462–467, \dodoi{10.1086/508739}

\bibitem[{T. Hastie \& R. Tibshirani(1987)Hastie \&
  Tibshirani}]{hastie1987generalized}
Hastie, T., \& Tibshirani, R. 1987, \bibinfo{title}{Generalized additive
  models: some applications,} Journal of the American Statistical Association,
  82, 371, \dodoi{10.2307/2289439}

\bibitem[{T. Hastie {et~al.}(2009)Hastie, Tibshirani, \&
  Friedman}]{hastie2009elements}
Hastie, T., Tibshirani, R., \& Friedman, J. 2009, The elements of statistical
  learning: data mining, inference, and prediction (Springer Science \&
  Business Media)

\bibitem[{P.~J. Huber(1964)Huber}]{huber1964}
Huber, P.~J. 1964, \bibinfo{title}{{Robust Estimation of a Location
  Parameter},} The Annals of Mathematical Statistics, 35, 73 ,
  \dodoi{10.1214/aoms/1177703732}

\bibitem[{P.~J. Huber(1996)Huber}]{huber1996robust}
Huber, P.~J. 1996, Robust statistical procedures (SIAM)

\bibitem[{P.~J. Huber \& E.~M. Ronchetti(2009)Huber \&
  Ronchetti}]{huber2009robust}
Huber, P.~J., \& Ronchetti, E.~M. 2009, Robust Statistics, 2nd edn. (Wiley),
  \dodoi{10.1002/9780470434697}

\bibitem[{ {IceCube Collaboration}(2025){IceCube Collaboration}}]{grbweb_cite}
{IceCube Collaboration}. 2025, {GRBweb: A Catalog of Gamma-Ray Bursts},,
  \url{https://user-web.icecube.wisc.edu/~grbweb_public/}

\bibitem[{P. Jakobsson {et~al.}(2006)Jakobsson {et~al.}}]{Jakobsson:2005jc}
Jakobsson, P., {et~al.} 2006, \bibinfo{title}{{A mean redshift of 2.8 for swift
  gamma-ray bursts},} Astron. Astrophys., 447, 897,
  \dodoi{10.1051/0004-6361:20054287}

\bibitem[{D.~A. {Kann} {et~al.}(2021){Kann}, {de Ugarte Postigo}, {Thoene},
  {Blazek}, {Agui Fernandez}, \& {Sota}}]{2021GCN.29296....1K}
{Kann}, D.~A., {de Ugarte Postigo}, A., {Thoene}, C.~C., {et~al.} 2021,
  \bibinfo{title}{{GRB 210112A: OSN Reddened Afterglow Detection},} GRB
  Coordinates Network, 29296, 1

\bibitem[{D.~A. {Kann} {et~al.}(2006){Kann}, {Klose}, \& {Zeh}}]{Kann2006}
{Kann}, D.~A., {Klose}, S., \& {Zeh}, A. 2006, \bibinfo{title}{{Signatures of
  Extragalactic Dust in Pre-Swift GRB Afterglows},} ApJ, 641, 993,
  \dodoi{10.1086/500652}

\bibitem[{M. Karson(1968)Karson}]{doi:10.1080/01621459.1968.11009335}
Karson, M. 1968, \bibinfo{title}{Handbook of Methods of Applied Statistics.
  Volume I: Techniques of Computation Descriptive Methods, and Statistical
  Inference. Volume II: Planning of Surveys and Experiments. I. M. Chakravarti,
  R. G. Laha, and J. Roy, New York, John Wiley; 1967, \$9.00.,} Journal of the
  American Statistical Association, 63, 1047,
  \dodoi{10.1080/01621459.1968.11009335}

\bibitem[{N.~S. Khatiya {et~al.}(2025)Khatiya, Dainotti, Narendra, Bal, Lenart,
  \& Hartmann}]{Khatiya_2025}
Khatiya, N.~S., Dainotti, M.~G., Narendra, A., {et~al.} 2025,
  \bibinfo{title}{Probing Evolution of Long Gamma-Ray Burst Properties through
  Their Cosmic Formation History,} The Astrophysical Journal, 990, 69,
  \dodoi{10.3847/1538-4357/adf219}

\bibitem[{F. Knust {et~al.}(2017)Knust {et~al.}}]{Knust:2017ysj}
Knust, F., {et~al.} 2017, \bibinfo{title}{{Long optical plateau in the
  afterglow of the short GRB 150424A with extended emission - Evidence for
  energy injection by a magnetar?},} Astron. Astrophys., 607, A84,
  \dodoi{10.1051/0004-6361/201730578}

\bibitem[{C. {Kouveliotou} {et~al.}(1993){Kouveliotou}, {Meegan}, {Fishman},
  {Bhat}, {Briggs}, {Koshut}, {Paciesas}, \& {Pendleton}}]{Kouveliotou1993}
{Kouveliotou}, C., {Meegan}, C.~A., {Fishman}, G.~J., {et~al.} 1993,
  \bibinfo{title}{{Identification of Two Classes of Gamma-Ray Bursts},} ApJL,
  413, L101, \dodoi{10.1086/186969}

\bibitem[{T. {Kr{\"u}hler} {et~al.}(2012){Kr{\"u}hler}, {Malesani},
  {Milvang-Jensen}, {Fynbo}, {Hjorth}, {Jakobsson}, {Levan}, {Sparre},
  {Tanvir}, \& {Watson}}]{2012ApJ...758...46K}
{Kr{\"u}hler}, T., {Malesani}, D., {Milvang-Jensen}, B., {et~al.} 2012,
  \bibinfo{title}{{The Optically Unbiased GRB Host (TOUGH) Survey. V.
  VLT/X-shooter Emission-line Redshifts for Swift GRBs at z
  \raisebox{-0.5ex}\textasciitilde 2},} \apj, 758, 46,
  \dodoi{10.1088/0004-637X/758/1/46}

\bibitem[{T. Kr{\"u}hler {et~al.}(2015)Kr{\"u}hler {et~al.}}]{Kruhler:2015ala}
Kr{\"u}hler, T., {et~al.} 2015, \bibinfo{title}{{GRB hosts through cosmic time
  - VLT/X-Shooter emission-line spectroscopy of 96 $\gamma$-ray-burst-selected
  galaxies at 0.1 $< z <$ 3.6},} Astron. Astrophys., 581, A125,
  \dodoi{10.1051/0004-6361/201425561}

\bibitem[{P. {Kumar} \& B. {Zhang}(2015){Kumar} \&
  {Zhang}}]{2015PhR...561....1K}
{Kumar}, P., \& {Zhang}, B. 2015, \bibinfo{title}{{The physics of gamma-ray
  bursts \& relativistic jets},} Physics Reports, 561, 1,
  \dodoi{10.1016/j.physrep.2014.09.008}

\bibitem[{D. {Levine} {et~al.}(2022){Levine}, {Dainotti}, {Zvonarek}, {Fraija},
  {Warren}, {Chandra}, \& {Lloyd-Ronning}}]{2022ApJ...925...15L}
{Levine}, D., {Dainotti}, M., {Zvonarek}, K.~J., {et~al.} 2022,
  \bibinfo{title}{{Examining Two-dimensional Luminosity-Time Correlations for
  Gamma-Ray Burst Radio Afterglows with VLA and ALMA},} ApJ, 925, 15,
  \dodoi{10.3847/1538-4357/ac4221}

\bibitem[{L. {Li} {et~al.}(2018){Li}, {Wu}, {Lei}, {Dai}, {Liang}, \&
  {Ryde}}]{Li2018b}
{Li}, L., {Wu}, X.-F., {Lei}, W.-H., {et~al.} 2018,
  \bibinfo{title}{{Constraining the Type of Central Engine of GRBs with Swift
  Data},} ApJ Supplement, 236, 26, \dodoi{10.3847/1538-4365/aabaf3}

\bibitem[{E.-W. {Liang} {et~al.}(2007){Liang}, {Zhang}, \& {Zhang}}]{Liang2007}
{Liang}, E.-W., {Zhang}, B.-B., \& {Zhang}, B. 2007, \bibinfo{title}{{A
  Comprehensive Analysis of Swift XRT Data. II. Diverse Physical Origins of the
  Shallow Decay Segment},} ApJ, 670, 565, \dodoi{10.1086/521870}

\bibitem[{R.~J. Little \& D.~B. Rubin(2019)Little \&
  Rubin}]{little2019statistical}
Little, R.~J., \& Rubin, D.~B. 2019, Statistical analysis with missing data,
  Vol. 793 (John Wiley \& Sons), \dodoi{10.1002/9781119013563}

\bibitem[{O. {Littlejohns} {et~al.}(2014){Littlejohns}, {Butler}, {Watson},
  {Kutyrev}, {Lee}, {Richer}, {Klein}, {Fox}, {Prochaska}, {Bloom},
  {Cucchiara}, {Troja}, {Ramirez-Ruiz}, {de Diego}, {Georgiev}, {Gonzalez},
  {Roman-Zuniga}, {Gehrels}, \& {Moseley}}]{2014GCN.16050....1L}
{Littlejohns}, O., {Butler}, N., {Watson}, A.~M., {et~al.} 2014,
  \bibinfo{title}{{GRB 140331A: RATIR Optical and NIR Observations.},} GRB
  Coordinates Network, 16050, 1

\bibitem[{D. {Malesani} {et~al.}(2013){Malesani}, {Xu}, {Fynbo}, {Kruehler},
  {Perley}, {Vergani}, \& {Goldoni}}]{2013GCN.14291....1M}
{Malesani}, D., {Xu}, D., {Fynbo}, J.~P.~U., {et~al.} 2013,
  \bibinfo{title}{{GRB 100424A: Keck host detection and VLT/X-shooter
  redshift.},} GRB Coordinates Network, 14291, 1

\bibitem[{A. {Melandri} {et~al.}(2006){Melandri}, {Grazian}, {Guidorzi},
  {Monfardini}, {Mundell}, \& {Gomboc}}]{2006GCN..4539....1M}
{Melandri}, A., {Grazian}, A., {Guidorzi}, C., {et~al.} 2006,
  \bibinfo{title}{{GRB060108: photometric redshift determination.},} GRB
  Coordinates Network, 4539, 1

\bibitem[{A.~N. {Morgan} {et~al.}(2012){Morgan}, {Long}, {Richards},
  {Broderick}, {Butler}, \& {Bloom}}]{morgan2012}
{Morgan}, A.~N., {Long}, J., {Richards}, J.~W., {et~al.} 2012,
  \bibinfo{title}{{Rapid, Machine-learned Resource Allocation: Application to
  High-redshift Gamma-Ray Burst Follow-up},} ApJ, 746, 170,
  \dodoi{10.1088/0004-637X/746/2/170}

\bibitem[{R. {Narayan} {et~al.}(1992){Narayan}, {Paczynski}, \&
  {Piran}}]{1992ApJ...395L..83N}
{Narayan}, R., {Paczynski}, B., \& {Piran}, T. 1992, \bibinfo{title}{{Gamma-ray
  bursts as the death throes of massive binary stars},} ApJL, 395, L83,
  \dodoi{10.1086/186493}

\bibitem[{A. Narendra {et~al.}(2022)Narendra, Gibson, Dainotti, Bogdan, Pollo,
  Liodakis, Poliszczuk, \& Rinaldi}]{narendra2022predicting}
Narendra, A., Gibson, S.~J., Dainotti, M.~G., {et~al.} 2022,
  \bibinfo{title}{Predicting the Redshift of Gamma-Ray Loud {AGNs} Using
  Supervised Machine Learning II,} ApJ Supplement Series, 259, 55,
  \dodoi{10.3847/1538-4365/ac545a}

\bibitem[{A. Narendra {et~al.}(2025)Narendra, Dainotti, Sarkar, Lenart, Bogdan,
  Pollo, Zhang, Rabeda, Petrosian, \& Iwasaki}]{Narendra_2025}
Narendra, A., Dainotti, M.~G., Sarkar, M., {et~al.} 2025,
  \bibinfo{title}{Gamma-ray burst redshift estimation using machine learning
  and the associated web app,} A\&A, 698, A92,
  \dodoi{10.1051/0004-6361/202452651}

\bibitem[{J.~A. Nelder \& R.~W.~M. Wedderburn(1972)Nelder \&
  Wedderburn}]{68aee965-a8a0-3e72-9f89-8d89ae91a62b}
Nelder, J.~A., \& Wedderburn, R. W.~M. 1972, \bibinfo{title}{Generalized Linear
  Models,} Journal of the Royal Statistical Society. Series A (General), 135,
  370.
\newblock \url{http://www.jstor.org/stable/2344614}

\bibitem[{J.~A. {Nousek} {et~al.}(2006){Nousek}, {Kouveliotou}, {Grupe},
  {Page}, {Granot}, {Ramirez-Ruiz}, {Patel}, {Burrows}, {Mangano}, {Barthelmy},
  {Beardmore}, {Campana}, {Capalbi}, {Chincarini}, {Cusumano}, {Falcone},
  {Gehrels}, {Giommi}, {Goad}, {Godet}, {Hurkett}, {Kennea}, {Moretti},
  {O'Brien}, {Osborne}, {Romano}, {Tagliaferri}, \& {Wells}}]{Nousek2006}
{Nousek}, J.~A., {Kouveliotou}, C., {Grupe}, D., {et~al.} 2006,
  \bibinfo{title}{{Evidence for a Canonical Gamma-Ray Burst Afterglow Light
  Curve in the Swift XRT Data},} ApJ, 642, 389, \dodoi{10.1086/500724}

\bibitem[{S.~R. {Oates} {et~al.}(2012){Oates}, {Page}, {De Pasquale}, {Schady},
  {Breeveld}, {Holland}, {Kuin}, \& {Marshall}}]{Oates2012}
{Oates}, S.~R., {Page}, M.~J., {De Pasquale}, M., {et~al.} 2012,
  \bibinfo{title}{{A correlation between the intrinsic brightness and average
  decay rate of Swift/UVOT gamma-ray burst optical/ultraviolet light curves},}
  MNRAS, 426, L86, \dodoi{10.1111/j.1745-3933.2012.01331.x}

\bibitem[{S.~R. Oates {et~al.}(2011)Oates, Page, Schady, De~Pasquale, Evans,
  Page, Chester, Curran, Koch, Kuin, Roming, Siegel, Zane, \&
  Nousek}]{Oates_2011}
Oates, S.~R., Page, M.~J., Schady, P., {et~al.} 2011, \bibinfo{title}{A
  statistical comparison of the optical/UV and X-ray afterglows of gamma-ray
  bursts using the Swift Ultraviolet Optical and X-ray Telescopes: Comparison
  of UVOT \&amp; XRT GRB light curves,} MNRAS, 412, 561–579,
  \dodoi{10.1111/j.1365-2966.2010.17928.x}

\bibitem[{P.~T. {O'Brien} {et~al.}(2006){O'Brien}, {Willingale}, {Osborne},
  {Goad}, {Page}, {Vaughan}, {Rol}, {Beardmore}, {Godet}, {Hurkett}, {Wells},
  {Zhang}, {Kobayashi}, {Burrows}, {Nousek}, {Kennea}, {Falcone}, {Grupe},
  {Gehrels}, {Barthelmy}, {Cannizzo}, {Cummings}, {Hill}, {Krimm},
  {Chincarini}, {Tagliaferri}, {Campana}, {Moretti}, {Giommi}, {Perri},
  {Mangano}, \& {LaParola}}]{OBrien2006}
{O'Brien}, P.~T., {Willingale}, R., {Osborne}, J., {et~al.} 2006,
  \bibinfo{title}{{The Early X-Ray Emission from GRBs},} ApJ, 647, 1213,
  \dodoi{10.1086/505457}

\bibitem[{G. Oganesyan {et~al.}(2019)Oganesyan, Nava, Ghirlanda, Melandri, \&
  Celotti}]{Oganesyan_2019}
Oganesyan, G., Nava, L., Ghirlanda, G., Melandri, A., \& Celotti, A. 2019,
  \bibinfo{title}{Prompt optical emission as a signature of synchrotron
  radiation in gamma-ray bursts,} A\&A, 628, A59,
  \dodoi{10.1051/0004-6361/201935766}

\bibitem[{A. {Panaitescu} \& W.~T. {Vestrand}(2008){Panaitescu} \&
  {Vestrand}}]{panaitescu2008taxonomy}
{Panaitescu}, A., \& {Vestrand}, W.~T. 2008, \bibinfo{title}{{Taxonomy of
  gamma-ray burst optical light curves: identification of a salient class of
  early afterglows},} MNRAS, 387, 497, \dodoi{10.1111/j.1365-2966.2008.13231.x}

\bibitem[{A. {Panaitescu} \& W.~T. {Vestrand}(2011){Panaitescu} \&
  {Vestrand}}]{panaitescu2011optical}
{Panaitescu}, A., \& {Vestrand}, W.~T. 2011, \bibinfo{title}{{Optical
  afterglows of gamma-ray bursts: peaks, plateaus and possibilities},} MNRAS,
  414, 3537, \dodoi{10.1111/j.1365-2966.2011.18653.x}

\bibitem[{D.~A. {Perley} {et~al.}(2014){Perley}, {Cao}, \&
  {Cenko}}]{2014GCN.17228....1P}
{Perley}, D.~A., {Cao}, Y., \& {Cenko}, S.~B. 2014, \bibinfo{title}{{GRB
  141221A: Keck spectroscopy.},} GRB Coordinates Network, 17228, 1

\bibitem[{D.~A. {Perley} \& S.~B. {Cenko}(2015){Perley} \&
  {Cenko}}]{2015GCN.17616....1P}
{Perley}, D.~A., \& {Cenko}, S.~B. 2015, \bibinfo{title}{{GRB 150323A: Keck
  redshift.},} GRB Coordinates Network, 17616, 1

\bibitem[{D.~A. Perley {et~al.}(2009)Perley, Cenko, Bloom,
  {et~al.}}]{Perley2009}
Perley, D.~A., Cenko, S.~B., Bloom, J.~S., {et~al.} 2009, \bibinfo{title}{The
  Hosts of Dark Gamma-Ray Bursts,} Astronomical Journal, 138, 1690,
  \dodoi{10.1088/0004-6256/138/6/1690}

\bibitem[{D.~A. Perley {et~al.}(2013)Perley, Kr{\"u}hler, Schulze,
  {et~al.}}]{Perley2013}
Perley, D.~A., Kr{\"u}hler, T., Schulze, S., {et~al.} 2013, \bibinfo{title}{The
  Swift Gamma-Ray Burst Host Galaxy Legacy Survey,} Astrophysical Journal, 778,
  128, \dodoi{10.1088/0004-637X/778/2/128}

\bibitem[{V. {Petrosian} \& M.~G. {Dainotti}(2024){Petrosian} \&
  {Dainotti}}]{2024ApJ...963L..12P}
{Petrosian}, V., \& {Dainotti}, M.~G. 2024, \bibinfo{title}{{Progenitors of
  Low-redshift Gamma-Ray Bursts},} The Astrophysical Journal Letters, 963, L12,
  \dodoi{10.3847/2041-8213/ad2763}

\bibitem[{L. {Piro} {et~al.}(1998){Piro}, {Amati}, {Antonelli}, {Butler},
  {Costa}, {Cusumano}, {Feroci}, {Frontera}, {Heise}, {in 't Zand}, {Molendi},
  {Muller}, {Nicastro}, {Orlandini}, {Owens}, {Parmar}, {Soffitta}, \&
  {Tavani}}]{Piro1998}
{Piro}, L., {Amati}, L., {Antonelli}, L.~A., {et~al.} 1998,
  \bibinfo{title}{{Evidence for a late-time outburst of the X-ray afterglow of
  GB970508 from BeppoSAX},} A\&A, 331, L41.
\newblock \doarXiv{astro-ph/9710355}

\bibitem[{E.~C. Polley {et~al.}(2011)Polley, Rose, \& van~der
  Laan}]{polley2010super}
Polley, E.~C., Rose, S., \& van~der Laan, M.~J. 2011, \bibinfo{title}{Super
  Learning,} in Targeted Learning: Causal Inference for Observational and
  Experimental Data (New York, NY: Springer New York), 43--66

\bibitem[{J.~X. Prochaska {et~al.}(2007)Prochaska, Chen, Bloom,
  {et~al.}}]{Prochaska2007}
Prochaska, J.~X., Chen, H.-W., Bloom, J.~S., {et~al.} 2007, \bibinfo{title}{The
  Host Galaxies of Gamma-Ray Bursts,} Astrophysical Journal Supplement Series,
  168, 231, \dodoi{10.1086/511026}

\bibitem[{J.~X. Prochaska {et~al.}(2008)Prochaska, Chen, Dessauges-Zavadsky, \&
  Bloom}]{Prochaska2008}
Prochaska, J.~X., Chen, H.-W., Dessauges-Zavadsky, M., \& Bloom, J.~S. 2008,
  \bibinfo{title}{The Galactic Environments of Gamma-Ray Bursts,} Astrophysical
  Journal, 672, 59, \dodoi{10.1086/523701}

\bibitem[{ {R Core Team}(2022){R Core Team}}]{R}
{R Core Team}. 2022, R: A Language and Environment for Statistical Computing, R
  Foundation for Statistical Computing, Vienna, Austria.
\newblock \url{https://www.R-project.org/}

\bibitem[{J.~C. Rastinejad {et~al.}(2022)Rastinejad, Gompertz, Levan, O'Connor,
  Fong, Perley, Troja, Tanvir, Schroeder, \& et~al.}]{rastinejad2022grb211211a}
Rastinejad, J.~C., Gompertz, B.~P., Levan, A.~J., {et~al.} 2022,
  \bibinfo{title}{A kilonova following a long-duration gamma-ray burst at 350
  Mpc,} Nature, 612, 223, \dodoi{10.1038/s41586-022-05465-6}

\bibitem[{A. {Rossi} {et~al.}(2019){Rossi}, {Heintz}, {Fynbo}, {Malesani}, {de
  Ugarte Postigo}, {Vergani}, {Kann}, {Thoene}, {Izzo}, {Campana}, {Pugliese},
  {Kaper}, {Kouveliotou}, {Tanvir}, \& {Stargate
  Collaboration}}]{2019GCN.25252....1R}
{Rossi}, A., {Heintz}, K.~E., {Fynbo}, J.~P.~U., {et~al.} 2019,
  \bibinfo{title}{{GRB 190719C: VLT/X-shooter spectroscopic redshift of the
  host galaxy},} GRB Coordinates Network, 25252, 1

\bibitem[{A. {Rowlinson} {et~al.}(2014){Rowlinson}, {Gompertz}, {Dainotti},
  {O'Brien}, {Wijers}, \& {van der Horst}}]{Rowlinson2014}
{Rowlinson}, A., {Gompertz}, B.~P., {Dainotti}, M., {et~al.} 2014,
  \bibinfo{title}{{Constraining properties of GRB magnetar central engines
  using the observed plateau luminosity and duration correlation},} MNRAS, 443,
  1779, \dodoi{10.1093/mnras/stu1277}

\bibitem[{D.~B. Rubin(1976)Rubin}]{rubin1976inference}
Rubin, D.~B. 1976, \bibinfo{title}{Inference and missing data,} Biometrika, 63,
  581, \dodoi{10.2307/2335739}

\bibitem[{T. {Sakamoto} {et~al.}(2007){Sakamoto}, {Hill}, {Yamazaki},
  {Angelini}, {Krimm}, {Sato}, {Swindell}, {Takami}, \&
  {Osborne}}]{Sakamoto2007}
{Sakamoto}, T., {Hill}, J.~E., {Yamazaki}, R., {et~al.} 2007,
  \bibinfo{title}{{Evidence of Exponential Decay Emission in the Swift
  Gamma-Ray Bursts},} ApJ, 669, 1115, \dodoi{10.1086/521640}

\bibitem[{B. {Sbarufatti} {et~al.}(2018){Sbarufatti}, {Bolmer}, {de Ugarte
  Postigo}, {Fynbo}, {Selsing}, {Heintz}, {Malesani}, {Tanvir}, {Levan},
  {Smette}, {Wiersema}, \& {Covino}}]{2018GCN.22484....1S}
{Sbarufatti}, B., {Bolmer}, J., {de Ugarte Postigo}, A., {et~al.} 2018,
  \bibinfo{title}{{GRB 180314A: VLT/X-shooter redshift.},} GRB Coordinates
  Network, 22484, 1

\bibitem[{P. Schady {et~al.}(2012)Schady, Dwelly, Page, Krühler, Greiner,
  Oates, De~Pasquale, Nardini, Roming, Rossi, \& Still}]{Schady_2012}
Schady, P., Dwelly, T., Page, M.~J., {et~al.} 2012, \bibinfo{title}{The dust
  extinction curves of gamma-ray burst host galaxies,} A\&A, 537, A15,
  \dodoi{10.1051/0004-6361/201117414}

\bibitem[{J.~L. Schafer \& J.~W. Graham(2002)Schafer \&
  Graham}]{schafer2002missing}
Schafer, J.~L., \& Graham, J.~W. 2002, \bibinfo{title}{Missing data: our view
  of the state of the art.,} Psychological methods, 7, 147,
  \dodoi{10.1037/1082-989X.7.2.147}

\bibitem[{S. {Schmidl} {et~al.}(2012){Schmidl}, {Nicuesa Guelbenzu}, {Klose},
  \& {Greiner}}]{2012GCN.13992....1S}
{Schmidl}, S., {Nicuesa Guelbenzu}, A., {Klose}, S., \& {Greiner}, J. 2012,
  \bibinfo{title}{{GRB 121123A: GROND observations.},} GRB Coordinates Network,
  13992, 1

\bibitem[{J. Selsing {et~al.}(2019)Selsing {et~al.}}]{Selsing:2018dwd}
Selsing, J., {et~al.} 2019, \bibinfo{title}{{The X-shooter GRB afterglow legacy
  sample (XS-GRB)},} Astron. Astrophys., 623, A92,
  \dodoi{10.1051/0004-6361/201832835}

\bibitem[{R.-F. Shen \& B. Zhang(2009)Shen \& Zhang}]{Shen_2009}
Shen, R.-F., \& Zhang, B. 2009, \bibinfo{title}{Prompt optical emission and
  synchrotron self-absorption constraints on emission site of GRBs,} MNRAS,
  398, 1936–1950, \dodoi{10.1111/j.1365-2966.2009.15212.x}

\bibitem[{G.~P. {Srinivasaragavan} {et~al.}(2020){Srinivasaragavan},
  {Dainotti}, {Fraija}, {Hernandez}, {Nagataki}, {Lenart}, {Bowden}, \&
  {Wagner}}]{Srinivasaragavan2020}
{Srinivasaragavan}, G.~P., {Dainotti}, M.~G., {Fraija}, N., {et~al.} 2020,
  \bibinfo{title}{{On the Investigation of the Closure Relations for Gamma-Ray
  Bursts Observed by Swift in the Post-plateau Phase and the GRB Fundamental
  Plane},} ApJ, 903, 18, \dodoi{10.3847/1538-4357/abb702}

\bibitem[{N.~R. {Tanvir} {et~al.}(2009){Tanvir}, {Fox}, {Levan}, {Berger},
  {Wiersema}, {Fynbo}, {Cucchiara}, {Kr{\"u}hler}, {Gehrels}, {Bloom},
  {Greiner}, {Evans}, {Rol}, {Olivares}, {Hjorth}, {Jakobsson}, {Farihi},
  {Willingale}, {Starling}, {Cenko}, {Perley}, {Maund}, {Duke}, {Wijers},
  {Adamson}, {Allan}, {Bremer}, {Burrows}, {Castro-Tirado}, {Cavanagh}, {de
  Ugarte Postigo}, {Dopita}, {Fatkhullin}, {Fruchter}, {Foley}, {Gorosabel},
  {Kennea}, {Kerr}, {Klose}, {Krimm}, {Komarova}, {Kulkarni}, {Moskvitin},
  {Mundell}, {Naylor}, {Page}, {Penprase}, {Perri}, {Podsiadlowski}, {Roth},
  {Rutledge}, {Sakamoto}, {Schady}, {Schmidt}, {Soderberg}, {Sollerman},
  {Stephens}, {Stratta}, {Ukwatta}, {Watson}, {Westra}, {Wold}, \&
  {Wolf}}]{Tanvir2009Nature}
{Tanvir}, N.~R., {Fox}, D.~B., {Levan}, A.~J., {et~al.} 2009,
  \bibinfo{title}{{A {\ensuremath{\gamma}}-ray burst at a redshift of
  z\raisebox{-0.5ex}\textasciitilde8.2},} Nature, 461, 1254,
  \dodoi{10.1038/nature08459}

\bibitem[{R. Tibshirani(1996)Tibshirani}]{TibshiraniLasso}
Tibshirani, R. 1996, \bibinfo{title}{Regression Shrinkage and Selection via the
  Lasso,} Journal of the Royal Statistical Society Series B, 58, 267

\bibitem[{T.~N. {Ukwatta} {et~al.}(2016){Ukwatta}, {Wo{\'z}niak}, \&
  {Gehrels}}]{ukwatta2016machine}
{Ukwatta}, T.~N., {Wo{\'z}niak}, P.~R., \& {Gehrels}, N. 2016,
  \bibinfo{title}{{Machine-z: rapid machine-learned redshift indicator for
  Swift gamma-ray bursts},} MNRAS, 458, 3821, \dodoi{10.1093/mnras/stw559}

\bibitem[{S. van Buuren \& K. Groothuis-Oudshoorn(2011)van Buuren \&
  Groothuis-Oudshoorn}]{van2011mice}
van Buuren, S., \& Groothuis-Oudshoorn, K. 2011, \bibinfo{title}{mice:
  Multivariate Imputation by Chained Equations in R,} JSS, 45, 1–67,
  \dodoi{10.18637/jss.v045.i03}

\bibitem[{M.~J. van~der Laan {et~al.}(2007)van~der Laan, Polley, \&
  Hubbard}]{vanderLaanPolleyHubbard+2007}
van~der Laan, M.~J., Polley, E.~C., \& Hubbard, A.~E. 2007,
  \bibinfo{title}{SuperLearner,} Statistical Applications in Genetics and
  Molecular Biology, 6, \dodoi{doi:10.2202/1544-6115.1309}

\bibitem[{J. {van Paradijs} {et~al.}(1997){van Paradijs}, {Groot}, {Galama},
  {Kouveliotou}, {Strom}, {Telting}, {Rutten}, {Fishman}, {Meegan}, {Pettini},
  {Tanvir}, {Bloom}, {Pedersen}, {N{\o}rdgaard-Nielsen}, {Linden-V{\o}rnle},
  {Melnick}, {Van der Steene}, {Bremer}, {Naber}, {Heise}, {in't Zand},
  {Costa}, {Feroci}, {Piro}, {Frontera}, {Zavattini}, {Nicastro}, {Palazzi},
  {Bennett}, {Hanlon}, \& {Parmar}}]{vanParadijs1997}
{van Paradijs}, J., {Groot}, P.~J., {Galama}, T., {et~al.} 1997,
  \bibinfo{title}{{Transient optical emission from the error box of the
  {\ensuremath{\gamma}}-ray burst of 28 February 1997},} Nature, 386, 686,
  \dodoi{10.1038/386686a0}

\bibitem[{W.~N. Venables \& B.~D. Ripley(2002)Venables \& Ripley}]{MASS}
Venables, W.~N., \& Ripley, B.~D. 2002, Modern Applied Statistics with S, 4th
  edn. (New York: Springer).
\newblock \url{https://www.stats.ox.ac.uk/pub/MASS4/}

\bibitem[{W.~T. {Vestrand} {et~al.}(2005){Vestrand}, {Wozniak}, {Wren},
  {Fenimore}, {Sakamoto}, {White}, {Casperson}, {Davis}, {Evans}, {Galassi},
  {McGowan}, {Schier}, {Asa}, {Barthelmy}, {Cummings}, {Gehrels}, {Hullinger},
  {Krimm}, {Markwardt}, {McLean}, {Palmer}, {Parsons}, \&
  {Tueller}}]{Vestrand2005Natur}
{Vestrand}, W.~T., {Wozniak}, P.~R., {Wren}, J.~A., {et~al.} 2005,
  \bibinfo{title}{{A link between prompt optical and prompt
  {\ensuremath{\gamma}}-ray emission in {\ensuremath{\gamma}}-ray bursts},}
  Nature, 435, 178, \dodoi{10.1038/nature03515}

\bibitem[{W.~T. {Vestrand} {et~al.}(2014){Vestrand}, {Wren}, {Panaitescu},
  {Wozniak}, {Davis}, {Palmer}, {Vianello}, {Omodei}, {Xiong}, {Briggs},
  {Elphick}, {Paciesas}, \& {Rosing}}]{2014Sci...343...38V}
{Vestrand}, W.~T., {Wren}, J.~A., {Panaitescu}, A., {et~al.} 2014,
  \bibinfo{title}{{The Bright Optical Flash and Afterglow from the Gamma-Ray
  Burst GRB 130427A},} Science, 343, 38, \dodoi{10.1126/science.1242316}

\bibitem[{S.~E. {Woosley} {et~al.}(1993){Woosley}, {Langer}, \&
  {Weaver}}]{Woosley1993}
{Woosley}, S.~E., {Langer}, N., \& {Weaver}, T.~A. 1993, \bibinfo{title}{{The
  Evolution of Massive Stars Including Mass Loss: Presupernova Models and
  Explosion},} ApJ, 411, 823, \dodoi{10.1086/172886}

\bibitem[{E. {Zaninoni} {et~al.}(2013){Zaninoni}, {Bernardini}, {Margutti},
  {Oates}, \& {Chincarini}}]{Zaninoni2013}
{Zaninoni}, E., {Bernardini}, M.~G., {Margutti}, R., {Oates}, S., \&
  {Chincarini}, G. 2013, \bibinfo{title}{{Gamma-ray burst optical light-curve
  zoo: comparison with X-ray observations},} A\&A, 557, A12,
  \dodoi{10.1051/0004-6361/201321221}

\bibitem[{A. {Zeh} {et~al.}(2006){Zeh}, {Klose}, \& {Kann}}]{Zeh2006}
{Zeh}, A., {Klose}, S., \& {Kann}, D.~A. 2006, \bibinfo{title}{{Gamma-Ray Burst
  Afterglow Light Curves in the Pre-Swift Era: A Statistical Study},} ApJ, 637,
  889, \dodoi{10.1086/498442}

\bibitem[{B. Zhang(2014)Zhang}]{2014IJMPD..2330002Z}
Zhang, B. 2014, \bibinfo{title}{GAMMA-RAY BURST PROMPT EMISSION,} International
  Journal of Modern Physics D, 23, 1430002, \dodoi{10.1142/S021827181430002X}

\bibitem[{B. Zhang(2018)Zhang}]{zhang_2018}
Zhang, B. 2018, The Physics of Gamma-Ray Bursts (Cambridge University Press),
  \dodoi{10.1017/9781139226530}

\bibitem[{B. {Zhang} {et~al.}(2006){Zhang}, {Fan}, {Dyks}, {Kobayashi},
  {M{\'e}sz{\'a}ros}, {Burrows}, {Nousek}, \& {Gehrels}}]{Zhang2006}
{Zhang}, B., {Fan}, Y.~Z., {Dyks}, J., {et~al.} 2006, \bibinfo{title}{{Physical
  Processes Shaping Gamma-Ray Burst X-Ray Afterglow Light Curves: Theoretical
  Implications from the Swift X-Ray Telescope Observations},} ApJ, 642, 354,
  \dodoi{10.1086/500723}

\bibitem[{B. {Zhang} \& P. {M{\'e}sz{\'a}ros}(2004){Zhang} \&
  {M{\'e}sz{\'a}ros}}]{Zhang2004IJMPA..19.2385Z}
{Zhang}, B., \& {M{\'e}sz{\'a}ros}, P. 2004, \bibinfo{title}{{Gamma-Ray Bursts:
  progress, problems \& prospects},} Int J Mod Phys A, 19, 2385,
  \dodoi{10.1142/S0217751X0401746X}

\end{thebibliography}
\bibliographystyle{aasjournalv7}

\end{document}